\documentclass[a4paper,11pt]{article}
\pdfoutput=1 

\usepackage{jcappub} 

\usepackage[T1]{fontenc} 

\usepackage{feynmp-auto,expdlist}
\usepackage{amsmath, amsfonts, amssymb}
\usepackage{graphicx}
\usepackage{enumerate}
\usepackage{hyperref}
\usepackage{latexsym}
\usepackage{dsfont}
\usepackage{hepnicenames}
\usepackage{enumerate}
\usepackage{soul}
\usepackage[normalem]{ulem}
\usepackage{bbold}
\usepackage[utf8]{inputenc}
\usepackage{float}
\usepackage{csquotes}

\usepackage{units}

\newcommand{\Sp}{\,{\rm Sp}}

\newcommand{\SU}{SU}
\newcommand{\SO}{\,{\rm SO}}

\newcommand{\dd}{\mathrm{d}}

\usepackage{cleveref}
\usepackage{float}
\usepackage{multirow}
\usepackage[normalem]{ulem}
\usepackage{comment}
\usepackage{slashed}
\usepackage{pifont}

\usepackage[dvipsnames]{xcolor}

\title{Heavy Dark Baryons as Self-Interacting Dark Matter: \\A GeV-Scale Coincidence}

\author[1,\,2]{Giovani Dalla Valle Garcia}
\author[3,\,4]{and Juan Herrero-Garc\'{\i}a}

\affiliation[1]{ARC Centre of Excellence for Dark Matter Particle Physics,
School of Physics, The University of Melbourne, Victoria 3010, Australia}

\affiliation[2]{Institut für Astroteilchen Physik, Karlsruher Institut für Technologie (KIT), Hermann-von-\\Helmholtz-Platz 1, 76344 Eggenstein-Leopoldshafen, Germany}

\affiliation[3]{Departament de Física Teòrica, Universitat de València, 46100 Burjassot, Spain}

\affiliation[4]{Instituto de Física Corpuscular (CSIC-Universitat de València),
Parc Científic UV, C/Catedrático José Beltrán, 2, E-46980 Paterna, Spain}

\emailAdd{giovani.dallavallegarcia@unimelb.edu.au}
\emailAdd{juan.herrero@ific.uv.es}

\abstract{
We study self-interacting dark matter composed of heavy baryons in a
confining dark $\SU(N)$ gauge theory with a single heavy vector-like
quark, $m_Q \gg \Lambda_d$. In this regime, the dark baryons are compact
Coulombic bound states whose long-distance interactions can be described
by a multipole expansion. We derive the leading baryon-baryon
interaction by adapting the quarkonium operator-product-expansion
formalism to heavy baryons. The resulting force contains non-retarded
London dispersion, retarded Casimir--Polder, and Yukawa contributions
from dark glueball exchange. We find that the self-interaction
phenomenology depends strongly on the number of colors. For $N=2,3$,
the attractive interaction is too weak to overcome the short-distance
Pauli repulsion, and no phenomenologically relevant velocity dependence
is obtained within our minimal set-up. For sufficiently large $N$, the attractive interaction remains competitive
with the repulsive core and can generate strong velocity dependent cross sections. Taking $\SU(10)$ as a representative
finite-$N$ realization, we find regions compatible with self-interaction
requirements inferred from dwarf galaxies and clusters, while allowing much
larger cross sections at low velocities relevant for gravothermal evolution. These regions correspond to GeV-scale dark baryons and a
confinement scale close to that of QCD, which naturally motivate
an asymmetric dark matter origin.
}

\begin{document}
	\maketitle
	\flushbottom

\section{Introduction}

Self-interacting dark matter (SIDM) provides a well-motivated framework for addressing possible discrepancies between collisionless cold dark matter (CDM) simulations and the observed structure of galaxies on small scales~\cite{Spergel:1999mh,Tulin:2017ara,Bullock:2017xww}. Scattering cross sections of order
\[
    \sigma/m_{\rm DM} \simeq 1-10~{\rm cm}^2/{\rm g}
\]
at dwarf-galaxy velocities $v_{\rm dwarf} \sim50$~km/s can modify halo cores and
substructure~\cite{Kaplinghat:2015aga,Tulin:2017ara}, while observations of
galaxy clusters require substantially smaller cross sections at velocities of
order \(v_{\rm cluster} \sim10^3~{\rm km/s}\)~\cite{Kaplinghat:2015aga,Sagunski:2020spe,Andrade:2020lqq}.
As a consequence, this separation of scales motivates particle-physics models in which
the dark matter (DM) self-interaction is both sizable and velocity
dependent~\cite{Tulin:2013teo,Tulin:2017ara}.

New confining sectors offer a natural origin for sizable DM self-interactions and arise in a variety of approaches to addressing outstanding problems of the Standard Model (SM), including the (flavor) hierarchy and strong CP problems~\cite{Asadi:2026mip,Bellazzini:2014yua,Witzel:2019jbe,Frigerio:2012uc,Foadi:2008qv,Csaki:2025ikr,Agrawal:2025mke,Cacciapaglia:2026yvm,Csaki:2008qq,Arkani-Hamed:1999ylh,Huber:2000ie,Kaplan:1991dc,Agashe:2025tge}. They also arise in mechanisms addressing the baryon--DM coincidence~\cite{Murgui:2021eqi,Chung:2024dgu} and in string-theory constructions such as hidden-valley scenarios~\cite{Strassler:2006im,Cvetic:2002qa,Arkani-Hamed:2005zuc}. In these theories, DM can be realized as a composite state whose interactions are set dynamically by the confinement scale, $\Lambda_d$, rather than by introducing an elementary light mediator. If DM is composed of dark glueballs, however, their self-interactions are set primarily by the confinement scale and are therefore approximately velocity independent~\cite{Boddy:2014yra,Yamanaka:2019gak,Yamanaka:2019yek}. Such interactions make it difficult to obtain the strong velocity dependence needed to affect dwarf-galaxy halos while remaining compatible with cluster constraints~\cite{Boddy:2014yra,Boddy:2014qxa}. Dark baryons and mesons containing light dark quarks~\cite{Hochberg:2014dra,Hochberg:2014kqa,Chu:2018fzy,Kribs:2016cew} can instead exhibit a richer velocity dependence, but typically rely on considering multiple light dark flavors and resonant dynamics to obtain the required phenomenology~\cite{Cline:2022leq,Garcia-Cely:2024ivo,Figueroa:2026bmi}.

In this work we explore a complementary minimal possibility: a confining \(\SU(N)\) gauge theory with a single heavy vector-like quark of mass
\[
    m_Q \gg \Lambda_d .
\]
In this regime, the stable DM candidate is a heavy dark baryon protected by dark baryon number conservation~\cite{Mitridate:2017oky,Assi:2023cfo}. Dark mesons  and glueballs, by contrast, are  generally not protected by a symmetry and may decay to the visible sector. Even if glueballs are cosmologically long-lived~\cite{Forestell:2016qhc,McKeen:2024trt}, a sufficiently cold dark sector can favor the heavier baryonic component as the dominant DM abundance. We therefore focus on the case in which the DM is composed of compact heavy dark baryons.

\begin{figure}
    \centering
    \includegraphics[width=0.85\linewidth]{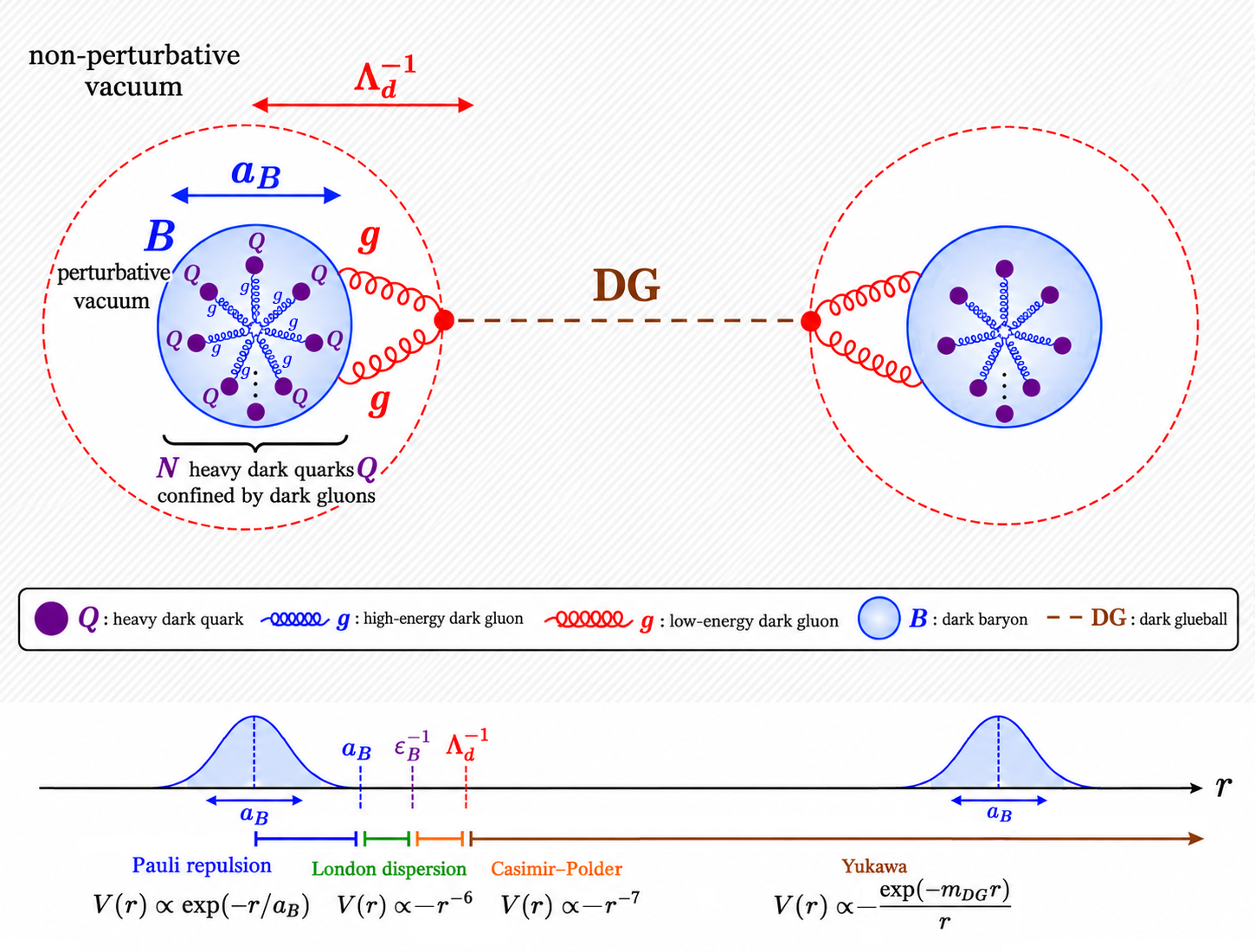}
   \caption{Illustration of the interaction between two dark baryons, each composed of $N$ confined heavy quarks $Q$. Heavy dark baryons are compact color-singlet bound states whose long-distance interactions arise from induced chromoelectric dipoles and Pauli repulsion at small scales. The corresponding long-range potential contains non-retarded London, retarded Casimir--Polder, and glueball-mediated Yukawa regimes, depending on the baryon separation \(r\).   }
    \label{fig:SIDM}
\end{figure}

The hierarchy \(m_Q \gg \Lambda_d\) makes the baryons Coulombic at short
distances and allows their long-distance interactions to be described by (chromo) Van der Waals forces in a multipole expansion. This setup is closely analogous to heavy quarkonium in QCD, where the leading interaction between color-singlet bound states is governed by
induced chromoelectric dipoles, as illustrated in
Fig.~\ref{fig:SIDM}~\cite{Peskin:1979va,Bhanot:1979vb,Fujii:1999xn}. We adapt the
quarkonium operator-product-expansion formalism~\cite{Peskin:1979va,Brambilla:2004jw}
to heavy dark baryons and derive the corresponding baryon--baryon potential. The
resulting force differs qualitatively from standard Yukawa-mediated
SIDM~\cite{Tulin:2013teo}: it is generated by the composite structure of the baryons and contains several distance regimes. In particular, the potential includes a non-retarded  London dispersive component at short distances, a perturbative Casimir--Polder contribution at distances below the
confinement length~\cite{Casimir:1947hx,Fujii:1999xn}, and a non-perturbative
Yukawa tail associated with the exchange of the lightest dark
glueballs.\footnote{For a modern discussion on Van der Waals forces also including the QCD case, see Ref.~\cite{Brambilla:2017ffe}. } Moreover, a repulsive potential due to Pauli
blocking must be included at distances of the size of the
baryon~\cite{Cline:2013pca,Assi:2025ysr}.

{We then compute the DM self-interaction cross section as a function of
velocity and determine how the resulting phenomenology depends on the
number of colors $N$. We find a qualitative distinction between small and
large $N$. For $\SU(2)$ and $\SU(3)$, the glueball-mediated attraction
is too weak to compete with the short-distance Pauli repulsion in the
minimal realization considered here, even for optimistic choices of
the short-distance suppression scale. Consequently, these theories do
not exhibit phenomenologically relevant velocity dependence. For
sufficiently large $N\sim 10$, in contrast, the attractive interaction remains
competitive with the repulsive core, and the velocity dependence
approaches a qualitative behavior which is approximately independent of $N$. For $N\gg10$, the
repulsive interaction eventually dominates again because
$V_{\rm Pauli}\propto N$ while $V_{\rm attractive}\propto N^0$ (for large $N$)
~\cite{Adhikari:2013dfa}. We therefore use $\SU(10)$ as a representative example of this large-$N$ behavior and find regions of parameter space that can
simultaneously satisfy the self-interaction requirements at dwarf,
galaxy, and cluster scales. These regions
correspond to GeV-scale dark baryon masses and confinement scales close
to the QCD one, $\Lambda_{\rm QCD}$.

We further investigate whether the same velocity dependence can accommodate the extremely large self-interaction cross sections potentially required for gravothermal core collapse, focusing, in particular, on the second perturber of the strong gravitational lens system JVAS B1938+666~\cite{Powell:2025rmj,Vegetti:2026mmx,Zhang:2026gur}. We find regions of parameter space that simultaneously satisfy constraints at galaxies and clusters while reaching $\mathcal{O}(100)~\mathrm{cm}^2/\mathrm{g}$ at $v\sim5~\mathrm{km/s}$, with the corresponding dark baryon and confinement scales again lying close to the ordinary baryon and QCD scales, respectively.  The specific study of self-interactions of heavy dark baryons in the large $N$ regime as a possible explanation of astrophysical observations of gravothermal core collapse halos is presented in the companion letter \cite{DallaValleGarcia:2026letter}.}

Finally, we discuss the cosmological production of these states and show that, in the parameter region favored by self-interactions, symmetric thermal freeze-out is typically insufficient. This points instead towards asymmetric production with a colder dark sector or unstable dark glueballs~\cite{Kaplan:2009ag,Petraki:2013wwa,Zurek:2013wia}.

{
A key difference from previous treatments of self-interactions in confining
dark sectors~\cite{Boddy:2014qxa,Boddy:2014yra} concerns the strength of the
glueball-mediated interaction. Those works model the glueball exchange with a
Yukawa potential and an $\mathcal{O}(1)$ coupling, motivated by the strong
coupling of the confining theory. In the heavy-quark regime considered here,
however, the coupling to glueballs is determined by the chromo-electric
polarizability of the compact baryon and is strongly suppressed. The resulting
Yukawa tail is therefore much weaker than in these earlier estimates. This
difference has important consequences for the resulting self-interaction
rates and, consequently, for the viable dark-matter mass scale and its
cosmological interpretation: in the phenomenologically relevant region
identified here, the dark matter consists of GeV-scale dark baryons with an
asymmetric origin, rather than the TeV-scale thermal relic scenario found
in those earlier treatments.
}

The rest of the paper is organized as follows. In \cref{sec:confining_ds} we introduce the confining dark sector, describe the heavy-baryon bound states, and identify the hierarchy of scales for which the Coulombic description is reliable. In \cref{sec:interaction_potential} we derive the baryon-baryon interaction potential and discuss its  London dispersive, Casimir--Polder, and glueball-mediated regimes, including the treatment of the short-distance regularization. In \cref{sec:self_interactions} we present the self-interaction formalism, the astrophysical targets, and the numerical method used to compute the viscosity cross section. In \cref{sec:results} we present the numerical self-interaction
results, emphasizing the distinction between the small-$N$ and
large-$N$ regimes and the regions of the large-$N$ parameter space
that yield phenomenologically relevant velocity dependence, including the possibility of gravothermal core collapse and the application to the second perturber of JVAS B1938+666. In \cref{sec:asymmetric_DM} we discuss the cosmological production of heavy dark baryons and the removal or dilution of dark glueballs, showing that standard symmetric freeze-out is generally insufficient in the phenomenologically preferred region. {We also briefly discuss the possibility of an observable nanohertz gravitational-wave signal from the confinement transition in light of the NANOGrav results~\cite{NANOGrav:2023gor}.} We conclude in \cref{sec:conclusion}.

Further technical details are collected in several appendices. In \cref{app:color-contraction} we derive the relevant color factors of the dipole operator for heavy baryons. \Cref{app:highern} discusses the transition from the non-retarded potential to the retarded regime, \cref{app:dipole-transition-integration} presents the calculation of the chromo-electric polarizability, and \cref{app:potentail_sum} justifies the potential-summation approximation. Although the main application of our results is to dark baryons, the chromo-electric polarizability derived in  \cref{app:dipole-transition-integration} also applies to heavy Coulombic  $SU(N)$ mesons, for which our two-body treatment is indeed exact. {\it We obtain in this case a closed analytical expression valid for arbitrary finite $N$.} The result can be generalized  to the two-quark singlet states of $SO(N)$ and $Sp(2N)$, making the formalism we developed also applicable to these systems.

\section{Confining Dark Sectors}\label{sec:confining_ds}

We consider DM arising from a confining hidden sector based on a non-Abelian gauge group $G_d$. Throughout this work we focus on
\begin{equation}
    G_d=\SU(N)\,,
\end{equation} both for simplicity and because these choices admit useful comparisons with known results from QCD.\footnote{A similar analysis could be performed for other confining gauge groups, such as $\SO(N)$ or $\Sp(N)$.}

The dark sector contains gauge bosons $A_\mu^a$ with gauge coupling $g_d$, and a single vector-like fermion $Q$ in the fundamental representation of $G_d$. The dark fermion is assumed to be neutral under the SM gauge group and has mass $m_Q$. We define the dark fine-structure constant as
\begin{equation}
    \alpha_d \equiv \frac{g_d^2}{4\pi}\,.
\end{equation}

At low energies the gauge coupling becomes strong, leading to confinement at a scale $\Lambda_d$ defined by the condition $\alpha_d(\Lambda_d)\gtrsim 1$. In this work we are interested in the heavy-quark regime,
\begin{equation}
    m_Q \gg \Lambda_d\,,
\end{equation}
for which the resulting dark baryons are compact compared to the confinement scale. In this limit non-perturbative effects are largely excluded from the baryon interior, allowing the bound states to be described using perturbative methods. A more quantitative criterion will be given below once the dark baryons are modeled explicitly.

The running of the dark gauge coupling is governed by
\begin{equation}
\frac{d\alpha_d}{d \ln\mu_s}
=
-\frac{\alpha_d^2}{2\pi}\beta_d\,,
\end{equation}
where   $\mu_s$ is an energy scale. At one loop, the beta-function coefficient (with only fermions) is
\begin{equation}
\beta_d=
\frac{11}{3}C_2(A)
-\frac{4}{3}\sum_F \kappa T_f
\,,
\end{equation}
where $C_2(A)$ is the quadratic Casimir of the adjoint representation, $T_f$ denotes the Dynkin index of a fermionic ($f$) representation (including multiplicity factors such as the number of flavors $n_f$) and $\kappa=1~(1/2)$ for Dirac (Weyl) fermions. The relevant group-theory factors for $\SU(N)$ are collected in Table~\ref{tab:groupth}. The values shown correspond to a single field in the indicated representation; for $n_f$ flavors one simply multiplies $T_f$ by $n_f$. In our case, $n_f=1$ for energy scales well above $m_Q$ and $n_f=0$ when $\mu_s\ll m_Q$ where $Q$ can be integrated out of the theory.

\begin{table}[t]
\begin{equation} \nonumber
\begin{array}{c|c|c|c}
\text{Group} & C_2(A) & C_2(F)\equiv C_N & T_f \\[0.2ex] \hline \rule{0pt}{4.2ex}

 \SU(N) & N & \dfrac{N^2-1}{2N} & \dfrac12 
\end{array}
\end{equation}
\caption{Group-theory factors relevant for $\SU(N)$ with additional fields in the fundamental representation $F$.}
\label{tab:groupth}
\end{table}

Integrating the renormalization-group equation yields the confinement scale
\begin{equation}
\Lambda_d
\simeq
\mu_0
\exp\!\left[
-\frac{2\pi}{\beta_d\,\alpha_d(\mu_0)}
\right],
\end{equation}
where $\mu_0$ is any energy scale at which the coupling remains perturbative.

Since our dark sector contains only a single vector-like fermion with mass $m_Q>\Lambda_d$, the dark quark decouples for scales $\mu_0\lesssim m_Q$. The low-energy theory is therefore pure Yang--Mills, for which
\begin{equation}
\beta_d\equiv b=\frac{11}{3}N\,.
\end{equation}
Consequently, once $\Lambda_d$ and the number of colors $N$ are specified, the running coupling at scales
\begin{equation}
\Lambda_d < \mu_0 < m_Q
\end{equation}
is completely determined,
\begin{equation}
\label{eq:alpha_per_energy}
\alpha_d(\mu_0)
=
\frac{2\pi}
{b\ln\!\left(\mu_0/\Lambda_d\right)}\,.
\end{equation}

It is convenient to characterize the hierarchy between the dark quark mass and the confinement scale through the dimensionless parameter
\begin{equation}
    \xi \equiv \frac{m_Q}{\Lambda_d}\,.
\end{equation}
As will become apparent below, many properties of the dark baryons and their self-interactions depend primarily on $\xi$, making it the key parameter controlling the phenomenology of the heavy-quark regime.

\subsection{The bound states}

Below the confinement scale $\Lambda_d$, the elementary degrees of freedom are no longer the dark gluons and quarks, but a spectrum of color-singlet bound states. The dark gluons form  dark glueballs (DG), while the heavy dark quarks bind into mesons and baryons. The lightest glueball has a mass~\cite{McKeen:2024trt}
\begin{equation}
    m_{\rm DG}\approx 6.4\,\Lambda_d\,,
\end{equation}
according to lattice simulations for $\SU(2)$ and $\SU(3)$~\cite{Teper:1998kw,Gockeler:2005rv}, with small corrections of order $\mathcal{O}(1/N^2)$ for larger values of $N$~\cite{Forestell:2016qhc,Teper:1998kw,Lucini:2012gg,McKeen:2024trt}. Heavier glueball states rapidly decay or are depleted through glueball number-changing processes (``cannibalism''), leaving the lightest state as the dominant glueball relic. Throughout this work, the term ``glueball'' therefore refers to the lightest glueball unless otherwise stated.

Following confinement, the dark quarks are likewise bound into color-singlet hadrons. The two relevant classes of states are mesons $\bar{Q}Q$ and baryons $NQ$. Dark baryons are automatically stable as a consequence of an accidental global $U(1)$ symmetry corresponding to dark baryon number. In contrast, dark mesons (M) generally carry no conserved quantum number and are therefore unstable.\footnote{For $\SU(2)$ with fundamental fermions, the pseudo-real representation enlarges the global symmetry to $\SU(2)_B$, under which the baryon, antibaryon, and a vector meson form a triplet. The corresponding vector mesons are therefore stable in the absence of $\SU(2)_B$-breaking interactions~\cite{Francis:2018xjd}. In the parameter space of interest here, however, quark and anti-quarks efficiently annihilate in the early Universe and the meson abundance is therefore negligible in the present-day Universe.} In the heavy-quark regime,
\begin{equation}
    m_M \approx 2m_Q > 2m_{\rm DG} \gtrsim 13\,\Lambda_d\,,
\end{equation}
allowing mesons to efficiently decay into glueballs.  Moreover, even if stable, in an asymmetric DM scenario (which will be preferred by small-structure observations),  quark and anti-quark pairs efficiently annihilate  leaving a negligible amount of relic mesons.

Dark baryons therefore emerge as the natural DM candidate of the theory.\footnote{Higher-dimensional operators can in principle violate dark baryon number and induce DM decay. However,  such operators first arise only at dimension six or higher \cite{Francis:2018xjd,Appelquist:2015yfa}.   } By contrast, mesons and glueballs are not generically protected by a stabilizing symmetry and can therefore decay through portals connecting the dark and visible sectors. Depending on the portal structure and its suppression scale, however, some of these states can be sufficiently long-lived to be cosmologically relevant.  Even in scenarios where glueballs are cosmologically stable, the larger mass of the baryons typically implies that they dominate the DM abundance in sufficiently cold dark sectors.

Before proceeding, it is worth commenting on the possibility of glueball DM  as an explanation of small-scale structure problems. Since glueball self-interactions are governed by the confinement scale alone, they are expected to exhibit only a mild velocity dependence~\cite{Boddy:2014yra,Boddy:2014qxa,Yamanaka:2019gak,Yamanaka:2019yek}. As a result, they do not readily realize the significant velocity-dependent self-interactions favored by astrophysical observations. Heavy baryons, on the other hand, are compact composite objects whose interactions arise from induced chromo-electric dipole forces. As we shall see, this leads to a much richer velocity dependence of the self-interaction cross section. For these reasons, we focus on dark baryons as the DM candidate in the remainder of this work.

\subsubsection{Baryon properties}
\label{sec:2p1p1_baryon_properties}

Throughout this work, we assume that the DM is entirely composed of the lightest dark baryon. This state consists of $N$ identical quarks (since there is only one flavour) in a color-singlet configuration. The ground state has vanishing orbital angular momentum, resulting in a symmetric spatial wave function. The color wave function of a singlet is fully antisymmetric, and therefore---since we consider a single quark flavor---Fermi statistics require the spin wave function to be fully symmetric, yielding a baryon spin of $N/2$.\footnote{The fully symmetric product of $N$ spin doublets has dimension $N + 1$. A spin-$S$ multiplet has dimension $2S + 1$. Therefore, $2S + 1 = N + 1$, so $S = N/2$.} We will consider the benchmark cases \(N=2,10\), corresponding to bosonic dark baryons, and \(N=3\), corresponding to fermionic dark baryons.\footnote{For identical-particle scattering, the cross section must be computed with the appropriate quantum-statistical symmetrization, which may enhance the bosonic case relative to the fermionic one by an \(\mathcal{O}(1)\)  factor.}

Neglecting non-perturbative effects, the baryon can be described in the non-relativistic limit by a Hamiltonian consisting of the kinetic energy of the constituent quarks together with the pairwise chromo-Coulomb interactions~\cite{Mitridate:2017oky},
\begin{equation}
    H=\sum_{i=1}^N K_i+\sum_{i<j}^NV(r_{ij})\,,
\end{equation}
where $r_{ij}$ is the distance between quarks $i$ and $j$, and the kinetic energy of quark $i$ is given by $K_i=p_i^2/(2m_Q)$. For two quarks inside a color-singlet baryon, the Coulomb potential is given by~\cite{Zhou:2025fpp,Brambilla:2009cd}
\begin{equation}
    V_s(r_{ij})=
    -\frac{C_N}{N-1}\frac{\alpha_d}{r_{ij}}\,.
\end{equation}
For comparison, the potential between a quark and antiquark in a singlet meson is
$V_{s,M}(r)=-C_N{\alpha_d}/{r}$.

As we will see later, transitions involving adjoint baryonic states play an important role in the baryon polarizability. In contrast to mesons (which present a unique adjoint state), there are $N-1$ possible adjoint representations for baryons, each associated with a different color structure. However, one can define an averaged potential $ V_{a}$ which is independent of the particular internal representation~\cite{Brambilla:2009cd}. Using color-current ``conservation'', \begin{equation}
   Q^b\equiv T^b(G)=\sum_{i=1}^{N} T_i^b(F)\,,
\end{equation} one can find $\sum_{i<j}T_i \cdot T_j$ from $Q^2$, and averaging over the $N_{\rm pairs}=N(N-1)/2$ different pairs of $T_i \cdot T_j$, one has~\cite{Brambilla:2009cd}
\begin{equation}\label{eq:potential_pairsie_adjoint}
   V_{a}(r_{ij})
   =
   \frac{C_2(G)-NC_2(F)}
        {N(N-1)}
   \frac{\alpha_d}{r_{ij}}
   =
   -\frac{C_N-1}{N-1}
   \frac{\alpha_d}{r_{ij}}  \,.
\end{equation} For adjoint mesons the corresponding potential is $
    V_{a,M}(r)={\alpha_d}/({2N\,r)}$,
which vanishes in the large-$N$ limit, reproducing the free adjoint behavior  exploited in the mesonic polarizability calculations~\cite{Peskin:1979va}.  Adjoint baryons may behave differently: the interaction can remain attractive and becomes equal to the singlet interaction at large $N$. Consequently, adjoint bound states can significantly contribute to the baryon polarizability. This observation suggests that, unlike the mesonic case, a proper treatment of baryons should include transitions between bound states rather than only transitions from the ground state to continuum states.

\paragraph{Quark--diquark approximation.}

Given the interactions above, one could in principle solve for the baryon wave function. However, the resulting $N$-body problem does not admit known analytic solutions. Ref.~\cite{Mitridate:2017oky} showed that the variational ansatz
\begin{equation}
    \psi_B \propto
    \exp\!\left(
    -\frac{1}{a_B}\sum_{i<j}^N r_{ij}
    \right),
\end{equation}
where $a_B$ is the effective baryon size, provides a good approximation to the ground state of the baryon wave function. For $\SU(3)$, this yields a binding energy $\epsilon_B\simeq0.47\,\alpha_d^2m_Q$, in remarkable agreement with the more accurate result $\epsilon_B\simeq0.46\,\alpha_d^2m_Q$ obtained in Ref.~\cite{Jia:2006gw}.

Motivated by the simple exponential behavior of the variational wave function, we adopt an even simpler approximation that permits analytic estimates for arbitrary $N$. Specifically, we model the baryon as an effective quark--``diquark'' bound state~\cite{Anselmino:1992vg,Mohan:2026rcg}. In this picture, a compact cluster of $N-1$ quarks plays the role of an effective color source around which the remaining quark orbits. The outer quark therefore experiences the coherent attraction generated by the other $N-1$ quarks. Approximating the $(N-1)$-quark cluster by a pointlike object of mass $m_{N-1}\simeq(N-1)m_Q$, the problem reduces to an effective two-body system with reduced mass
\begin{equation}
    \mu\simeq\frac{N-1}{N}m_Q\,.
\end{equation}

The corresponding singlet interaction is
\begin{equation}
    V_s(r)
    =
    -C_N\frac{\alpha_d}{r}
    \equiv
    -\frac{\alpha_B}{r}\,,
\end{equation}
which is obtained by coherently summing the interactions of the outer quark with the remaining $N-1$ constituents. Equivalently, since the $(N-1)$-quark cluster transforms uniquely as an anti-fundamental representation for a singlet baryon, the effective interaction is exactly that of a quark with an antiquark in a color-singlet meson. Likewise, the adjoint baryon is described by
\begin{equation}
    V_a(r)
    =
    -(C_N-1)\frac{\alpha_d}{r}\,,
\end{equation} which may be interpreted as the effective quark--cluster potential obtained by averaging over the $N-1$ possible color representations of the $(N-1)$-quark cluster that couple with the remaining quark to form the adjoint baryon.

Interestingly, the singlet interaction is therefore identical to that of a heavy meson. In contrast, the adjoint baryon interaction remains attractive and approaches the singlet strength in the large-$N$ limit, whereas the corresponding mesonic adjoint interaction scales as $1/N$ and therefore vanishes. For $N=2$, however, the baryonic and mesonic adjoint potentials coincide exactly.

\paragraph{Binding energy and baryon size.}

Using the hydrogenic solution, the baryon binding energy is estimated as\footnote{Note that this is the averaged binding energy per quark for large $N$ while for small $N$ it is closer to the total binding energy. To compute it exactly, one would have to add this binding energy to the di-quark binding energy which is smaller for $N=3$ since the attraction of a quark-quark pair is weaker than that of a quark-diquark one.}
\begin{equation}
    \epsilon_B
    =
    \frac{1}{2}\mu\alpha_B^2
    \approx
    \frac{N-1}{2N}C_N^2\alpha_d^2 m_Q \,,
\end{equation} which is much smaller than $m_Q$, justifying the approximation in the diquark mass $m_{N-1}$ and taking the baryon mass to be $m_B \approx N m_Q$. This simple expression reproduces the variational results of Ref.~\cite{Mitridate:2017oky}, for $N\leq 6$ within roughly $30\%$. For $\SU(3)$, it gives $\epsilon_B\simeq0.59\,\alpha_d^2m_Q$, notably close to the accurate numerical result of Ref.~\cite{Jia:2006gw}, $\epsilon_B\simeq0.46\,\alpha_d^2m_Q$, despite the simplicity of the approximation. For large $N$, we have simply $\epsilon_B\simeq\,0.5\,\alpha_B^2m_Q$ while the exact result computed in Ref.~\cite{Cohen:2011cw} is $\epsilon_B\simeq\,0.217\,\alpha_B^2m_Q$.

The corresponding Bohr radius is
\begin{equation}
 a_B^{-1}
    =
     \alpha_B\mu
    =
    \frac{N-1}{N}C_N\alpha_dm_Q \,.
\end{equation}  For comparison, the variational wave function $\exp(-\sum_{i<j}^N r_{ij}/a_B)$ employed in Ref.~\cite{Mitridate:2017oky} yields
\begin{equation}
a_B^{-1}
=
\frac{C_V}{2C_K}
{C_N\alpha_dm_Q}\,,
\end{equation}
where $C_V, C_K$ are given in Table 2 of Ref.~\cite{Mitridate:2017oky}. Their result reproduces exactly ours for $\SU(2)$ (since baryons are two-body systems in this case). For $\SU(3)$ (and $\SU(4)$), Ref.~\cite{Mitridate:2017oky} finds  $
{a_B^{-1}}/{(\alpha_Bm_Q)}
\simeq
0.255\,(0.17)$
while our quark--diquark approximation gives $
{a_B^{-1}}/{(\alpha_Bm_Q)}
\simeq
0.666\,(0.75) $. In fact, for $\SU(4)$, their second variational ansatz  $\sum_{i=1}^N\exp(-\sum_{j=1}^N r_{ij}/a_B)$  performs better and yields a smaller baryon size value $a_B^{-1}/{(\alpha_Bm_Q)} \simeq 0.338$. Additionally, the QCD calculations of Ref.~\cite{Jia:2006gw} find $a_B^{-1}/{(\alpha_Bm_Q)} \simeq 0.382$ for $N=3$ which is closer to our results. Thus, our estimate of the inverse radius differs from the variational calculations by nearly a factor of two (or three depending on the methods) for $N=3,4$, while reproducing exactly the $\SU(2)$ result.

For large $N$ the baryon wave function is expected to approach a factorized Gaussian form~\cite{Albertus:2015fca} at short-distances while retaining an exponential fall asymptotically~\cite{Cohen:2011cw}. Although this profile is more complex and more localized  around its characteristic radius (at short scales) than the exponential ansatz adopted here, it predicts a larger Gaussian baryon size, with $a_B^{-1}\simeq0.39\,\alpha_B m_Q$~\cite{Cohen:2011cw}, compared with our estimate $a_B^{-1}\simeq\alpha_B m_Q$, which delays the gaussian fall. Additionally, at large radius, the exponential falls with a baryon radius of $a_B^{-1} \approx 1.14\, \alpha_Bm_Q$~\cite{Cohen:2011cw} which is very similar to our results.  Thus, the simplified treatment is also justified for large $N$.

Taken together with the factor of two agreement for binding energies, these results indicate that the quark--diquark picture captures the main features of the baryon despite reducing the original $N$-body problem to an effective two-body system. We therefore expect it to provide a reasonable framework for estimating the baryon polarizability, while keeping in mind an uncertainty of order unity in quantities that are particularly sensitive to the baryon size and binding energy.

\paragraph{Regime of validity of the Coulombic description.}

The dark coupling $\alpha_d$ appearing in the previous expressions must be evaluated at the characteristic momentum scale of the bound state. In general this scale is set by the inverse size of the system,\footnote{For $N=3,10$, we use this scale, $\mu_0=a_B^{-1}$, since we take the results for $a_B$ present in the literature~\cite{Jia:2006gw,Cohen:2011cw}. The changes compared to a hydrogenic case ($\mu_0=\alpha_B \mu$) are less than 20\% (10\%) for $N=3(10)$ and  small $\xi\lesssim100$, while for  large $\xi$ they drop further down to   10\% (5\%).} \begin{equation}\label{eq:momentum_scale_general}
    \mu_0 \sim \dfrac{1}{a_B}\,.
\end{equation} For a Coulombic bound state, $a_B = (\alpha_B \mu)^{-1}$, and therefore this reduces to \begin{equation}\label{eq:momentum_scale_bohr_atoms}
    \mu_0=\alpha_B\mu\,,
\end{equation} consistent with the momentum estimate $k= \mu v$  where $v$ is the relative velocity between constituents in a Coulombic system $v\approx \alpha_B$. This hierarchy justifies retaining only chromo-electric interactions, while chromo-magnetic contributions are suppressed by powers of the velocity, $\mathcal{O}(\alpha_B^2)$~\cite{Brambilla:2015rqa}.

When the baryon size $a_B$ becomes comparable to the confinement length, $a_B\gtrsim \Lambda_d^{-1}$, confinement effects can no longer be neglected. In this regime the effective interaction potential between the quark and the diquark ``nucleus'' may be approximated   by~\cite{Mitridate:2017oky,Koma:2006si,McKeen:2024trt}
\begin{equation}
    V_s(r)\approx -\frac{\alpha_B}{r}+3.6\Lambda_d^2 r\,.
\end{equation}
Thus, the Coulombic contribution dominates provided that
\begin{equation}\label{eq:coulombic_condition}
    3.6{\Lambda_d^2 a_B^2} <\alpha_B\,.
\end{equation}
This provides a consistency condition for the  Coulombic description.

\begin{figure}[t]
    \centering
    \includegraphics[width=0.475\linewidth]{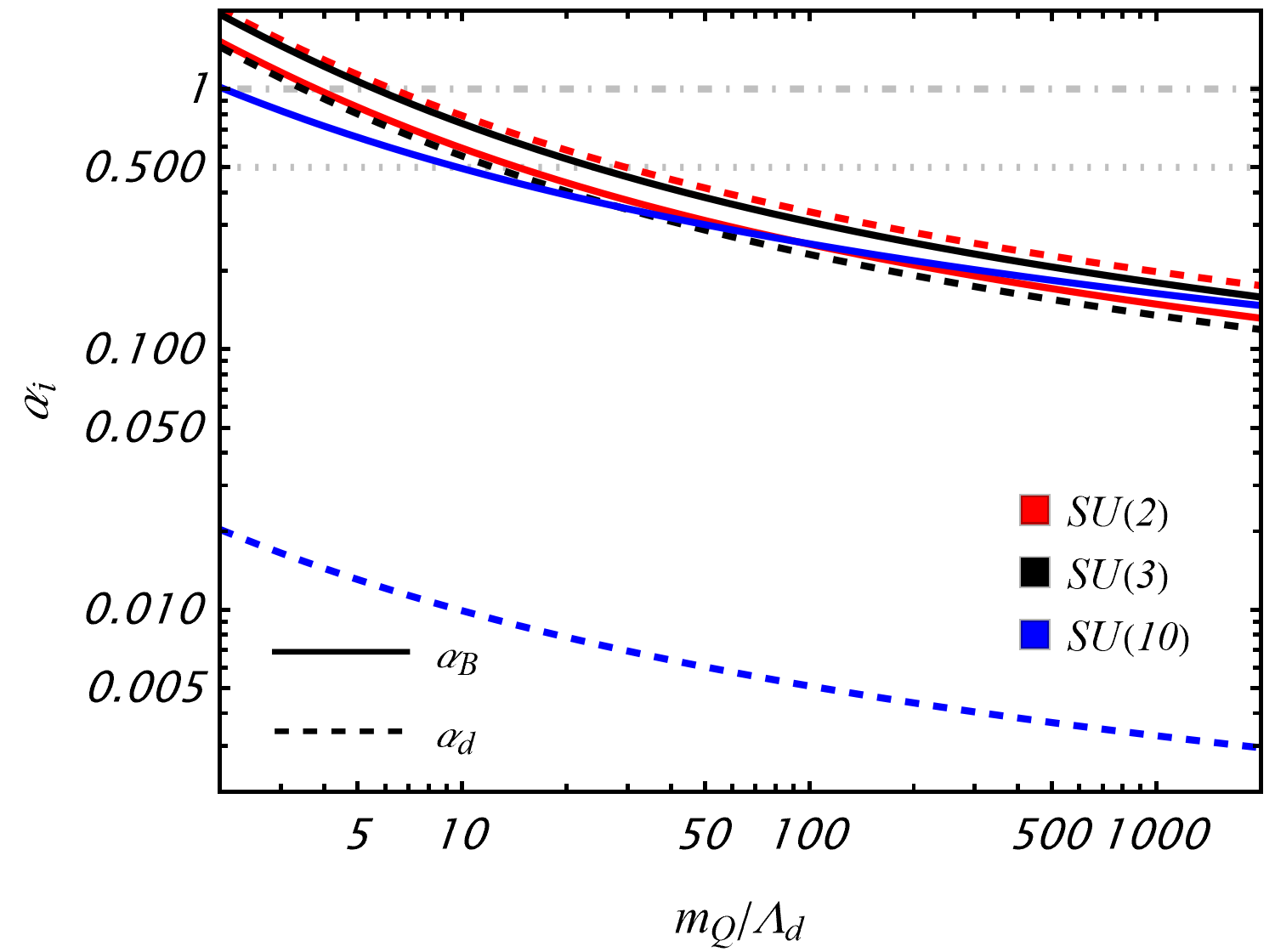}\hspace{0.1cm}
    \includegraphics[width=0.475\linewidth]{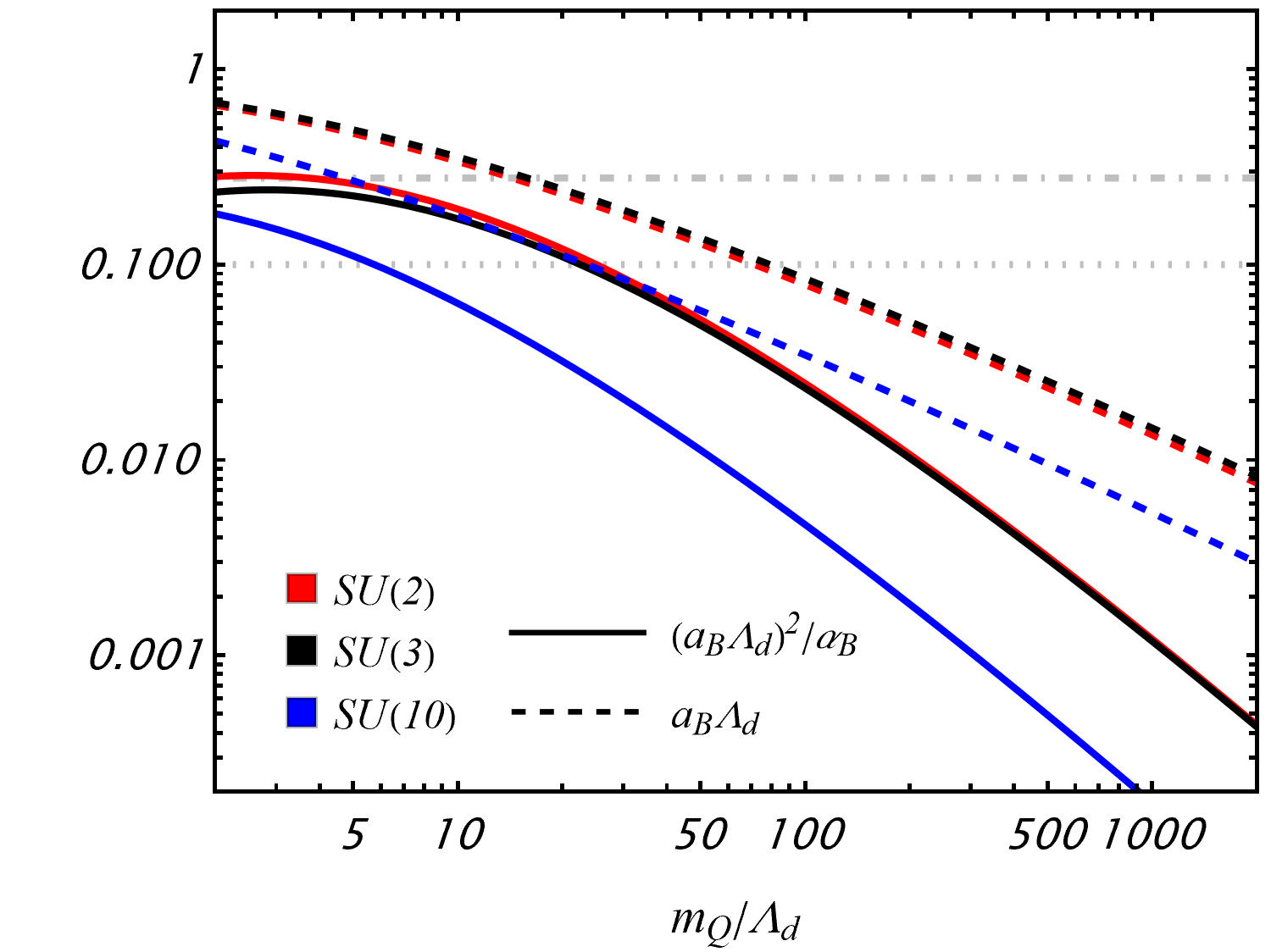}
\caption{
Left: Dark baryon $\alpha_B$ (in solid) and dark gauge $\alpha_d$ (in dashed) evaluated at the characteristic Bohr momentum scale $\mu_0=\alpha_B \mu$, as a function of the hierarchy parameter $\xi=m_Q/\Lambda_d$. Results are shown for $N=2,3,10$ using red, black and blue colors. The gray horizontal lines correspond to high coupling values: $\alpha_i=0.5$ (solid) and $\alpha_i=1$ (dashed). Right: Corresponding values of the dimensionless quantity $a_B^2\Lambda_d^2/\alpha_B$, which measures Coulomb dominance over confinement effects closely related to the separation between the baryon size and the confinement length scale. The regime $a_B^2\Lambda_d^2/\alpha_B\ll 1/3.6$ corresponds to compact Coulombic baryons whose internal dynamics are largely insensitive to confinement effects. The gray horizontal lines correspond to high values at which these effects cease to be negligible: $0.1$ (dotted), and $1/3.6\approx0.278$ (dot-dashed).}
    \label{fig:couplingXratio}
\end{figure}

Using  \cref{eq:alpha_per_energy} and given a fixed ratio $\xi$, the coupling $\alpha_d$ at the Bohr scale (\ref{eq:momentum_scale_bohr_atoms}) can be determined numerically as a function of $\xi$ and $N$, as shown in the left panel of  \cref{fig:couplingXratio}. The right panel displays the Coulombic consistency condition in \cref{eq:coulombic_condition} alongside the quantity $a_B\Lambda_d$, which measures the degree of scale separation between the baryon size and the confinement length.

As illustrated in \cref{fig:couplingXratio}, the Coulombic description becomes increasingly accurate as $\xi=m_Q/\Lambda_d$ grows. In particular, the condition $a_B\ll \Lambda_d^{-1}$ is satisfied for $\xi\gtrsim10$, indicating that the baryon is largely contained within a perturbative region. This criterion is consistent with Ref.~\cite{Mitridate:2017oky}, where confinement effects become important for $\alpha_d\gtrsim0.4$ and $m_Q\lesssim10\Lambda_d$. For $N=2$, the confining linear potential contributes only about $10\%$ ($30\%$) of the attractive interaction at $r=a_B$ for $\xi\gtrsim80$ ($\xi\gtrsim20$), while the effective coupling remains below $\alpha_B\sim v\lesssim0.5$ for $\xi\gtrsim14$. In this regime, relativistic (in particular, chromo-magnetic) corrections are expected to remain sub-leading, scaling as $\mathcal{O}(v^2)\lesssim25\%$. For larger $N$, even smaller $\xi$ are allowed by the Coulombic consistency condition. However, the small velocity approximation ($\alpha_B \lesssim0.5$) still keeps $\xi\gtrsim24$   and $\xi\gtrsim10$ for $N=3$ and 10, respectively. If one considers neglecting corrections up to $\mathcal{O}(v^2)\lesssim50\%$, one could in principle go to $\xi\gtrsim7$ for $\SU(2)$, $\xi\gtrsim11$ for $\SU(3)$ and $\xi\gtrsim4$ for $\SU(10)$.

\emph{In the remainder of this work, we therefore regard the non-relativistic Coulombic approximation as quantitatively reliable for:}
\begin{equation}
    \xi \equiv \dfrac{m_Q}{\Lambda_d} \gtrsim 10\,.
\end{equation}
Results up to $\xi\sim5$ should be interpreted only qualitatively, since confinement corrections can reach $50\%$ and chromomagnetic effects approximately $60\%$. For even smaller values of $\xi$,    non-perturbative and relativistic effects are expected to play a dominant role and our treatment ceases to be trustworthy.

\section{The interaction potential}
\label{sec:interaction_potential}

Having established the conditions under which dark baryons are compact, non-relativistic Coulombic bound states, we now turn to their long-distance interactions. Since ground state baryons are spherical color singlets, the leading interaction is not mediated by one-gluon exchange. Instead, the dominant force arises from induced chromo-electric dipole moments, in   direct analogy with the interactions between heavy quarkonia in QCD~\cite{Appelquist:1977tw,Appelquist:1977es,Gottfried:1977gp,Appelquist:1978rt,Voloshin:1978hc,Peskin:1979va,Bhanot:1979vb,Leutwyler:1980tn,Fujii:1999xn}.

A more standard analogue, at least in the perturbative regime, is provided by ordinary hydrogen atoms. Since the hydrogen ground state carries neither net charge nor a permanent electric dipole moment, the interaction between two neutral atoms is generated by induced dipoles and gives rise to the familiar London and Casimir-Polder potentials~\cite{Casimir:1947hx,Dzyaloshinskii1956}. As we will show, heavy dark baryons exhibit precisely the same qualitative behavior. The analogy with hydrogen atoms is particularly useful because although the underlying gauge interactions are very different, the long-distance physics is controlled only by the existence of compact neutral bound states with finite polarizability.

\begin{figure}[t]
    \centering
    \includegraphics[width=0.56\linewidth]{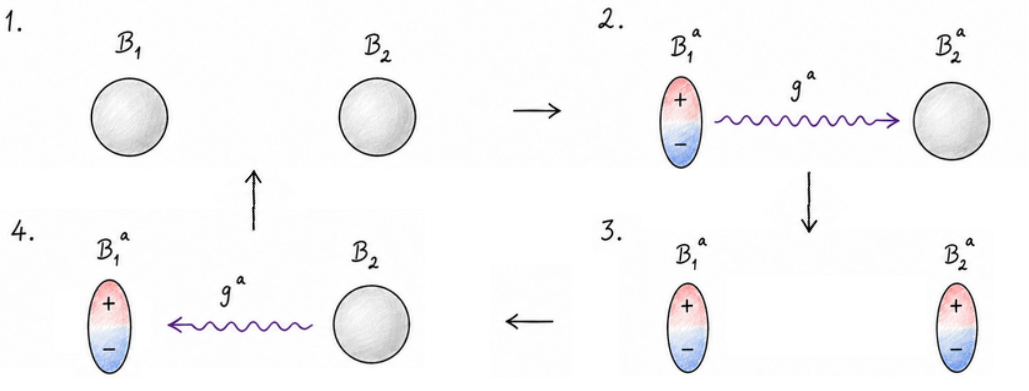} \includegraphics[width=0.43\linewidth]{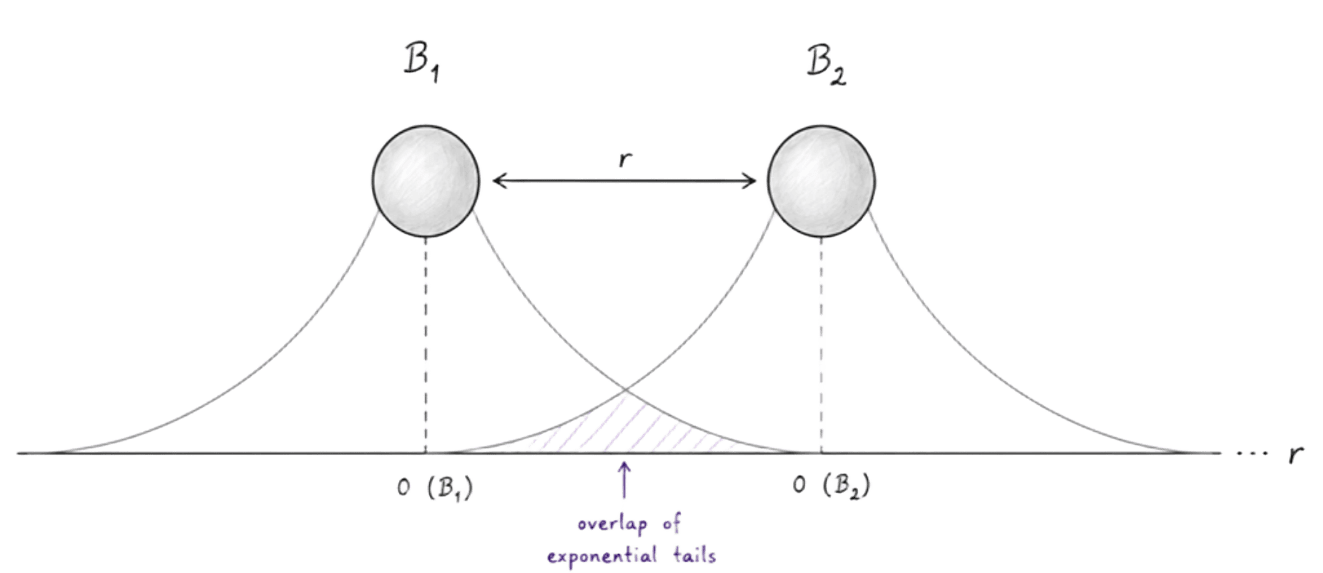}
    \caption{Left: Schematic illustration of the instantaneous dipole–induced dipole interaction between two color-singlet baryons. A quantum fluctuation promotes the first baryon, $\mathcal B_1$, into a color-adjoint excited state with a non-zero chromo-electric dipole moment through gluon emission. The resulting chromo-electric field polarizes a nearby baryon $\mathcal B_2$, inducing its own dipole moment and color excitation. A second gluon exchange returns both baryons to their original color-singlet ground states, producing an effective elastic interaction analogous to the Van der Waals interaction between neutral atoms. Right: Schematic illustration of the effective Pauli repulsion at short distances. As the exponentially decaying quark wave functions of two baryons begin to overlap, antisymmetrization of the fermionic many-body wave function becomes important. This raises the energy of the system, generating an effective repulsive interaction that prevents the baryons from occupying the same region of space.}
    \label{fig:schematic_view_of_the_forces}
\end{figure}

In particular, the long-distance ($r\gg a_B$) baryonic interaction potential displays three distinct regimes,
\begin{equation}
V(r)\propto
\begin{cases}
r^{-6}\,, & a_B\ll r\ll \epsilon_B^{-1}\,,\\[1mm]
r^{-7}\,, & \epsilon_B^{-1}\ll r\ll \Lambda_d^{-1}\,,\\[1mm]
e^{-m_{\rm DG}r}/r\,, & r\gg \Lambda_d^{-1}\,,
\end{cases}
\end{equation} corresponding respectively to a non-retarded London dispersion force, a retarded Casimir-Polder interaction, and a Yukawa tail generated by glueball exchange.\footnote{For hadrons in the SM, two-pion exchange is actually the lightest, and therefore dominant, scalar interaction channel due to their pseudo-Goldstone boson nature. Therefore, the potential tail will be dominated by the exchange of these particles.} Again similarly to (triplet) hydrogen, at short distances, $r\lesssim a_B$, the overlap of the baryon wave functions gives rise to an effective Pauli repulsion. Since the quark wave functions decay exponentially outside the baryon, this repulsive interaction is itself expected to decrease approximately as
\begin{equation}
V_{\rm rep}(r)\propto e^{-r/a_B},
\end{equation}
although its precise form depends on the short-distance dynamics beyond our effective description. A schematic representation of the different interactions discussed in this section is shown in \cref{fig:schematic_view_of_the_forces}.

Before deriving the interaction potential explicitly, we note that general arguments presented in Ref.~\cite{Assi:2025ysr}, based only on general symmetry considerations, already suggest that interactions between compact color-singlet baryons must be extremely weak and can only generate long-range potentials of the form $1/r^n$ with $n\ge3$ (or weaker, such as the Yukawa potential). The explicit calculation below confirms this expectation while determining both the precise power-law behavior and its normalization.

\subsection{Derivation of the baryon--baryon potential}

The interaction between compact heavy hadrons has been studied extensively in the context of heavy quarkonia in QCD, where the long-distance force is generated by induced chromo-electric dipole interactions~\cite{Appelquist:1977tw,Appelquist:1977es,Peskin:1979va,Bhanot:1979vb}. Since heavy dark baryons are likewise compact color-singlet objects, the same effective field theory techniques can be applied here with only minor modifications. We therefore closely follow the derivation developed for heavy quarkonia, emphasizing the differences arising from the baryonic structure.

 In particular, following the multipole, or operator product, expansion formalism of Refs.~\cite{Peskin:1979va,Bhanot:1979vb,Fujii:1999xn}, the heavy dark baryon interaction potential can be obtained from the Born approximation of the scattering amplitude, neglecting the small energy transfer in the non-relativistic collision process,\footnote{The leading-order interaction arises from the exchange between two single clusters, each containing two gluonic legs. Contributions with a larger number of gluonic legs are systematically suppressed by powers of $\alpha_d$, which explains why we only retain operators containing two chromo-electric fields.}
\begin{equation}
\label{eq:general_potential}
V({\bf r})=-i\int_{-\infty}^{\infty}dt\,
\langle 0 |{\rm T}
\left(\sum_p c_p\mathcal{O}_p(0,{\bf 0})\right)
\left(\sum_q c_q\mathcal{O}_q(t,{\bf r})\right)
|0\rangle \, ,
\end{equation} where $|0\rangle$ denotes the vacuum state,  $\mathcal{O}_p(t,\vec {\bf r})$ local operators evaluated at time $t$ and position ${\bf r}$, and $c_p$ their corresponding Wilson coefficients.   The operators have been computed in Ref.~\cite{Peskin:1979va} and can be expressed as  (assuming  rotational invariance)
\begin{equation}
\label{eq:operator_definition}
\sum_p c_p\mathcal{O}_p(t,r)
=
-\sum_{\substack{n=2\\ \mathrm{even}}}^{\infty}
d_n a_B^3 \epsilon_B^{2-n}\,
{\rm tr}
\!\left[E^k
(-iD^0)^{n-2}
E^k
\right]\,,
\end{equation}   with  dimensionless coefficients\begin{equation}\label{eq:dn_definition}d_n\equiv\dfrac{\epsilon_B^{n-2}}{a_B^3} \frac{4\pi\alpha_d}{N(N-1)}
\dfrac{1}{3}\sum_{i<j}^N
\langle \phi |
r_{ij}^k
\frac{1}{(H_a+\epsilon_B)^{n-1}}
r_{ij}^k
|\phi\rangle \,,
\end{equation}  where one explicitly factors out the dependence on the dimensional parameters $\epsilon_B$ and $a_B$. Here,  $\phi$ denotes the  baryon  wave function (in our case, it corresponds to the singlet ground-state $1S$ wave function which is indeed rotationally invariant), $k=1,2,3$ is a spatial index, $r^k_{ij}$ is the $k$-th component of the relative separation between the $i$-th and $j$-th quarks, $H_a$ is the effective Hamiltonian describing the intermediate color-adjoint state, $E^k$ is the $k$-th component of the chromo-electric field, and $D^0$ is the time component of the covariant derivative acting on $E^k$.\footnote{The trace over the color indices of the chromo-electric field operator guarantees gauge invariance~\cite{Fujii:1999xn}. Additionally, note that the trace of the non-Abelian gauge fields leads to a factor of $1/2$ from the normalization ${\rm tr}[T^aT^b]=\delta^{ab}/2$.  } We have also replaced the meson color factor $1/N$ by the baryon one $1/[N(N-1)]$, derived in \cref{app:color-contraction}.

When the gluon fields vary slowly compared with the characteristic time scale of dipole fluctuations,
\begin{equation} \label{eq:correlation_time}
t_c\sim\epsilon_B^{-1},
\end{equation}
which corresponds to the time associated with the emission and reabsorption of a gluon converting a singlet $1S$ state into an adjoint $kP$ state, the operator expansion is dominated by the $n=2$ contribution~\cite{Kharzeev:1995ij}. This corresponds to a retarded potential, as the dipole fluctuations cannot be ignored.  In the following, we therefore retain only the $n=2$ term whenever$$
r\gg t_c\sim \epsilon_B^{-1},$$
and quantify the associated error in \cref{app:highern}. In the perturbative regime, we additionally perform the full calculation and show that the perturbative potential remains under theoretical control down to separations $r\gg a_B$.

Focusing on the regime $r\gg \epsilon_B^{-1}$, Ref.~\cite{Fujii:1999xn} has shown that  the potential can be written as a superposition of Yukawa interactions associated with the exchange of scalar states of mass $m_\sigma$~\cite{Fujii:1999xn},
\begin{equation}
\label{eq:spectral_function_potential_def}
V(r)=-
\alpha_{\mathcal B}^2
\left(
\frac{4\pi^2}{b}
\right)^2
\int d m_\sigma^2\,
\rho_\theta(m_\sigma^2)\,
\frac{e^{-m_\sigma r}}{4\pi r}\,,
\end{equation}
where
\begin{equation} \label{eq:polarizability_baryon_main}
\alpha_{\mathcal B}\equiv
\dfrac{d_2 a_B^3 }{4\pi \alpha_d}
\end{equation}
is the \emph{chromo-electric polarizability}~\cite{Brambilla:2015rqa} of the baryon~\cite{Dong:2022rwr}, in direct analogy with the electric polarizability of the hydrogen atom,  and $\rho_\theta$   is the spectral density of the correlator of the trace of the energy-momentum tensor $\theta_\mu^\mu$.

To derive the above spectral representation, one rewrites $\mathbf E^2$ in terms of the gluon field strength tensor as a combination of the trace of the energy-momentum tensor $\theta_\mu^\mu$ and of $\alpha_d\theta_{00}^{(G)}$; the latter describes the tensorial $2^{++}$ channel associated with two-gluon exchange. For $r\gg m_Q^{-1}$, the heavy-quark contributions to both operators can be neglected, and they reduce to their pure Yang-Mills expressions. In the non-perturbative regime, lattice calculations show that the tensorial $2^{++}$ glueball is significantly heavier than the lightest scalar $0^{++}$ state associated with $\theta_\mu^\mu$~\cite{Godfrey:2002ay}. Consequently, its contribution is exponentially suppressed at large distances, $r\gg\Lambda_d^{-1}$, and may be neglected. In the perturbative regime, however, both operators correspond to pairs of massless gluons and contribute at the same order in $\alpha_d$, implying that the tensor channel cannot be ignored. In practice, its effect can be incorporated through a simple multiplicative factor of $23/15\simeq1.5$ in the perturbative spectral density $\rho_\theta$, such that \cref{eq:spectral_function_potential_def} reproduces the full perturbative potential in the regime $\epsilon_B^{-1}\ll r\ll\Lambda_d^{-1}$.

For the exchange of the lightest glueball, lattice calculations find~\cite{Batz:2023zef,Chen:2005mg,Meyer:2008tr}
\begin{equation} \label{eq:spectral_density_glueball}
\rho_\theta(m_\sigma^2)\sim
0.04\left(\frac{N}{3}\right)^2
m_{\rm DG}^6
\delta(m_\sigma^2-m_{\rm DG}^2) \,,
\end{equation} with uncertainties of about 30\% for $\SU(3)$ while for $N\neq3$ we expect order one uncertainties as we are simply using a large-\(N\) treatment to predict the $N$-dependence.\footnote{Low energy theorems also predict $\rho_\theta(m_\sigma^2)\sim
-16 \epsilon_{\rm vac}
m_{\rm DG}^2
\delta(m_\sigma^2-m_{\rm DG}^2)$ where $\epsilon_{\rm vac}$ is the  energy density of non-perturbative $\SU(N)$ vacuum which is proportional to $\Lambda_d^4$~\cite{Fujii:1999xn}.} The exchange of the lightest glueball, therefore, generates a simple Yukawa interaction. We neglect heavier scalar resonances of $\SU(N)$ since, at sufficiently large distances $r\gg \Lambda_d^{-1}$, the interaction is dominated by the lightest state.  We note, however, that heavier resonances may still affect the potential around $r\sim\Lambda_d^{-1}$, corresponding to momentum transfers $ k={m_Bv}/{2}\sim\Lambda_d $. This region is typically unimportant for the parameter space of interest, namely $m_Q/\Lambda_d\lesssim10^3$, where the self-interaction cross section is largest.

 In \cref{app:dipole-transition-integration}, within the quark-diquark approximation, we obtain\footnote{Since the total baryon polarizability $\alpha_\mathcal{B}$ is highly sensitive to the baryon size, scaling as $a_B^3$, and the potential depends on $\alpha_\mathcal{B}^2$, even a factor of 2 difference in the baryon size implies cross section changes that could be larger than three orders of magnitude. Therefore, we will use the most accurate values found in the literature for $N>2$ and our exact results for $N=2$.}
\begin{equation}
d_2=
\pi
\frac{9C_N+5}
{(C_N+1)(2C_N+1)}\, ,
\end{equation} which includes all discrete and continuum intermediate states.
In particular, for $\SU(3)$, this gives $d_2\simeq6.242$, which  is considerably larger than an estimate obtained from only the bound state contributions. The estimate retaining only the leading bound state $1S\rightarrow2P$ transition gives $d_2=0.092$, while if one includes all bound state transitions, the result is $d_2=0.16$. The enhancement originates from the comparatively large size of the adjoint baryon at small $N$, which makes transitions to continuum states dominate the dipole polarizability. This leads to baryon results which are similar to the mesonic case for $N\lesssim5$. Specifically, for $N=2$ our result exactly reproduces the mesonic one $d_2=28\pi/9$ when neglecting the adjoint potential. This was expected since $\SU(2)$ baryons are themselves two-body bound states with the same interaction strength as mesons. In the opposite large-$N$ limit, the $1S\rightarrow2P$ transition dominates (contributing to 65\%, while summing over all bound states increases this percentage to 81\%) and one recovers the familiar hydrogenic result, $
C_Nd_2 \to {9\pi}/{2} $. The values of the other coefficients $d_n$  can also be found in \cref{app:dipole-transition-integration}.

The distinction between continuum-dominated and bound-state-dominated polarizabilities has important implications for the higher-order coefficients $d_n$. For intermediate bound states one simply finds
\begin{equation}
d_n\propto
\Delta\epsilon^{-(n-2)}\,,
\end{equation}
where $\Delta\epsilon$ denotes the binding energy difference between the corresponding adjoint bound state and the singlet ground state. By contrast, continuum contributions exhibit a considerably more complicated dependence, preventing a simple geometrical resummation of all higher-$n$ terms. In \cref{app:highern}, within the perturbative regime, we perform the full summation of \cref{eq:general_potential} in  two limiting cases: the (free) continuum-dominated regime relevant for $N=2,3$ and the bound-state-dominated regime relevant at larger $N$.   In both cases we recover a London potential scaling as $r^{-6}$ for $r\lesssim\epsilon_B^{-1}$, albeit with different coefficients.

The transition between these two regimes occurs at intermediate values
of $N$, where neither limiting description is fully controlled. This is
particularly relevant for $\SU(5)$ and nearby groups, which are also of
interest from the perspective of grand-unified constructions. In our
quark--diquark approximation, the attractive interaction remains close
to, but does not fully overcome, the Pauli repulsion in this range,
yielding a velocity dependence that is too weak to establish a
phenomenologically relevant SIDM regime. Applying the large-$N$
approximation instead leads to a substantially stronger attraction and
non-trivial velocity dependence. Since these two treatments do not yet
agree quantitatively, we refrain from drawing conclusions for
intermediate values of $N$. A reliable determination will require improved
baryonic wave functions and chromo-electric polarizabilities, as well as
a more complete treatment of the higher-order $r^{-6}$ potential.
Numerical lattice results beyond $\SU(3)$ would provide an important bridge
between the small-$N$ calculations and the large-$N$ limit.

Finally, in \cref{app:potentail_sum}, we perform the full spectral-density integration and determine the matching energy scale $\sqrt s_0$  separating the perturbative and non-perturbative spectral densities using quark-hadron duality. We find that the dependence on $s_0$ is negligible and that, throughout the intermediate region,
$$
\epsilon_B^{-1}\ll r\lesssim\Lambda_d^{-1}, $$
the full potential is very well approximated by the sum of a Casimir-Polder $r^{-7}$ interaction and a Yukawa contribution, while for$$
r\gtrsim \Lambda_d^{-1} $$
the Yukawa  term fully dominates.\footnote{A more precise treatment would include the running of $\alpha_d$ in the perturbative potentials, however, the matching scale satisfies $s_0>m_{\rm DG }^2$  and, since $m_{\rm DG}^2\gg\Lambda_d^2$, the running is numerically negligible, as explicitly shown in \cref{app:potentail_sum}.}

\subsection{Structure and regimes of the potential}

Combining the perturbative and non-perturbative contributions, the interaction potential is therefore well described by
\begin{equation} \label{eq:Vr_definition}
V(r)=-\frac{d_2^2}{16\pi^2}
\begin{cases}
f(y)C_6\,r^{-6}\,, & a_B\ll r<r_{\rm ret}\,,\\[1mm]
f(y)C_7\,r^{-7}+C_Y\,e^{-m_{\rm DG}r}/r\,, & r_{\rm ret}\le r\,,
\end{cases}
\end{equation}
where \begin{align}
C_6
&=
\frac{N^2-1}{8}\,
\Delta\epsilon\,
a_B^6\,,
\qquad
\text{non-retarded (London)}\,,
\\
C_7
&=
\frac{N^2-1}{8}\,
\frac{46}{\pi}\,
a_B^6\,,
\qquad
\text{retarded (Casimir-Polder)}\,,
\\
C_Y
&=
\frac{1}{\alpha_d^2}
\frac{4\pi^3}{3025}
(m_{\rm DG}a_B)^6\,,
\qquad
\text{dark glueball Yukawa}\,,
\end{align} $y= r \sqrt s_0=r 12 \Lambda_d$ and $f(y)$ is a suppression function defined as \begin{equation}
    f(y)=e^{-y} \left(1+y+\frac{y^2}{2}+\frac{y^3}{6} +\frac{y^4}{24} +\frac{y^5}{120}\right)
\end{equation}  such that $f(y\ll1)\approx1$ and $f(y\gg 1)\approx 0$ (for further details see \cref{app:potentail_sum}).
For large $N$, where the coefficients $d_n$ are dominated by intermediate bound states, one has $\Delta\epsilon\simeq 21\epsilon_B/4$, while for small $N=2,3$, where continuum states dominate, one finds $\Delta\epsilon\simeq 9.72\epsilon_B$. The coupling $\alpha_d$ that appears above is understood to be evaluated at the Bohr scale, i.e. it is the same coupling entering the definitions of $a_B$ and $\epsilon_B$. The transition scale between the non-retarded and retarded regimes is\footnote{For sufficiently small $m_Q/\Lambda_d $, one can have
$r_{\rm ret}>\Lambda_d^{-1}$, in which case the Casimir--Polder regime is
absent.} \begin{equation}
r_{\rm ret}\equiv\frac{46}{\pi\Delta\epsilon}\,.
\end{equation}

Note that for small ratios $m_Q/\Lambda_d\lesssim\mathcal O(10-100)$, significant non-perturbative effects can penetrate below $r_{\rm ret}$. Since we do not include higher-$n$ terms in the non-perturbative calculation, we actually truncate the Yukawa contribution at $r_{\rm NP}\simeq 0.2 /\epsilon_B$ for $N\leq5$ and  $r_{\rm NP}\simeq 0.4 /\epsilon_B$ for $N>5$, where retaining only the $d_2$ term in \cref{eq:general_potential,eq:operator_definition,eq:dn_definition} remains accurate well within an order of magnitude, see \cref{app:highern}.

In the perturbative regime, we have also included the color factor $(N^2-1)/8$ multiplying $C_{6,7}$, which arises from the perturbative expectation value of the squared chromo-electric field $\mathbf{E}^2$~\cite{Bhanot:1979vb}. We further note that the interaction strength scales in all cases as $a_B^6$ and therefore vanishes, as it should, in the point-like limit,
\begin{equation}
V(r)\longrightarrow0\,,
\qquad
a_B\longrightarrow0\,.
\end{equation}

An important simplification occurs in the large-$N$ limit. The running
of the coupling implies $\alpha_d\propto 1/N$ at fixed $\xi$, while
$C_N\propto N$, such that the effective baryonic coupling
$\alpha_B=C_N\alpha_d$ approaches an $N$-independent value. At the
same time, $d_2\propto 1/N$, while the factors of $N$ entering the
perturbative and non-perturbative potentials compensate this
suppression.\footnote{This further motivates the $d_2\propto 1/N$
behavior we find for baryons, in contrast to $d_2\propto 1/N^2$ for
mesons~\cite{Peskin:1979va}.}
Consequently, the dimensionless baryon-baryon attractive potential ($a_B V$)
approaches an $N$-independent limit~\cite{Cohen:2011cw,Adhikari:2013dfa}. This motivates
using a single sufficiently large-$N$ theory as a representative
realization of this regime. In the numerical analysis below, we use
$\SU(10)$ for this purpose. However, the Pauli repulsion grows with $N$
due to the increasing number of overlapping constituents. Therefore, in Ref.~\cite{DallaValleGarcia:2026letter}
we have also studied much larger values of $N$
in order to explore the transition to the regime where repulsion dominates over
the attractive induced-dipole interaction.

In practice, we use our potential parameters directly for $N=2$, where they
are exact, while for $N=3$ and $N=10$ we adopt more accurate estimates
available in the literature~\cite{Dong:2022rwr,Cohen:2011cw,Jia:2006gw},
which we summarize in \cref{tab:dark_sector_parameters}.
The calculation of the chromo-electric polarizability is detailed in
\cref{app:dipole-transition-integration}.\footnote{For $N=10$, the chromo-electric polarizability is inferred from the
second moment of the baryon density distribution in the large-$N$ result of
Ref.~\cite{Cohen:2011cw}, rather than being given explicitly in that work.}

\begin{table}[t]
\centering
\caption{Summary of the dark sector parameters for different values of  $N$. For $N=2$ we show our own computation while for $N=3$ and $N=10$ the results are quoted from Refs.~\cite{Dong:2022rwr,Jia:2006gw} and Ref.~\cite{Cohen:2011cw}, respectively.}
\label{tab:dark_sector_parameters}
\begin{tabular}{ccccc}
\hline
$N$ & $d_2$ & $a_B$ & $\epsilon_B$ & $\alpha_{\mathcal{B}}$\\
\hline
2   & 8.437  & $2/(\alpha_B m_Q)$   & $0.25\,\alpha_B^2 m_Q$ & $12.7/(\alpha_d^4 m_Q^3)$\\
3   & 3.97  & $2.62/(\alpha_B m_Q)$   & $0.26\,\alpha_B^2 m_Q$ & $2.4/(\alpha_d^4 m_Q^3)$\\
10   & $16$ & $ 0.876/(\alpha_B m_Q)$   &  $0.217\,\alpha_B^2 m_Q$ & $0.007   /(\alpha_d^4 m_Q^3)$\\
\hline
\end{tabular}
\end{table}

A final remark concerns the origin of the $r^{-7}$ behavior. At separations $r\gg\epsilon_B^{-1}$, the finite propagation time of the chromo-electric fluctuations becomes important. The induced dipole moments therefore possess a finite correlation time (\ref{eq:correlation_time}), leading to retardation effects that further suppress the interaction and transform the non-retarded London potential into the retarded Casimir-Polder potential. Therefore, our perturbative calculation reproduces the same transition observed for ordinary hydrogen atoms, reinforcing the analogy between the two systems.

 \subsection{Regularizing the potential}\label{sec:regularizing_potential}

The long-distance potential derived above cannot be extrapolated down to arbitrarily short separations. In particular, retaining the London dispersive form all the way to $r\rightarrow0$ leads to  a nonphysical divergence of the wave function at the origin. A short-distance regularization is therefore required.

Before introducing a phenomenological regularization, it is useful to assess whether the details of the potential at scales $r\lesssim \epsilon_B^{-1}$ are expected to significantly affect the self-interaction cross section. The maximum DM velocities observed in galaxy clusters are of order $v_{\rm max}\sim3\times10^{-3}$, corresponding to a maximal momentum probe $
k_{\rm max}={m_{B}v_{\rm max}}/{2}$. Comparing the $k_{\rm max}^{-1}$ resolution scale to the baryon size, $a_B$, one finds
\begin{equation}
a_Bk_{\rm max}
\sim
1.5\times10^{-3} \left(
\frac{N}{\alpha_B}\right)\, .
\end{equation}
For the region of interest in this work, $\alpha_B\gtrsim0.1$ and $N\leq10$, this ratio remains smaller than unity. Therefore, the scattering process is not expected to efficiently resolve the detailed structure of the interactions at distances $r\lesssim 10 \,a_B$.

One might then expect the precise form of the potential in this region to be largely irrelevant, provided the long-distance attractive tail dominates. It will turn out, however, that the attractive potential can be extremely weak, leading to cross sections of the order of the geometric one or smaller. In such cases, the inner structure of the potential can still significantly influence the scattering through modifications of the phase shifts entering the partial-wave decomposition of the scattering cross section, effectively acting as a short-distance boundary condition. Since a first-principles determination of the inner potential  goes beyond the scope of the current work, we adopt phenomenological regularizations inspired by atomic and hadronic physics.

\paragraph{Phenomenological regularization.}

For the region where the condition $r\gg a_B$ is no longer satisfied, we follow approaches developed in atomic physics~\cite{KOLOS1974457,SILVERA1986139,RevModPhys.52.393}. At short distances, one expects both a suppression of the induced-dipole interaction, since the dipole expansion ceases to be valid, and the appearance of a repulsive exchange interaction associated with the Pauli exclusion principle as the constituent wave functions begin to overlap.\footnote{Regularization and contact terms of this kind are not merely borrowed from atomic physics—they are also required in purely SM QCD calculations. For example, Ref.~\cite{Dong:2021lkh} derives the $J/\psi$-$J/\psi$ potential from correlated $\pi\pi/K\bar K$ exchange using a dispersive representation, and finds that the resulting momentum-space integral diverges logarithmically, so that a regulator (there, a Gaussian form factor) must be introduced together with an accompanying contact term (a Gaussian repulsion in momentum space).} We neglect possible further small-scale effects, whose treatment would require a considerably more sophisticated many-body analysis.\footnote{This includes the possible existence of dibaryon (hexaquark) bound states~\cite{Lu:2022myk,Alcaraz-Pelegrina:2022fsi,Huang:2020bmb,Weng:2022ohh,Assi:2025ysr}. Existing studies indicate that fully-heavy hexaquarks are not deeply bound states~\cite{Alcaraz-Pelegrina:2022fsi,Lu:2022myk}. Thus, one could expect weak effects related to such resonances if not accessible due to repulsion. We also neglect potential terms with higher powers of $\alpha_d$ such as the NNLO $\mathcal{O}(\alpha_d^3)$ potential found in Ref~\cite{Assi:2025ysr} which is non-zero for $r\lesssim a_B$.}

Following Ref.~\cite{Cline:2013pca} which studied atomic DM, we regularize the short-distance interaction through
\begin{equation}
V_{\rm reg}(r)
= V(r)\times\begin{cases}
    e^{-\left(\frac{r_{\rm in}}{r}-1\right)^2}
\,,
&
\text{ for } r \leq r_{\rm in}\,, \\
1
\,,
&\text{ for } r > r_{\rm in}\,,
\end{cases} \end{equation}
where $V(r)$ is given in \cref{eq:Vr_definition} and $r_{\rm in}\sim a_B$ (for the hydrogen atom $r_{\rm in}\approx 10\,a_B$~\cite{Cline:2013pca}).  This exponentially suppresses the attractive potential once the multipole expansion ceases to be valid. We define for later convenience the normalized regularization scale $x_{\rm in}\equiv r_{\rm in}/a_B$.

The suppression of the attractive interaction at short distances is additionally motivated by Ref.~\cite{Assi:2025ysr}, which found that one-gluon-exchange potentials between any constituents of different color-singlet hadrons each vanish
individually. They also found that these potentials can only arise at NNLO $\mathcal{O}(\alpha_d^3)$ and their sum vanishes for $r\gtrsim a_B$.  This suggests a strong cancellation of short-range interactions due to a loss of coherence, analogous to the suppression induced by form factors in DM--nucleus scattering at large momentum transfer.

We also add to the total potential a phenomenological repulsive interaction~\cite{Cline:2013pca}
\begin{equation}
V_{\rm tot}(r)=V_{\rm reg}(r)+V_{\rm rep}(r)\,,\quad \text{with } \,\, V_{\rm rep}(r)=C\,\epsilon_B e^{-r/a_B}\,,
\end{equation}
where $C=0.75$,  the minimum energy to excite one quark from the ground $1S$ state to the excited $2S$ or $2P$ states.\footnote{The overlap of two hydrogenic $1S$ wave functions,
$
S(x)=
e^{-x}
\left(
1+x
+x^2/3
\right)$, implies an exponentially localized exchange interaction since the probability of attempting to occupy the same quantum state scales approximately as $S^2$.} This Born--Mayer-type potential is widely used to model short-distance exchange repulsion in molecular and condensed-matter systems~\cite{10.1063/1.1712098,ZAKIRYANOV2023113951}.\footnote{See Refs.~\cite{PhysRev.134.A362,smirnov1965electron,https://doi.org/10.1002/cphc.202400887} for detailed calculations in the hydrogen case. Unlike quarks in different baryons, electrons and protons in distinct hydrogen atoms directly interact already at leading order $\mathcal{O}(\alpha_{\rm em})$   in non-relativistic QED. Therefore, hydrogenic results can only serve as qualitative guidance and cannot be straightforwardly mapped onto the present baryonic system.} Moreover, Ref.~\cite{Cohen:2011cw} finds that the Pauli repulsion between heavy baryons in the large-$N$ limit also falls exponentially with the inter-baryon separation, providing further support for such repulsion. { In particular, the large-$N$ repulsion is estimated as~\cite{Adhikari:2013dfa}
 \begin{equation}\label{eq:largeN_repulsion}
    V_{\rm rep}^{(N)}(r) \approx  11.6 N \epsilon_B  \left(\dfrac{r}{a_B}\right)^7
 e^{-6.4r/a_B} \,,
\end{equation} where we included a factor of 1/2 due to the two spin states available for quarks within the baryon  and neglected a contact term depending on the width of the baryon charge $\mathcal{O}(a_B)$, which is parametrically small compared to the Pauli repulsion and the induced-dipole attraction. For compactness, the total potential with the large $N$ repulsion given above will be denoted by $V_{\rm tot}^{(N)}$.}

For small $N$, we also consider smaller values,
$C=0.4,\,0.23$ 
for $N=2,3$, 
respectively, chosen such that the repulsive potential reproduces the geometrical cross section
\begin{equation}\label{eq:geo_cross_seciton}
    \sigma_{\rm geo}=4\pi a_B^2
\end{equation}
at low velocities, $v\sim10~\mathrm{km/s}$. In particular, in the limit of a large hierarchy,
$\xi\gg10^3$, the attractive interaction becomes negligible for small $N\lesssim5$, so that the scattering approaches the geometrical cross section by construction. For larger values of $N$, however, the attractive contribution remains comparable to the repulsive one, allowing for a partial cancellation between the two interactions and consequently modifying the resulting self-interaction cross section.

The strength of the exchange repulsion is expected to depend on the total spin configuration of the two-baryon system.\footnote{Color degrees of freedom may also contribute to exchange effects. However, changing the color configuration necessarily excites the system out of the two color-singlet baryon configuration. Indeed, previous studies indicate that fully heavy dibaryons (hexaquarks) are resonances that decay into two baryons~\cite{Alcaraz-Pelegrina:2022fsi,Lu:2022myk}, while nonsinglet baryons are less tightly bound than singlet ones. Consequently, color rearrangements are also expected to lead to a repulsive core. By contrast, neglecting fine-structure effects, the baryon Hamiltonian is approximately spin independent. The lowest-energy exchange excitations are therefore expected to arise predominantly from rearranging the spin wave function.} For hydrogen--hydrogen scattering, for example, the spin-triplet channel is repulsive due to the Pauli exclusion principle, whereas the singlet channel experiences essentially no exchange repulsion. In the present case, the situation is considerably more involved since the relevant Hilbert space contains the spins of all $2N$ constituent quarks. The corresponding spin wave function decomposes into several irreducible symmetry sectors, rather than the simple singlet--triplet decomposition of two spin-$1/2$ particles. Similar exchange effects arise in many-electron atoms and molecules, where Slater determinants provide the standard framework for describing exchange interactions and the resulting short-range repulsion~\cite{doi:10.1021/acs.jctc.6b00209,wing2026transferablemodelmolecularexchangerepulsion}. Fully symmetric spin configurations are expected to maximize the exchange repulsion induced by the Pauli principle, while mixed-symmetry sectors should lead to weaker exchange effects. Since the quark spin Hilbert space is only two-dimensional, the spin wave function of more than two quarks occupying the same spatial orbital cannot be completely antisymmetric, and therefore one does not expect completely vanishing exchange effects analogous to the hydrogen singlet channel.

A fully quantitative treatment would require constructing spin-dependent short-distance potentials $V_J(r)$ for each total-spin channel, computing the corresponding scattering cross sections, and performing the appropriate spin-statistical average over the initial states. Such an analysis lies beyond the scope of the present work. Instead, we employ the Born--Mayer-type spin-independent effective potential introduced above and interpret the coefficient $C$ as an effective parameter encoding the spin-statistical average over unresolved short-distance channels. The choice $C=3/4$ corresponds to an estimate for the minimal orbital-excitation energy of the maximally Pauli-blocked configuration, while the smaller values adopted to reproduce the geometrical cross section can be interpreted as effective averages over the various spin-symmetry sectors, resulting in a weaker net repulsive interaction.\footnote{The characteristic energy associated with color rearrangement is  expected to be of order $\epsilon_B/C_N$, corresponding to promoting a singlet baryon into an adjoint baryon. This provides an additional qualitative motivation for considering effective values $C<3/4$, particularly for $N>3$.}

\medskip

In order to estimate the theoretical uncertainty associated with the poorly known short-distance baryon-baryon interaction, we will compare $V_{\rm tot}$ results with the case without a repulsive core $V_{\rm reg}$. In addition to varying among our two choices of coefficient $C$, we also vary the matching scale $r_{\rm in}$ from $10 a_B$ down to $5\,a_B$ to further assess the robustness of our predictions.  However, for large $N$, we will only consider the exact large $N$ repulsion given in \cref{eq:largeN_repulsion}.

Notably,  the dimensionless combination $a_BV(r)$ depends only on the normalized distance $x=r/a_B$, the ratio $\xi=m_Q/\Lambda_d$, and the number of colors $N$.  It follows that the quantity $m_Q^2\sigma$ is also a function only of $\xi$ and $N$, see for instance \cref{eq:total_xs}.   Consequently, the general properties of the self-interaction cross section can be obtained by scanning $\xi$ for fixed $N$, while the absolute value of $m_Q$ merely determines the overall normalization of the cross section.

\begin{figure}[t]
    \centering \includegraphics[width=0.4825\linewidth]{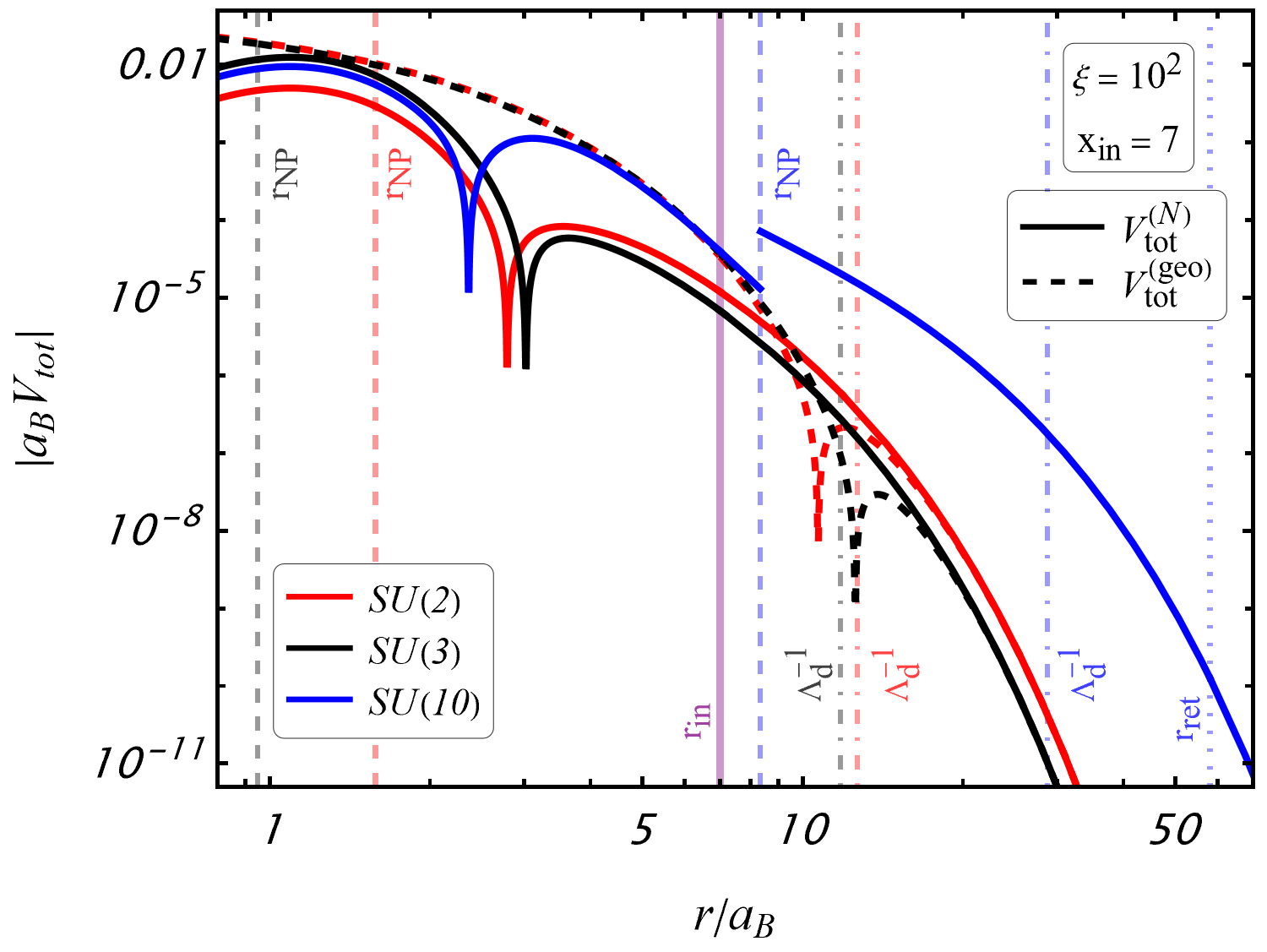} \includegraphics[width=0.5\linewidth]{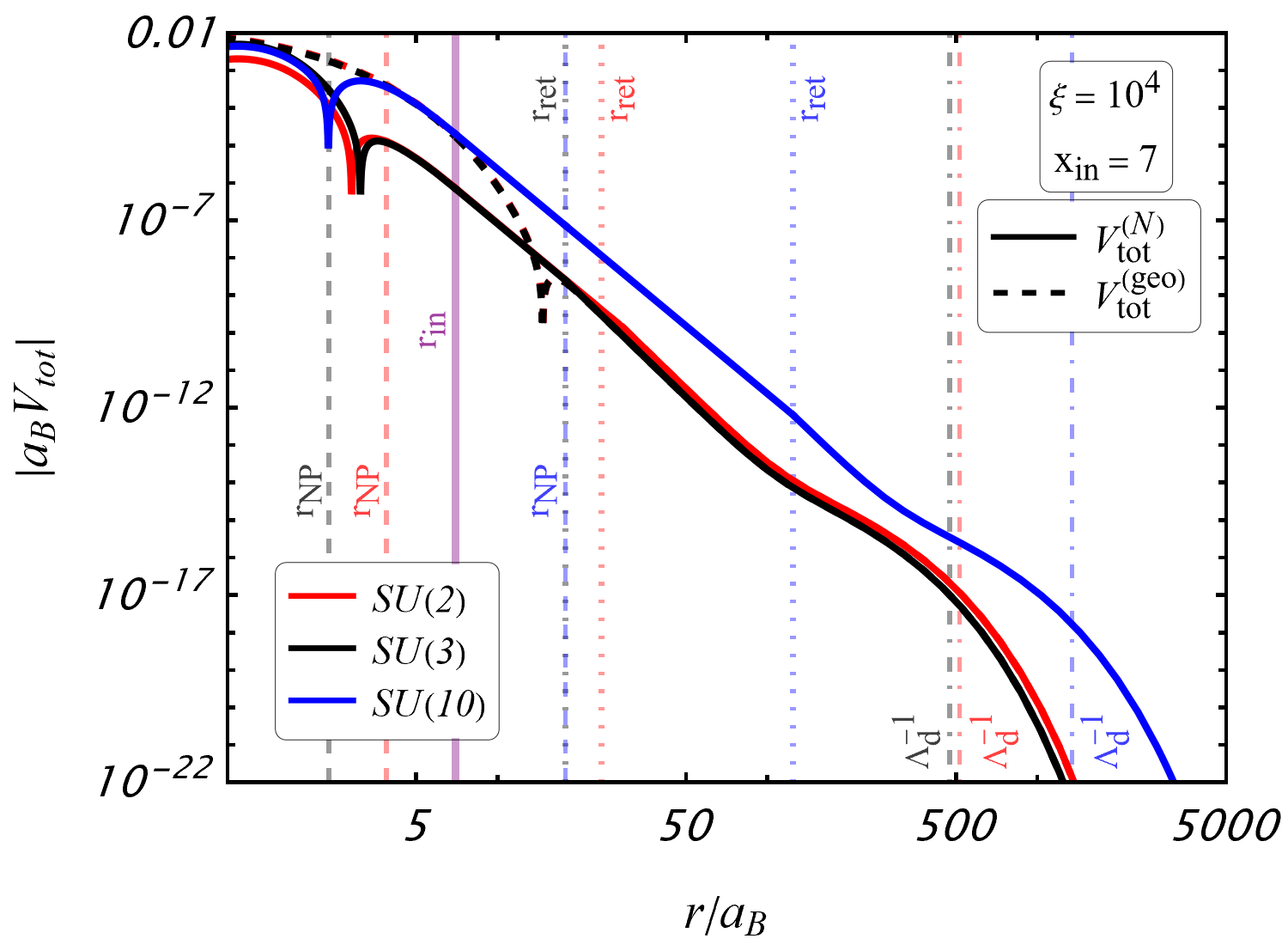}
\caption{The dark baryons interaction potential times the baryon size $|a_BV(x)|$ as a function of normalized distance $x=r/a_B$ for different numbers of colors $N=2,3,10$ with  $x_{\rm in}=7$ and a larger-$N$ repulsion (solid) as well as a Born–Mayer potential with  $C$ fixed by $\sigma_{\rm geo}$ (dashed).  Left and right panels correspond to $\xi=10^2$ and $10^4$, respectively---note the very different scales on the $x-$ and $y-$axes. The vertical lines correspond: solid purple to the small scale regularization $r_{\rm in}=7a_B$, colorful dashed to the non-perturbative truncation $r_{\rm NP}$, colorful dotted to the $r^{-6}\leftrightarrow r^{-7}$ matching point $r_{\rm ret}$, and colorful dot-dashed to the non-perturbative scale $r=\Lambda_d^{-1}$.
Note the turning points where the potential changes sign as it converts from attractive to repulsive, and the discontinuities for $\SU(10)$ which are due to our truncation of the non-perturbative contribution at small radii.  }
    \label{fig:potential}
\end{figure}

In particular, \cref{fig:potential} shows $|a_BV(x)|$ versus the normalized distance $x=r/a_B$ for $N=2,3,10$ with  a regularizing distance $r_{\rm in}=7 a_B$ and the larger-$N$ repulsive potential as well as the Born–Mayer potential with a value of $C$ fixed to reproduce the geometric cross section . We see that for small hierarchies $\xi \lesssim10^3$ the potential is strongly dominated by the non-perturbative Yukawa contribution, while for larger hierarchies the perturbative $r^{-6}$ and $r^{-7}$ are essential for accurately describing the attraction. For $\SU(10)$, we see the discontinuity due to our truncation of the non-perturbative result at small radii, being a conservative estimate as the potential is expected to get weaker smoothly.   This effect is not relevant for small $N$ as the truncation scale  $r_{\rm NP}$  is smaller than our regularization scale $r_{\rm in}$. Moreover, for large ratios $\xi$, the non-perturbative contributions are completely under control and we see no discontinuity for $\xi\gtrsim10^3$.

Alternative approaches to heavy hadron--hadron interactions have also been developed in the literature. For example, the Quark Delocalization Color Screening Model (QDCSM)~\cite{Huang:2020bmb,Wang:1992wi,Huang:2019esu} and dispersive approaches~\cite{Dong:2021lkh} rely on phenomenological ingredients calibrated to hadron spectroscopy and hadron--hadron data in QCD. In this work, we instead employ the operator product expansion (OPE) framework~\cite{Peskin:1979va,Bhanot:1979vb,Kharzeev:1995ij,Fujii:1999xn}, which is particularly well suited to the Coulombic regime $m_Q\gg\Lambda_d$ considered here. In this limit, the interaction admits a systematic multipole expansion and depends only on a small set of low-energy matrix elements, without requiring phenomenological input from hadron spectroscopy. This makes the formalism readily applicable to generic confining dark sectors. A closely related effective-field-theory framework is weakly coupled potential nonrelativistic QCD (pNRQCD), which derives heavy hadron--hadron interactions~\cite{Brambilla:2015rqa,Dong:2022rwr,Assi:2025ysr} through a systematic EFT expansion.

 Finally, having established both the long-distance induced-dipole interaction and phenomenological prescriptions for the short-distance behavior, we now possess a complete effective potential for dark baryon scattering. This potential constitutes the sole dynamical input required to determine the self-interaction cross section, to which we now turn.

\section{Dark matter self-interactions}
\label{sec:self_interactions}

Sizable DM self-interactions have been proposed as a possible solution to some
small-scale structure tensions between collisionless CDM simulations and
observations~\cite{Tulin:2013teo,Tulin:2017ara,Adhikari:2022sbh}, including the inferred diversity
of galactic rotation curves, the possible presence of cored density profiles in dwarf and low-surface-brightness (LSB) galaxies,
and the abundance and internal structure of satellite halos. The relevant quantity controlling
the impact of self-interactions on halo dynamics is the scattering rate, which depends on the cross section per dark
matter mass. Phenomenologically, cross sections of order~\cite{Kaplinghat:2015aga,Tulin:2017ara}
\begin{equation}
    {\sigma}/{m}
    \sim
    (1-10)~{\rm cm}^2/{\rm g}
    \label{eq:sidm_dwarfs}
\end{equation}
at dwarf-galaxy velocities, $v_{\rm dwarf}\sim 30~{\rm km/s}$, can transport
heat into the inner halo over a Hubble time, flattening the cusp into a core and erasing the
densest subhalos. By contrast, observations of
galaxy clusters, at $v_{\rm cluster}\sim 10^3~{\rm km/s}$, require substantially smaller effective cross sections, roughly~\cite{Markevitch:2003at,Balan:2024cmq}
\begin{equation}
    {\sigma}/{m}
    \lesssim
    {\cal O}(1)~{\rm cm}^2/{\rm g}\,.
    \label{eq:sidm_clusters}
\end{equation}
This is the main reason why velocity-dependent
self-interactions are required.

For distinguishable particles, the most commonly used quantity to describe momentum
redistribution in halos is the transfer cross section,
\begin{equation}
    \sigma_T
    =
    \int d\Omega\,
    (1-\cos\theta)\,
    \frac{d\sigma}{d\Omega}\,.
    \label{eq:transfer_xs}
\end{equation}
The angular weight \(1-\cos\theta\) suppresses forward scattering, which does not efficiently
exchange momentum between particles. However, for identical particles, backward scattering
is physically equivalent to exchanging the labels of the two particles. In that case the more
appropriate quantity is the viscosity cross section,\footnote{The transfer weight $1-\cos\theta$ is \emph{not} invariant under $\theta\to\pi-\theta$. For identical particles the forward and backward hemispheres
are the same physical final state, so this weight double-counts and is ambiguous. The
\emph{viscosity} weight $\sin^2\theta=1-\cos^2\theta$ \emph{is} invariant, and is the
standard transport cross section for identical particles \cite{Tulin:2017ara,Kahlhoefer:2013dca}.}
\begin{equation}
    \sigma_V
    =
    \int d\Omega\,
    \sin^2\theta\,
    \frac{d\sigma}{d\Omega}\,,
    \label{eq:viscosity_xs}
\end{equation}
which suppresses both forward scattering, \(\theta\simeq0\), and backward scattering,
\(\theta\simeq\pi\). Since dark baryon--dark baryon scattering corresponds to the scattering
of identical composite particles, the viscosity cross section is the natural quantity to use for
an asymmetric dark-baryon abundance. If instead the cosmology is symmetric, both identical
processes, \(BB\) and \(\bar B\bar B\), and distinguishable processes, \(B\bar B\), are present.
The former should be described by \(\sigma_V\), while the latter should be described by
\(\sigma_T\).

In the remainder of this section we first review the partial-wave formalism used to compute
the scattering cross section from the dark baryon potential derived in Sec.~\ref{sec:interaction_potential}.
We then describe the numerical implementation and the validity checks used in our analysis.

\subsection{Partial-wave formalism}
\label{subsec:partial_waves}

Our numerical partial-wave implementation follows the standard treatment of velocity-dependent SIDM scattering developed in Refs.~\cite{Tulin:2013teo,Kaplinghat:2013yxa,Tulin:2017ara}. For identical-particle scattering, we follow the symmetrized partial-wave treatment and the viscosity cross sections discussed in Refs.~\cite{Colquhoun:2020adl,Girmohanta:2022giy}.

The scattering of two dark baryons of mass \(m_B\) interacting through a central potential
\(V(r)\) can be reduced to a one-body problem with reduced mass $\mu = m_B/2$ and
relative momentum $k = \mu v$, where \(v\) is the relative velocity. For a static central potential $V(r)$ the scattering problem separates in partial waves.
Writing the reduced radial function $u_\ell(r)=r\,R_\ell(r)$, for each partial wave \(\ell\) the radial Schr\"odinger
equation in the centre-of-mass frame reads
\begin{equation}
    u_\ell''(r)
    +
    \left[
        k^2
        -
        \frac{\ell(\ell+1)}{r^2}
        -
        2\mu V(r)
    \right]
    u_\ell(r)
    =
    0\,,
\label{eq:radial_schr}
\end{equation}
with $u_\ell(0)=0$. At sufficiently large radius, where the potential is negligible, the solution approaches
\begin{equation}
    u_\ell(r) {\longrightarrow} \;r\left[\cos \delta_\ell j_\ell(kr) - \sin \delta_\ell n_\ell(kr)
    \right],
    \label{eq:phase_shift}
\end{equation} where here $j_\ell\,(n_\ell)$ is the spherical Bessel (Neumann) function. This solution defines the phase shift \(\delta_\ell(k)\). For distinguishable particles, the scattering amplitude is
\begin{equation}
    f(\theta)
    =
    \frac{1}{k}
    \sum_{\ell=0}^{\infty}
    (2\ell+1)
    e^{i\delta_\ell}
    \sin\delta_\ell\,
    P_\ell(\cos\theta)\,,
    \label{eq:partial_wv}
\end{equation}
and the differential cross section is
\begin{equation}
    \frac{d\sigma}{d\Omega}
    =
    |f(\theta)|^2\,.
\end{equation}
The total scattering cross section is then
\begin{equation}
    \sigma
    =
    \frac{4\pi}{k^2}
    \sum_{\ell=0}^{\infty}
    (2\ell+1)
    \sin^2\delta_\ell \,.
    \label{eq:total_xs}
\end{equation}
The transfer cross section can be written directly in terms of phase-shift differences as ~\cite{Tulin:2013teo}
\begin{equation}
    \sigma_T
    =
    \frac{4\pi}{k^2}
    \sum_{\ell=0}^{\infty}
    (\ell+1)
    \sin^2\left(\delta_{\ell+1}-\delta_\ell\right)\,.
    \label{eq:transfer_xs_pw}
\end{equation}
For identical particles the amplitude must be symmetrized or antisymmetrized,
\begin{equation}
    f_\pm(\theta)
    =
    f(\theta)
    \pm
    f(\pi-\theta)\,,
    \label{eq:symmetrized_amplitude}
\end{equation}
where \(+\) corresponds to a spatially symmetric state and \(-\) to a spatially antisymmetric
state. Using $P_\ell(-\cos\theta)=(-1)^\ell P_\ell(\cos\theta)$, it is easy to check that the symmetric amplitude contains only even partial waves, while the
antisymmetric amplitude contains only odd partial waves:
\begin{equation}
    f_+(\theta)
    =
    \frac{2}{k}
    \sum_{\ell~{\rm even}}
    (2\ell+1)
    e^{i\delta_\ell}
    \sin\delta_\ell\,
    P_\ell(\cos\theta)\,,
    \label{eq:even_amplitude}
\end{equation}
\begin{equation}
    f_-(\theta)
    =
    \frac{2}{k}
    \sum_{\ell~{\rm odd}}
    (2\ell+1)
    e^{i\delta_\ell}
    \sin\delta_\ell\,
    P_\ell(\cos\theta)\,.
    \label{eq:odd_amplitude}
\end{equation}
In the one-flavor theory considered here, the lightest dark baryon has spin $S = N/2$. Thus \(N=2,10\) correspond to bosonic dark baryons, while \(N=3\) corresponds to fermionic
dark baryons. Denoting the number of spin states by $g = 2S+1 = N+1$, the spin-averaged weights of the even and odd spatial channels of bosons and fermions are
\begin{align}
    w_{\rm even}^{\rm boson}&=w_{\rm odd}^{\rm fermion}
    =
    \frac{g+1}{2g}\,,\\
    \qquad
    w_{\rm odd}^{\rm boson}
    &= w_{\rm even}^{\rm fermion}=
    \frac{g-1}{2g}\,.
\label{eq:boson_even_odd_weights}
\end{align}

\begin{table}[t]
    \centering
    \caption{Spin, statistics, and even- and odd-channel weights for different
    values of \(N\).}
    \label{tab:spin-statistics-weights}
    \vspace{0.3cm}
    \begin{tabular}{cccc}
        \hline
        \(N\) & \(S=N/2\) & Statistics & \((w_{\rm even},w_{\rm odd})\) \\
        \hline
        \(2\)  & \(1\)   & Boson   & \((2/3,\,1/3)\) \\
        \(3\)  & \(3/2\) & Fermion & \((3/8,\,5/8)\) \\
        \(10\) & \(5\)   & Boson   & \((6/11,\,5/11)\) \\
        \hline
    \end{tabular}
\end{table}
For reference, we show in Table~\ref{tab:spin-statistics-weights} the relative weights depending on the value of \(N\). Thus the spin-averaged identical-particle differential cross section is
\begin{equation}
    \frac{d\sigma}{d\Omega}
    =
    w_{\rm even}
    \left|f_+(\theta)\right|^2
    +
    w_{\rm odd}
    \left|f_-(\theta)\right|^2,
    \label{eq:spin_averaged_dsigma}
\end{equation}
with the standard identical-particle convention of integrating over a hemisphere, or equivalently including the corresponding double-counting factor.

For a pure $s$-wave ($\ell=0$, even, isotropic $f_0$) one has $f_-=0$ and $f_+=2f_0$, so
\begin{equation}
\frac{\dd\sigma}{\dd\Omega}=4\,w_{\rm even}\,|f_0|^2\,,
\qquad
\sigma_{\rm indist}=2\,w_{\rm even}\,\sigma_{\rm dist}\,,
\end{equation}
i.e.\ identical \emph{bosons enhance} the $s$-wave (spin-$0$ limit: $2\times$) while identical
\emph{fermions suppress} it (spin-$\tfrac12$ limit: $\tfrac12\times$). For the dark baryons this
$O(1)$ factor acts in opposite directions for even vs.\ odd $N$ and must be included.

The spin-averaged viscosity cross section, used for identical dark baryon scattering, is therefore
\begin{equation}
    \sigma_V
    =
    w_{\rm even}\,\sigma_V^{\rm even}
    +
    w_{\rm odd}\,\sigma_V^{\rm odd}.
    \label{eq:spin_averaged_viscosity}
\end{equation}
with
\begin{equation}
    \sigma_V^{\rm even}
    =
    \frac{8\pi}{k^2}
    \sum_{\ell~{\rm even}}
    \frac{(\ell+1)(\ell+2)}{2\ell+3}
    \sin^2\left(\delta_{\ell+2}-\delta_\ell\right),
    \label{eq:viscosity_even}
\end{equation}
\begin{equation}
    \sigma_V^{\rm odd}
    =
    \frac{8\pi}{k^2}
    \sum_{\ell~{\rm odd}}
    \frac{(\ell+1)(\ell+2)}{2\ell+3}
    \sin^2\left(\delta_{\ell+2}-\delta_\ell\right).
    \label{eq:viscosity_odd}
\end{equation} Note that for large $N$, both weights converge to 1/2 such that $\sigma_V $ simplifies to \begin{equation}
     \sigma_V
    \approx
    \frac{4\pi}{k^2}
    \sum_{\ell}
    \frac{(\ell+1)(\ell+2)}{2\ell+3}
    \sin^2\left(\delta_{\ell+2}-\delta_\ell\right).
\end{equation}

One should also be careful about the difference between symmetric and asymmetric dark baryon abundances. In the symmetric case, both identical-particle scatterings, \(BB\) and \(\bar B\bar B\), and distinguishable-particle scatterings, \(B\bar B\), are present. The former should be described using \(\sigma_V\), while the latter should use \(\sigma_T\), with \(n_B=n_{\bar B}=n/2\). In the asymmetric case, the relevant process is only \(BB\) scattering, so one should use \(\sigma_V\) with the full DM number density \(n\). {\it In the remainder, we will focus on the asymmetric paradigm,  justifying this choice of scenario in \cref{sec:asymmetric_DM}.}

\subsection{Numerical implementation}
\label{subsec:numerical_implementation}

To numerically solve \cref{eq:radial_schr}, it is useful to define the  dimensionless  variables~\cite{Buckley:2009in,Tulin:2013teo} \begin{align}
   x\equiv r/a_B\, ,&& a\equiv k a_B
\end{align} such that the radial equation reduces to \begin{equation}\label{eq:radial_schr_normalized}
    u_\ell''(x)
    +
    \left(
        a^2
        -
        \frac{\ell(\ell+1)}{x^2}
        -
        2\,\mu\, a_B[a_B V(x)]
    \right)
    u_\ell(x)
    =
    0\,.
\end{equation} Note that $a$, $\mu\, a_B$  and $ a_B V(x)$ are independent of the quark mass $m_Q$ such that the velocity evolution of the cross section is not affected by the specific choice of the dark quark mass scale (which only affects the overall normalization of the cross section).

For each point in parameter space, we compute the phase
shifts \(\delta_\ell\) by solving \cref{eq:radial_schr_normalized} for the
regularized baryon-baryon potential derived in Sec.~\ref{sec:interaction_potential}. The integration is
started at a radius \(x_{\rm min}\) where the potential is negligible compared to the angular momentum term (for $\ell=0$, we start at $x_{\rm min}=\min[3x_{\rm in},1/a]$). The initial condition is chosen to match the regular solution near the origin,
\begin{equation}
    u_\ell(x_{\rm min}) = x_{\rm min}^{\ell+1}\qquad \text{and}\qquad u'_\ell(x_{\rm min}) = (\ell +1)x_{\rm min}^{\ell}\,.
\end{equation}
The solution is then evolved to a radius \(x_{\rm max}\) where
\begin{equation}
    |a_B V(x_{\rm max})| \ll \frac{a^2}{2\mu a_B}\,,
\end{equation}
so that the wave function can be matched to the free asymptotic form in
\cref{eq:phase_shift}.   This yields the phase shift~\cite{Tulin:2013teo} \begin{equation}
\tan\delta_\ell
=
\frac{a x_{\rm max} j_\ell'(a x_{\rm max}) - \beta_\ell\, j_\ell(a x_{\rm max})}
     {a x_{\rm max} n_\ell'(a x_{\rm max}) - \beta_\ell\, n_\ell(a x_{\rm max})}
\qquad \text{with }
\beta_\ell = x_{\rm max} \frac{u_\ell'(x_{\rm max})}{u_\ell(x_{\rm max})} - 1\,,
\label{eq:phase_shift2}
\end{equation} where  $u_\ell(x_{\rm max})$ is our numerical solution to \cref{eq:radial_schr_normalized}  at $x_{\rm max}$ and $u_\ell'(x_{\rm max})$ is its derivative. In practice, \(x_{\rm min(max)}\) is chosen  several times, each time we decrease (increase) its value and recompute $\delta_\ell$ until the change in the resulting $\delta_\ell$ is smaller than 1\%.

The partial-wave sum is truncated at a finite \(\ell_{\rm max}\). We increase \(\ell_{\rm max}\)
until the cross section changes by less than 1\%.\footnote{ In case of slow convergence, we in fact truncate the $\ell$-sum at $\ell=40$   and perform at most 50 simultaneous variations of $x_{\rm min}$ and $x_{\rm max}$ for the computation $\delta_\ell$.}  A useful estimate
for the required number of modes is
\begin{equation}
    \ell_{\rm max}
    \gtrsim
    k r_{\rm eff}\,,
    \label{eq:lmax_estimate}
\end{equation}
where   \(r_{\rm eff}\) is the effective range over which the potential produces a significant
phase shift.

As a consistency check, we compare the partial-wave result with the Born approximation whenever the potential is weak. A useful criterion is
\begin{equation}
    2\mu
    \int dr\, r\, |V(r)|
    \ll
    1\,.
    \label{eq:born_validity}
\end{equation}
In the Born regime, the wave function is only weakly distorted
from a free wave, and the differential cross section is approximately
\begin{equation}
    \frac{d\sigma}{d\Omega}
    \simeq
    \frac{\mu^2}{4\pi^2}
    \left|
        \widetilde V(q)
    \right|^2\,,
    \qquad
    q = 2k\sin\frac{\theta}{2}\,,
    \label{eq:born_differential_cross_section}
\end{equation}
where
\begin{equation}
    \widetilde V(q)
    =
    \int d^3r\, e^{i\mathbf q\cdot\mathbf r} V(r)
\end{equation}
is the Fourier transform of the potential,  with $\mathbf q$ the momentum transfer and $\theta$ the centre-of-mass angle. As an analytic anchor, we verified that our numerical methods result in cross sections matching the $s$- and higher-wave Born sum for a Yukawa potential
which has an analytic closed form. We also checked that our numerical implementation of the partial waves method reproduces standard Yukawa results for $\sigma_T$ and $\sigma_V$ of Refs.~\cite{Tulin:2013teo,Figueroa:2026bmi}.

Outside of the Born regime, the full partial-wave computation is required. This is especially important in the Yukawa-dominated regime, where attractive potentials can support quasi-bound states and generate resonant enhancements or Ramsauer--Townsend suppressions of the scattering cross section. We also cross-check our results against known Yukawa fitting formulae in the classical regime~\cite{Tulin:2013teo,Tulin:2017ara}. \medskip

\subsection{Velocity-averaged cross sections}
Astrophysical systems probe a distribution of relative velocities rather than a single
velocity. Therefore, we compute the velocity-averaged cross section
\begin{equation}
    \overline{\sigma}_X
    =
    \frac{\langle \sigma_X v^\gamma\rangle}{\langle v^\gamma\rangle}
    =
    \frac{
    \int_0^\infty dv\,
    f_{\rm rel}(v)\,
    v^\gamma\,\sigma_X(v)
    }{
    \int_0^\infty dv\,
    f_{\rm rel}(v)\,
    v^\gamma
    }\,,
\label{eq:velocity_averaged_cross_section}
\end{equation}
where $\gamma=1$ is used for rate-weighted averaging, while  $\gamma=3$ is employed for energy-transfer weighting, often employed in studies of core formation.
Here \(X=T,V\) refer to the transfer and viscosity cross-sections, respectively, and \(f_{\rm rel}(v)\) is the relative-velocity distribution of the system, which we take to be a Maxwell--Boltzmann distribution,
\begin{equation}
    f_{\rm rel}(v)
    =
    \frac{4}{\sqrt{\pi}}
    \frac{v^2}{v_0^3}
    e^{-v^2/v_0^2}\,,
\label{eq:maxwell_boltzmann_relative_velocity}
\end{equation}
with \(v_0\) the most probable relative speed. Since
\begin{equation}
    \langle v\rangle
    =
    \int_0^\infty dv\,
    f_{\rm rel}(v)\,v
    =
    \frac{2v_0}{\sqrt{\pi}}\,,
\end{equation}
\cref{eq:velocity_averaged_cross_section} can be evaluated directly once \(\sigma_X(v)\) is known.

In the absence of narrow resonances, using
\(\sigma_X(v)\) at a representative velocity gives a reasonable estimate. However, in the
resonant regime the velocity average can differ substantially from the single-velocity result,
and \cref{eq:velocity_averaged_cross_section} should be used when comparing to
dwarf, galactic, and cluster systems. As we will check, the qualitative conclusions remain unchanged under either averaging prescription.

\subsection{Validity region}

Before presenting our numerical results, we summarize the conditions under which the effective description developed in the previous sections is expected to be reliable.

First, the scattering process should not resolve the internal structure of the dark baryons. Since the de Broglie wavelength of the relative motion is of order $k^{-1}$, this requires that the interaction is dominated by distances larger than the baryon size. In our framework this is ensured by the short-distance repulsive core, which prevents the wave function from significantly probing the region $r\lesssim a_B$, provided
\begin{equation}
    V(a_B)\gg \frac{\mu v^2}{2}.
\end{equation}
Using our phenomenological repulsive potential with the smallest value adopted in this work, $C=0.23$, one finds
\begin{equation}
    4V(a_B)=4C\,\epsilon_B\,e^{-1}
    >
    0.3\,\epsilon_B
    \gg
    m_Bv_{\rm max}^2,
\end{equation}
where $v_{\rm max}=v_{\rm cluster}\simeq1500~{\rm km/s}$ is the largest velocity relevant for the astrophysical systems under consideration. Using the binding energies derived in Sec.~\ref{sec:2p1p1_baryon_properties}, this condition becomes
\begin{equation}
    \alpha_B^2(N-1)\gg6N^2v_{\rm max}^2,
\end{equation}
which is comfortably satisfied throughout the parameter space explored in this work (see \cref{fig:couplingXratio}).

A related requirement is that the shortest wavelength probed during the collision remains  larger than the radius where the phenomenological regularization is introduced,
\begin{equation}
    k_{\rm max}^{-1}>r_{\rm in}.
\end{equation}
This condition excludes only a small corner of parameter space corresponding to very large hierarchy ratios $\xi$, where the baryons become extremely compact. As will be shown in Sec.~\ref{sec:results}, these regions are not relevant for obtaining phenomenologically interesting self-interaction cross sections.

A second consistency requirement is that the scattering remains purely elastic. The maximum kinetic energy available in the center-of-mass frame,
\begin{equation}
    E_{\rm max}\simeq\frac{\mu v_{\rm max}^2}{2},
\end{equation}
must therefore be smaller than both the baryon binding energy and the lightest glueball mass,
\begin{equation}
    E_{\rm max} <\frac{3}{4}\epsilon_B,
    \qquad
    E_{\rm max}< m_{\rm DG}.
\end{equation}
These conditions ensure that collisions cannot excite internal baryonic states, dissociate the baryons, or produce dark glueballs. Again, thanks to the smallness of $v_{\rm max}^2$, the elasticity conditions are satisfied over the parameter space considered throughout our self-interaction analysis.

Finally, our entire description relies on the heavy-quark hierarchy discussed in Sec.~\ref{sec:2p1p1_baryon_properties}. Quantitatively, the Coulombic description is expected to be reliable for
\begin{equation}
    \xi\gtrsim10,
\end{equation}
while $\xi\gtrsim5$ should still provide qualitatively meaningful results. We therefore do not consider smaller values of $\xi$, where confinement and relativistic effects become comparable to the Coulombic dynamics and the approximations underlying the present analysis cease to be controlled.

\section{Numerical results }\label{sec:results}

Having derived the effective dark baryon interaction potential and described the computation of the viscosity cross section along with the domain of validity of our calculations, we now present our numerical results. We begin by validating our numerical implementation, demonstrating the convergence of the partial-wave expansion and its agreement with the Born approximation in the perturbative regime. We also show that the inclusion of the phenomenological repulsive core naturally  reproduces the expected geometrical scattering limit when the interaction is dominated by  the  baryon overlap.

We then investigate the velocity dependence of the self-interaction
cross section, studying its sensitivity to the phenomenological
short-distance parameters. Although quantitative differences arise,
the large-$N$ regime exhibits regions of parameter space where the
cross section is significantly enhanced at dwarf-galaxy velocities
while remaining compatible with cluster-scale bounds. For intermediate
values of $N$, the two approximations discussed in Sec.~3 do not give
quantitatively consistent predictions, so we focus here on $N=2,3$ and
the large-$N$ regime, in particular the $\SU(10)$ case. The former serve as a comparison showing the
limitations of the small-$N$ realization, while the latter provides the
phenomenologically relevant velocity dependence.

Finally, fixing representative values of the phenomenological parameters, we identify the dark quark masses and dark confinement scales that can solve small-scale structure problems and briefly comment on their implications for heavy dark baryons as self-interacting Asymmetric Dark Matter (ADM)~\cite{Kaplan:2009ag}. We also explore the very low-velocity regime, $v\sim5~{\rm km/s}$, where recent observations~\cite{Powell:2025rmj} have motivated considerably larger self-interaction cross sections~\cite{Vegetti:2026mmx,Zhang:2026gur}.

\subsection{Cross section overview}

\begin{figure}[t]
        \centering
    \includegraphics[width=0.475\linewidth]{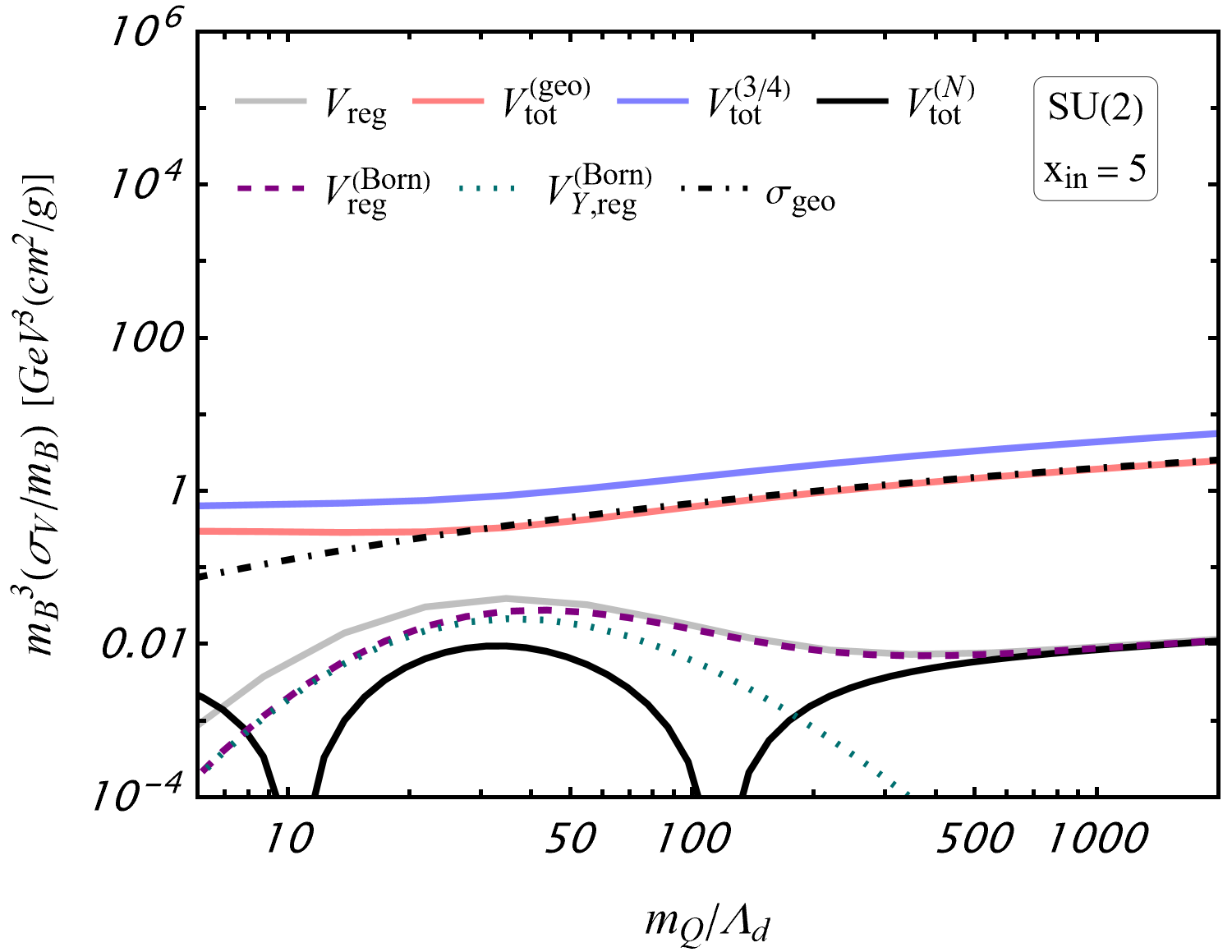}\hspace{0.1cm}
    \includegraphics[width=0.475\linewidth]{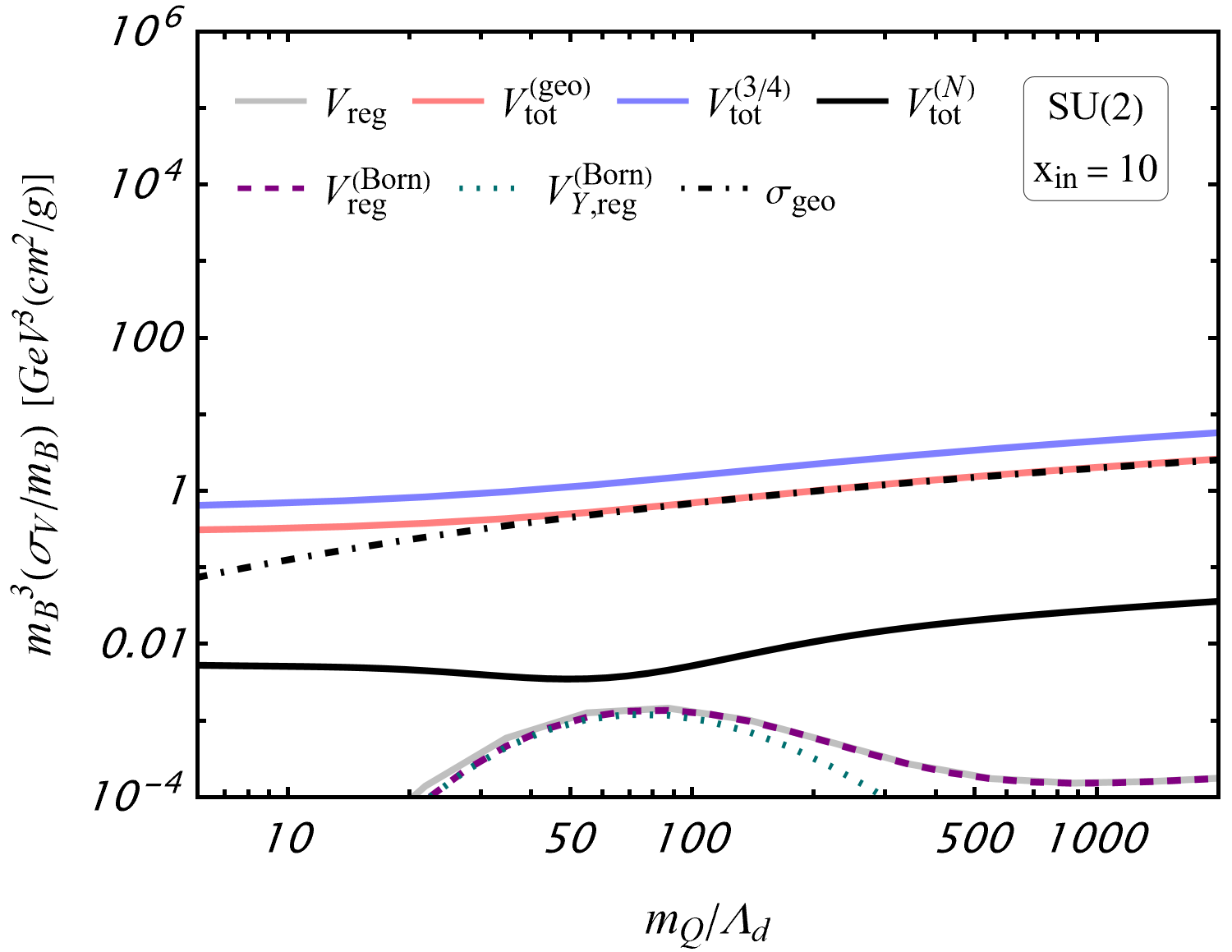}

    \includegraphics[width=0.475\linewidth]{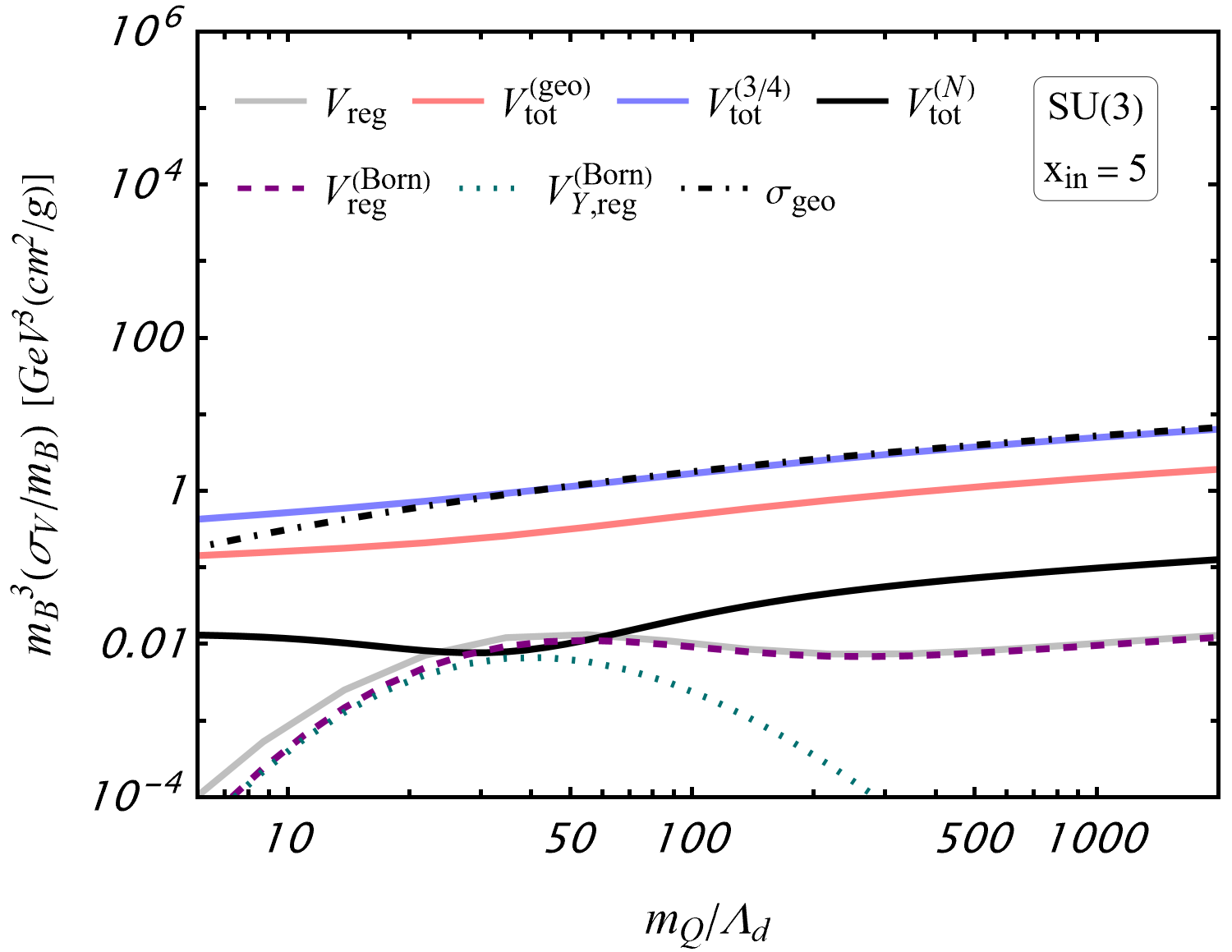}\hspace{0.1cm}
    \includegraphics[width=0.475\linewidth]{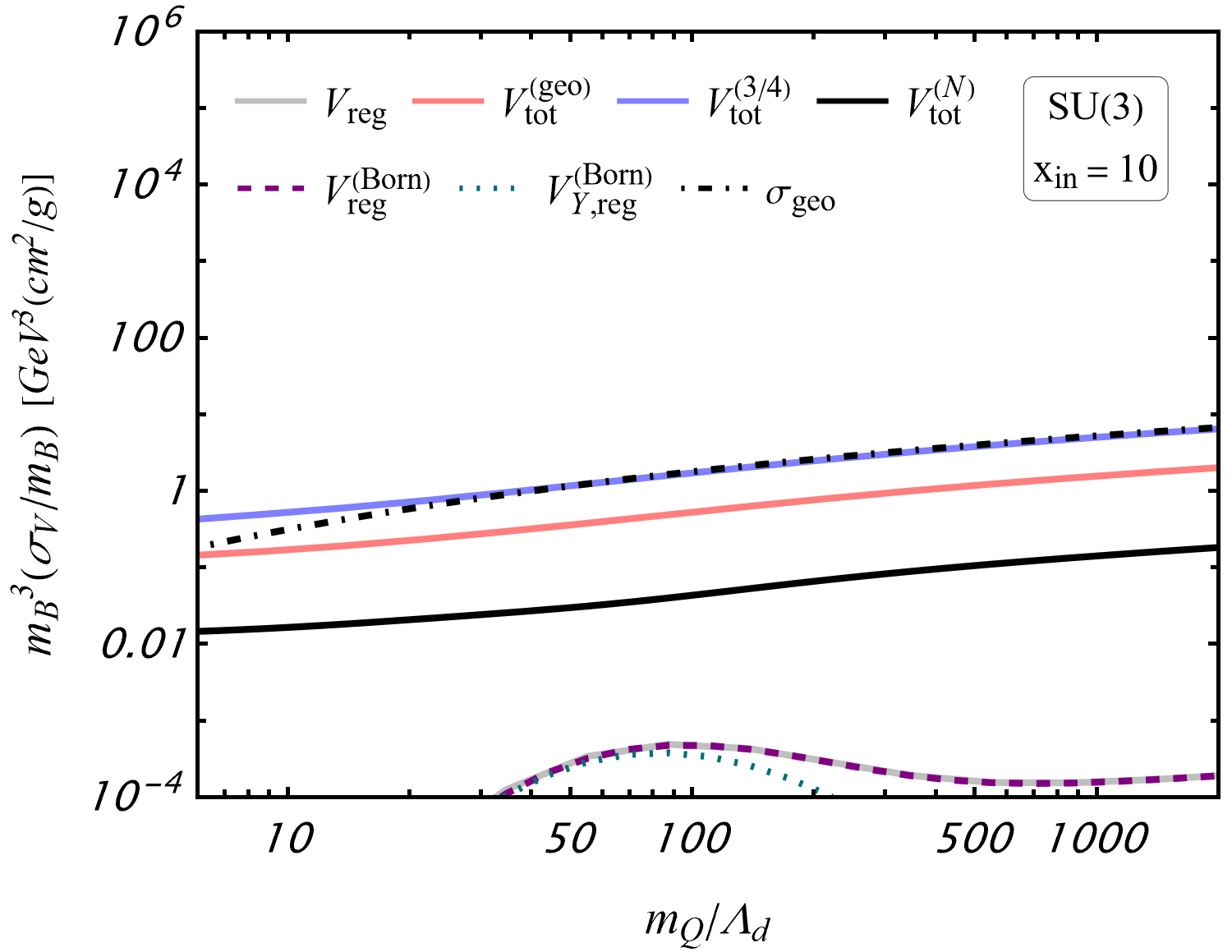}

    \includegraphics[width=0.475\linewidth]{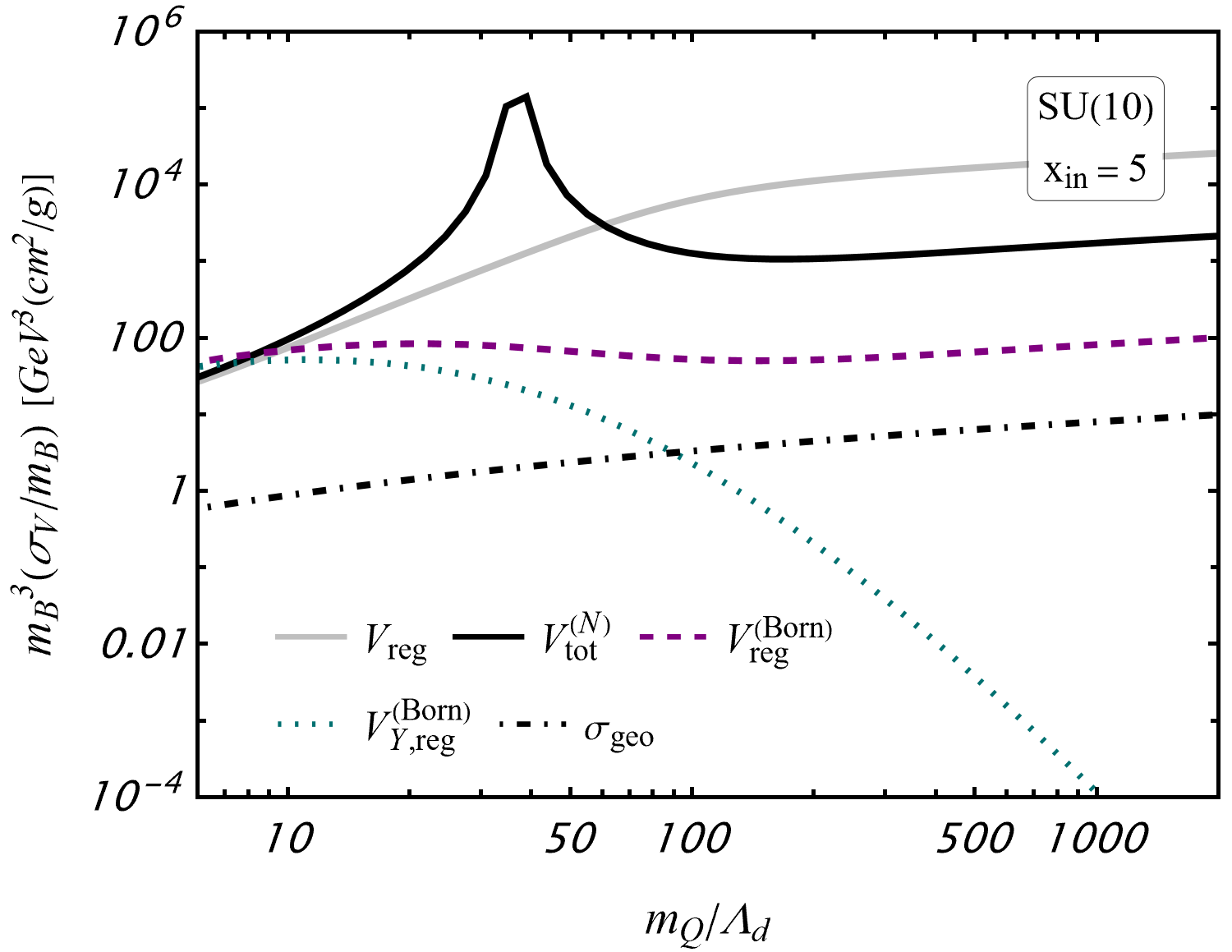}\hspace{0.1cm}
    \includegraphics[width=0.475\linewidth]{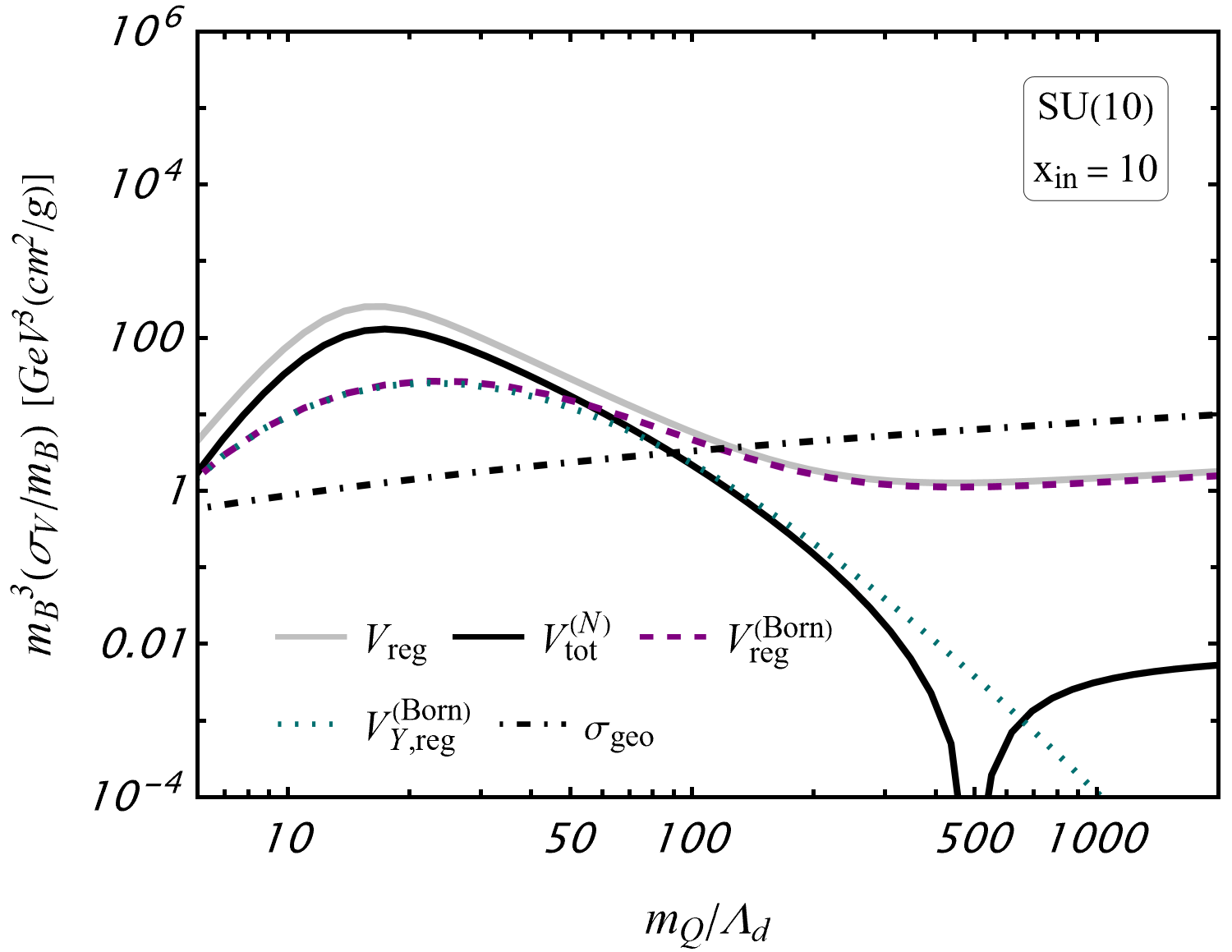}

    \caption{ Evolution of the self-interaction cross section over DM mass $\sigma_V/m_{B}$ (normalized by $m_B^{3}$)  as a function of the hierarchy parameter $\xi=m_Q/\Lambda_d$ for low velocities, $v_{\rm dwarf}=30$~km/s. The gray, red, blue and black lines correspond to cross sections computed via the partial wave method using the respective potentials: $V_{\rm reg}$, $V_{\rm tot}$ with $C$ fixed to reproduce the geometric cross section $\sigma_{\rm geo}$ (in black dot-dashed) denoted by $V_{\rm tot}^{(\rm geo)}$,  $V_{\rm tot}$ with $C=3/4$ denoted by $V_{\rm tot}^{(3/4)}$, and $V_{\rm tot}$ with the large-$N$ repulsive potential denoted by $V_{\rm tot}^{(N)}$; while the dashed purple and dotted teal  correspond to cross sections computed via the Born approximation using $V_{\rm reg}$ and $V_{\rm Y, reg}$, respectively.  Top, middle and bottom correspond respectively to $N=2,3,10$ while left and right panels to $x_{\rm in}=5,10$.  }
    \label{fig:cross_section_flow}
\end{figure}

\Cref{fig:cross_section_flow} shows the dimensionless combination
$m_B^3(\sigma_V/m_B)$ as a function of the hierarchy parameter
$\xi\equiv m_Q/\Lambda_d$ for $N=2,3,10$ at the representative
dwarf-galaxy velocity $v_{\rm dwarf}=30~\mathrm{km/s}$. Besides the full
regularized potential $V_{\rm reg}$ and the total potential $V_{\rm tot}$,
which includes the phenomenological repulsive core, we also compute the
cross section using the regulated Yukawa potential alone,
\begin{equation}
V_{\rm Y, reg}
=
-
\frac{d_2^2}{16\pi^2}\,
C_Y\,
\frac{e^{-m_{\rm DG}r}}{r}\times\begin{cases}
    e^{-\left(\frac{r_{\rm in}}{r}-1\right)^2}
\,,
&
\text{ for } r \leq r_{\rm in}\,, \\
1
\,,
&\text{ for } r > r_{\rm in}\,,
\end{cases}
\end{equation}
which isolates the purely non-perturbative contribution to the
attractive interaction. For compactness, we denote $V_{\rm tot}$ with
$C$ fixed to reproduce the geometric cross section
$\sigma_{\rm geo}$ by $V_{\rm tot}^{(\rm geo)}$, and $V_{\rm tot}$ with
$C=3/4$ by $V_{\rm tot}^{(3/4)}$.

The figure illustrates the transition between the different dynamical
regimes discussed in Sec.~\ref{sec:interaction_potential}. For smaller
hierarchies, $\xi\lesssim10^3$, the interaction is almost entirely
controlled by the non-perturbative Yukawa exchange, so $V_{\rm Y,reg}$ and
$V_{\rm reg}$ give nearly identical cross sections. At larger
hierarchies, the Yukawa contribution becomes strongly suppressed and
the perturbative induced-dipole interaction dominates the attraction.
Consequently, for $\xi\gg100$, the cross section obtained from
$V_{\rm Y,reg}$ rapidly falls below that obtained from the complete
regularized potential.

Once the phenomenological repulsive core is included, the behavior changes qualitatively.  For $N=2$ and $3$, the short-distance repulsion
dominates the scattering, driving the cross section close to the
geometrical limit. In contrast, for $N=10$ the induced-dipole attraction
remains comparable to the repulsion even at large $\xi$, owing to the
significantly larger chromo-electric polarizability coefficient $d_2$.
The attractive contribution remains relevant for $x_{\rm in}\lesssim10$,
whereas for larger suppression radii the cross section progressively
approaches a geometrical behavior. As discussed in
\cref{sec:regularizing_potential}, the physically motivated range of
the regularization prescription is expected to lie within
$x_{\rm in}\lesssim10$.

\Cref{fig:cross_section_flow} also provides a useful validation of our numerical implementation. As expected, the Born approximation accurately reproduces the partial-wave calculation whenever the interaction potential is weak. This occurs naturally at large hierarchy ratios, while deviations
appear as $\xi$ decreases and the Yukawa interaction becomes
sufficiently strong to invalidate the Born approximation. We have additionally verified that our partial-wave implementation reproduces  standard Yukawa results for both the transfer and viscosity cross sections presented in Refs.~\cite{Tulin:2013teo,Figueroa:2026bmi}. We restrict the comparison between the Born and partial-wave methods to the attractive interaction alone, since the phenomenological repulsive core with $C\gtrsim0.1$ already lies outside the weak potential regime.

Finally, we note that when adopting the benchmark value $C=3/4$, the repulsive potential produces cross sections typically within a factor of a few of the geometrical limit. Nevertheless, the discussion of \cref{sec:regularizing_potential} suggests that the physically relevant spin-averaged repulsion should generally correspond to smaller effective values of $C$. Indeed, the hydrogen atom provides a simple analogy: the triplet
channel experiences exchange repulsion whereas the singlet channel does
not, so that spin averaging reduces the effective interaction relative
to the maximally repulsive triplet channel. Likewise, averaging over
the different spin (and potentially color) configurations of two dark
baryons is expected to reduce the effective strength of the repulsive
core, motivating the smaller values of $C$ considered throughout this
work.

\subsection{Velocity dependence}

Having established the behavior of the different contributions to the potential and validated our numerical approach, we now turn to its velocity dependence, which is the key ingredient required for simultaneously addressing the small-scale structure anomalies while satisfying galaxy-cluster constraints.

\begin{figure}
        \centering
    \includegraphics[width=0.475\linewidth]{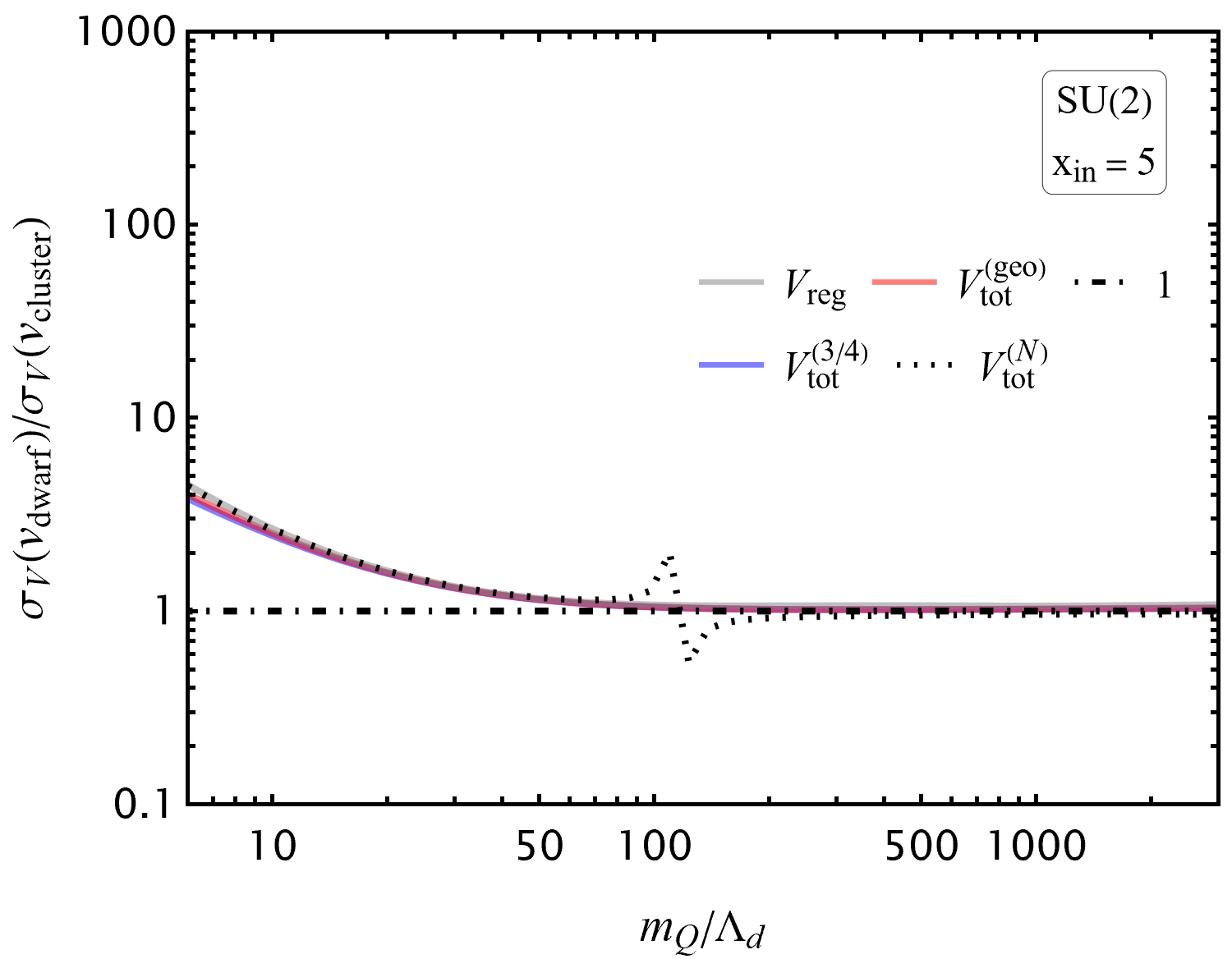}\hspace{0.1cm}
    \includegraphics[width=0.475\linewidth]{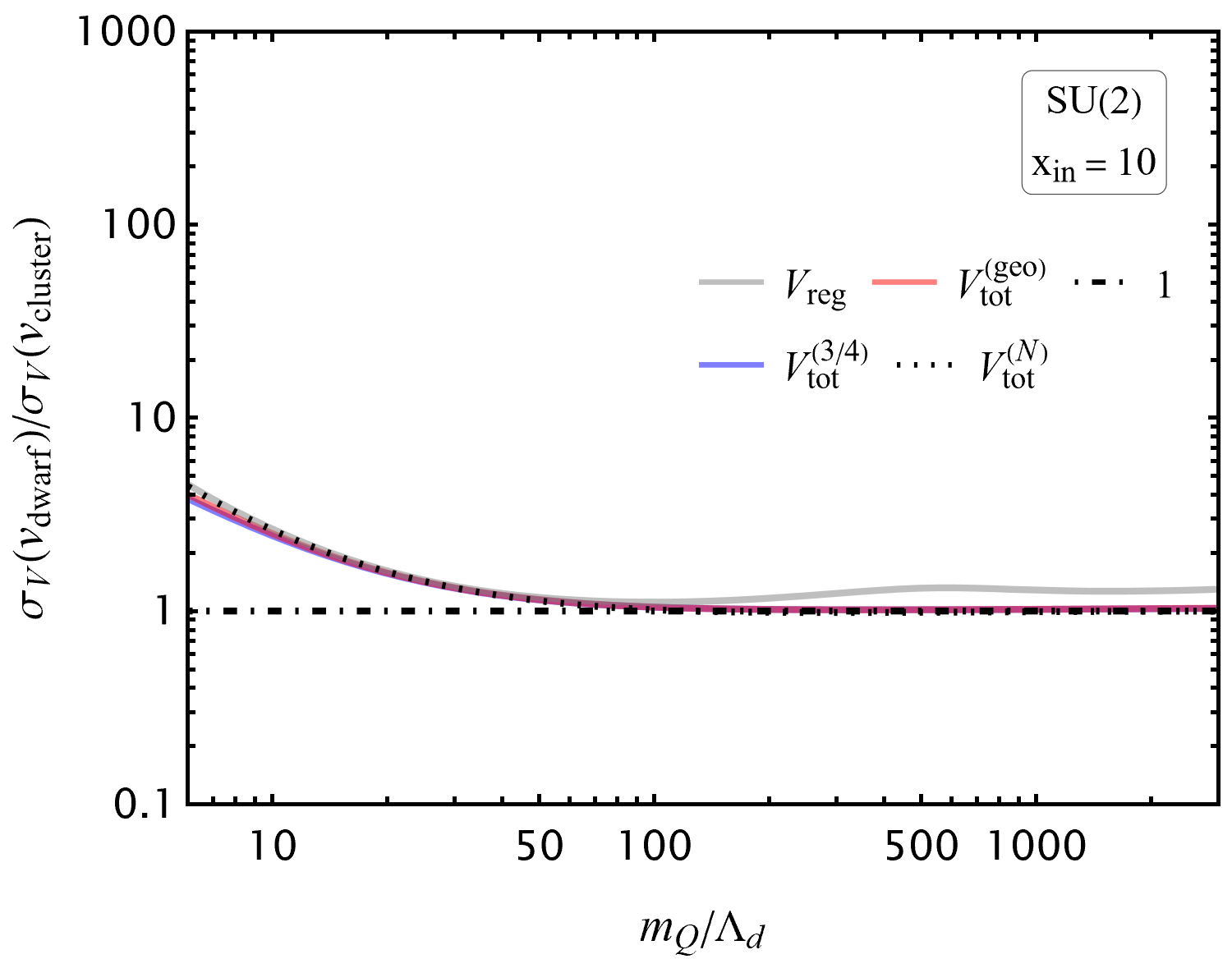}

    \includegraphics[width=0.475\linewidth]{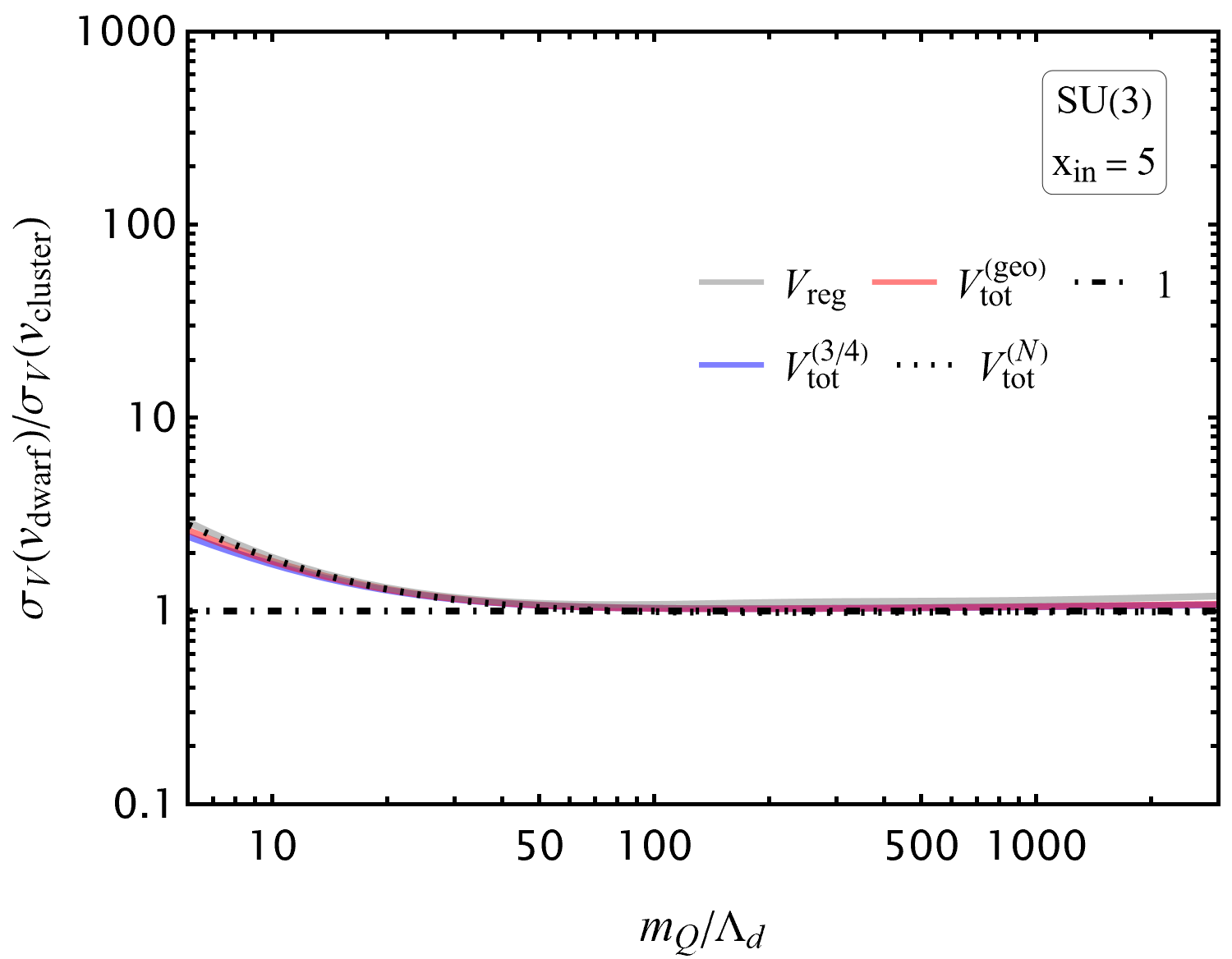}\hspace{0.1cm}
    \includegraphics[width=0.475\linewidth]{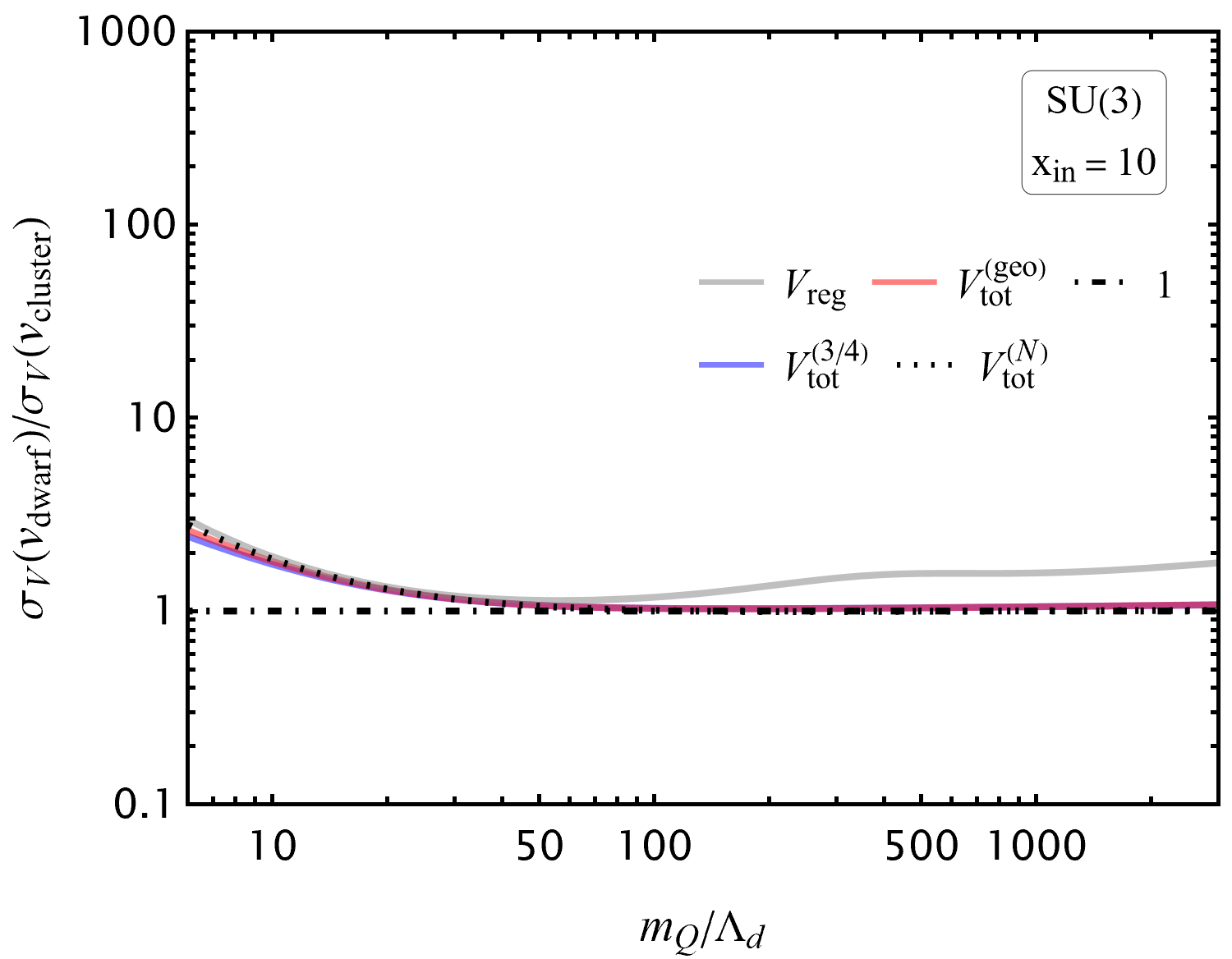}.

    \includegraphics[width=0.475\linewidth]{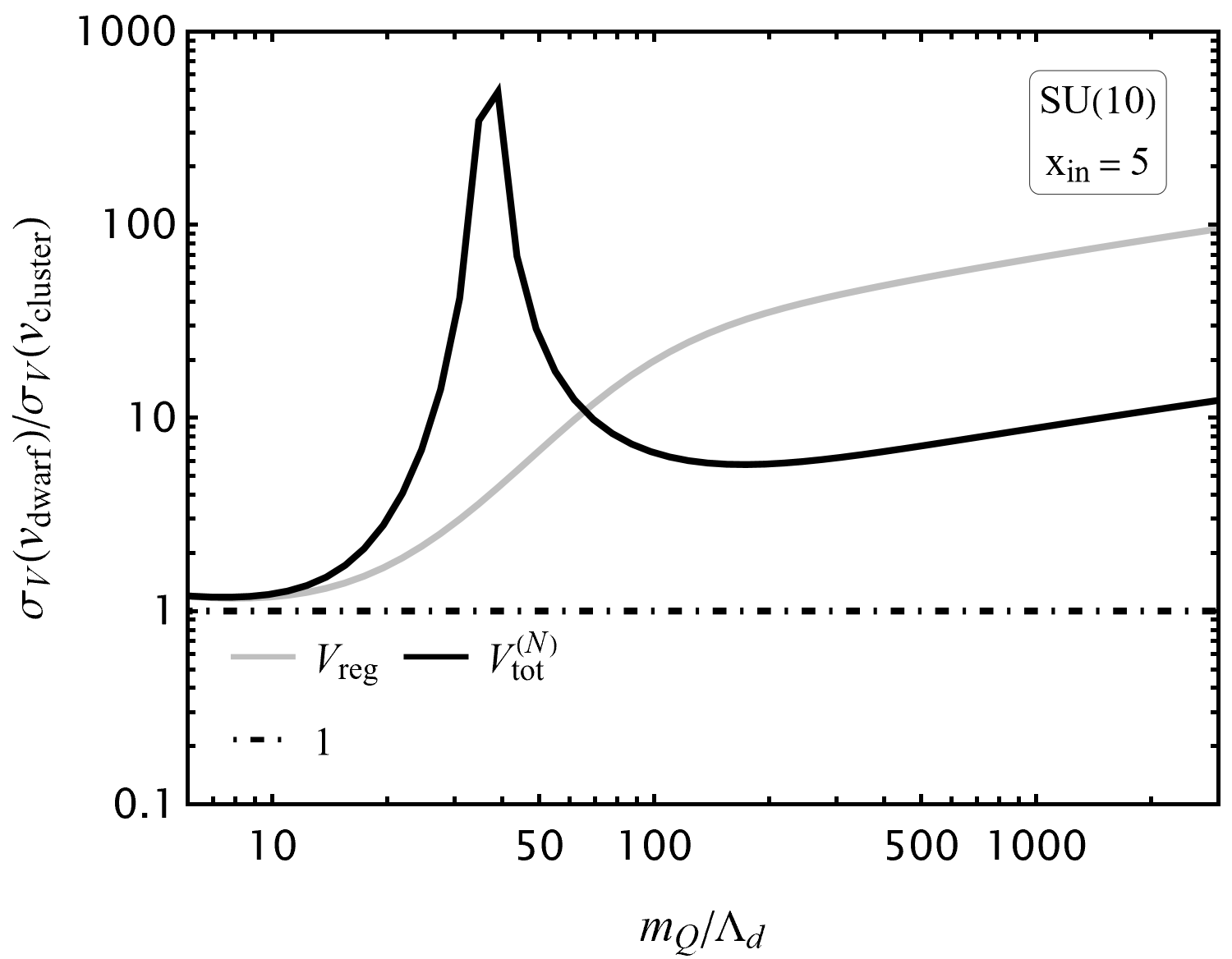}\hspace{0.1cm}
    \includegraphics[width=0.475\linewidth]{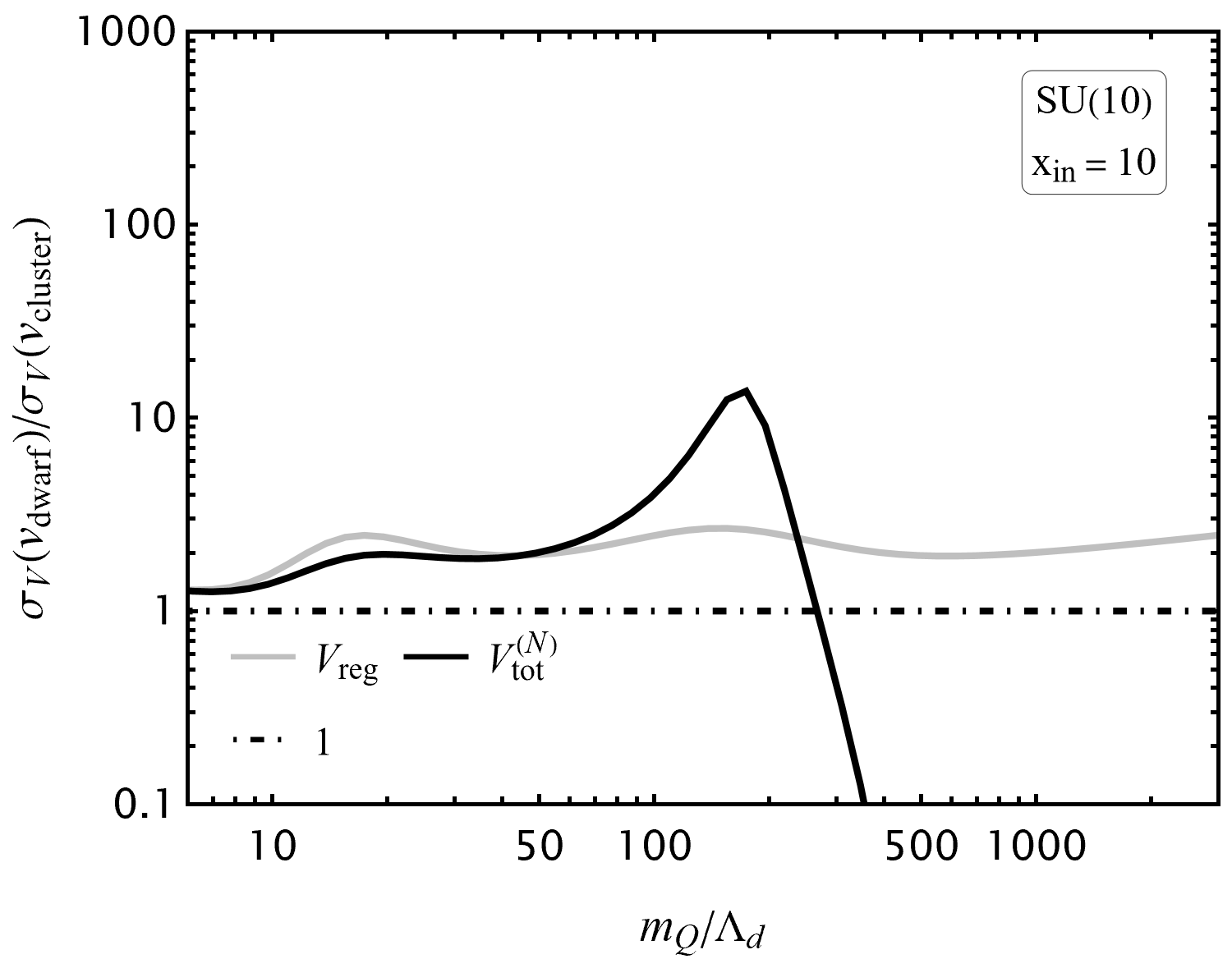}

    \caption{Ratio between the self-interaction cross section $\sigma_V(v_{\rm dwarf})$ at dwarfs ($v_{\rm dwarf}=30$~km/s) over $\sigma_V(v_{\rm cluster})$ at clusters ($v_{\rm cluster}=1500$~km/s) as a function of the hierarchy parameter $\xi=m_Q/\Lambda_d$. The gray, red, blue and black  lines correspond to cross sections computed using $V_{\rm reg}$,  $V_{\rm tot}^{(\rm geo)}$, $V_{\rm tot}^{(3/4)}$ and $V_{\rm tot}^{(N)}$, respectively.  The dot-dashed horizontal black line denotes a ratio of unity, meaning that there is no increase or decrease for the corresponding  velocities.  Top, middle and bottom correspond respectively to $N=2,3,10$ while left and right panels to $x_{\rm in}=5,10$.  }
    \label{fig:cross_section_velocity_ratio}
\end{figure}

In \cref{fig:cross_section_velocity_ratio}, we present the ratio of the viscosity cross section at dwarf-scale velocities to that at cluster velocities,
\begin{equation}
    \frac{\sigma_V(v_{\rm dwarf})}
         {\sigma_V(v_{\rm cluster})}\,,
\end{equation} with  $v_{\rm dwarf}=30~{\rm km/s}$ and $v_{\rm cluster}=1500~{\rm km/s}$, as a function of the hierarchy parameter
$\xi=m_Q/\Lambda_d$.
Since both the baryon potential and the Schr\"odinger equation become independent of the overall quark-mass scale after expressing distances in units of the Bohr radius, this ratio depends only on the number of colors $N$ and on $\xi$. The dark-quark mass $m_Q$ simply fixes the overall normalization of the cross section.

The velocity evolution exhibits a significant dependence on the phenomenological parameters describing the poorly known short-distance interaction, particularly the matching scale
$r_{\rm in}=a_Bx_{\rm in}$.
This behavior is expected, since the attractive potential grows rapidly toward small radii.
At distances of order the baryon size, however, the multipole expansion breaks down as the internal structure of the baryons begins to be resolved.
The coherent induced-dipole interaction should therefore gradually disappear, as the scattering starts probing the individual quark constituents rather than the baryon as a whole (see \cref{sec:regularizing_potential}).
Our choices $x_{\rm in}=5$ and $10$ should therefore be interpreted as slightly optimistic and conservative estimates, respectively, of the radius at which this loss of coherence becomes important.

The repulsive-core parameter $C$ introduces an additional source of uncertainty, although its impact is generally milder than that of $x_{\rm in}$ (typically, this merely shifts the position of the cross-section peaks slightly).
A first-principles determination of the short-distance interaction would require a dedicated treatment of the spin and color structure of overlapping baryons and lies beyond the scope of the present work.
Nevertheless, an important qualitative result already emerges from
\cref{fig:cross_section_velocity_ratio}: independently of the precise
values of $x_{\rm in}$ and $C$ (within the probed ranges), the large-$N$
cases exhibit regions of parameter space where the self-interaction
cross section at dwarf velocities is substantially larger than at
cluster velocities. In contrast, the small-$N$ cases do not develop
phenomenologically relevant velocity dependence. The relative evolution
seen for $\xi\lesssim10$ occurs only where the baryonic description
itself begins to break down.

We have also checked that this conclusion is not an artifact of the
choice of regularization radius $x_{\rm in}=5$--$10$. Repeating the calculation for the more
optimistic choices $x_{\rm in}=3$ and $x_{\rm in}=1$ does not restore
a phenomenologically relevant velocity dependence for $N=2,3$: the
Yukawa attraction remains too weak to overcome the repulsive core.

\begin{figure}
             \centering \includegraphics[height=5.25cm]{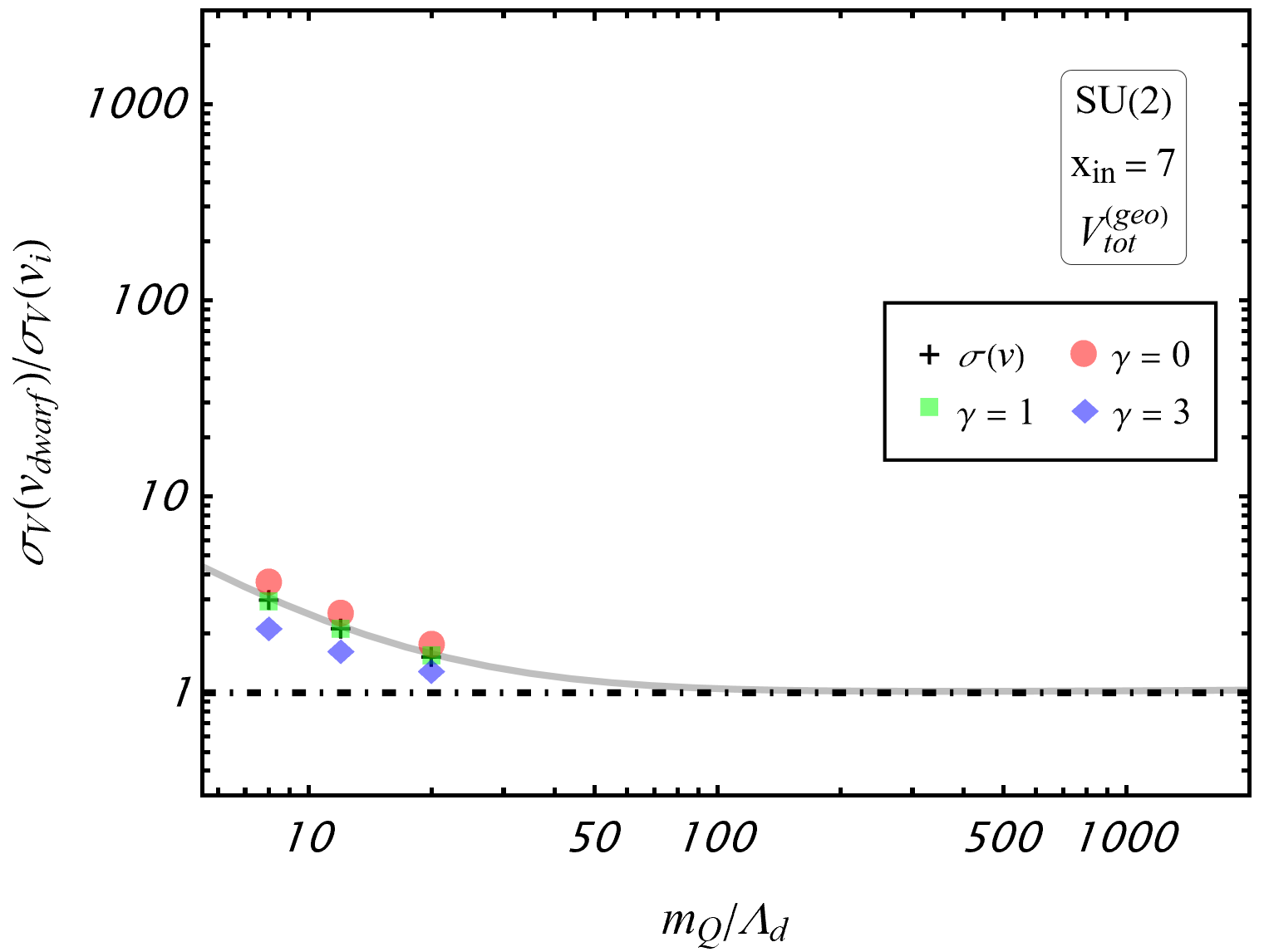} \includegraphics[height=5.25cm]{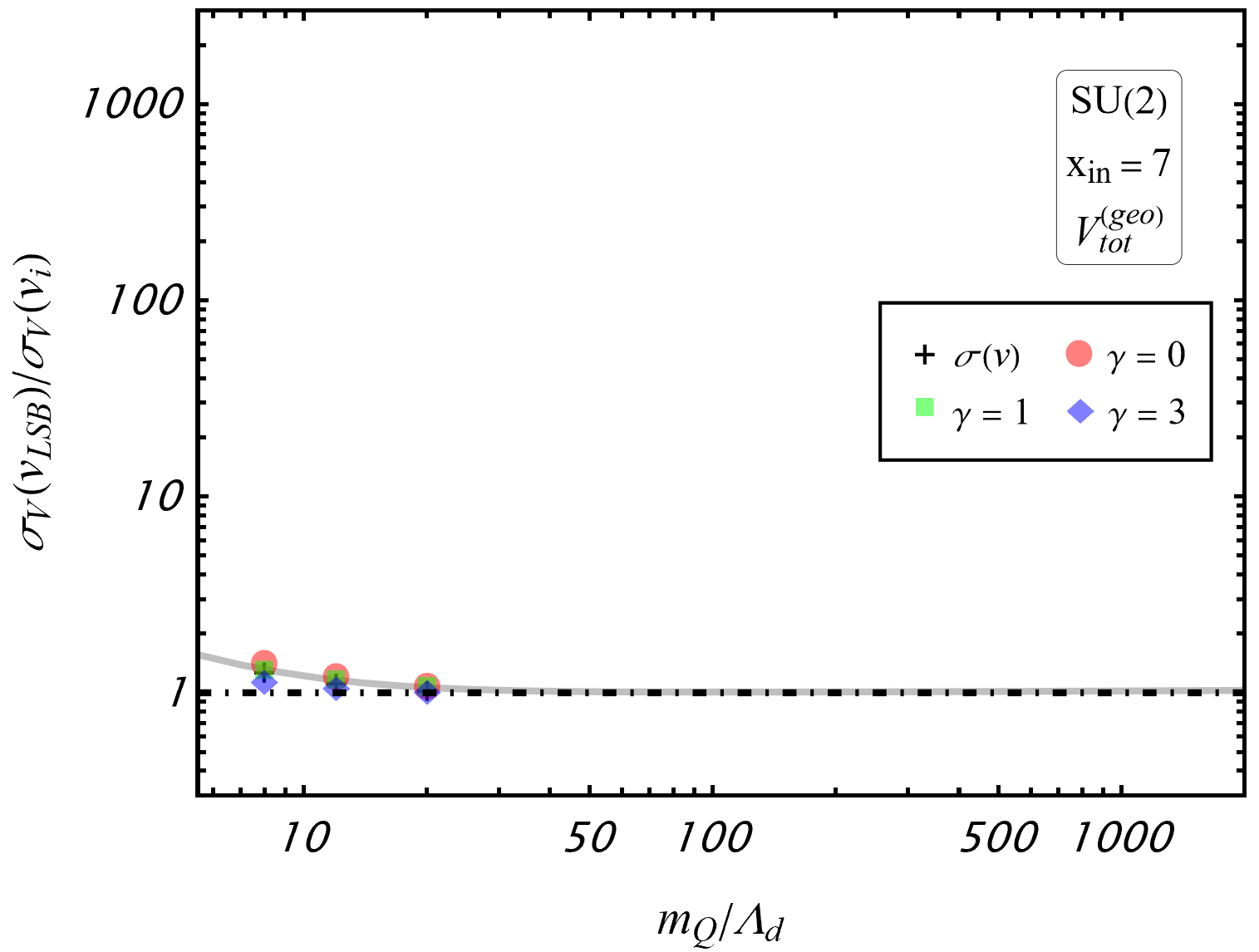}

             \includegraphics[height=5.25cm]{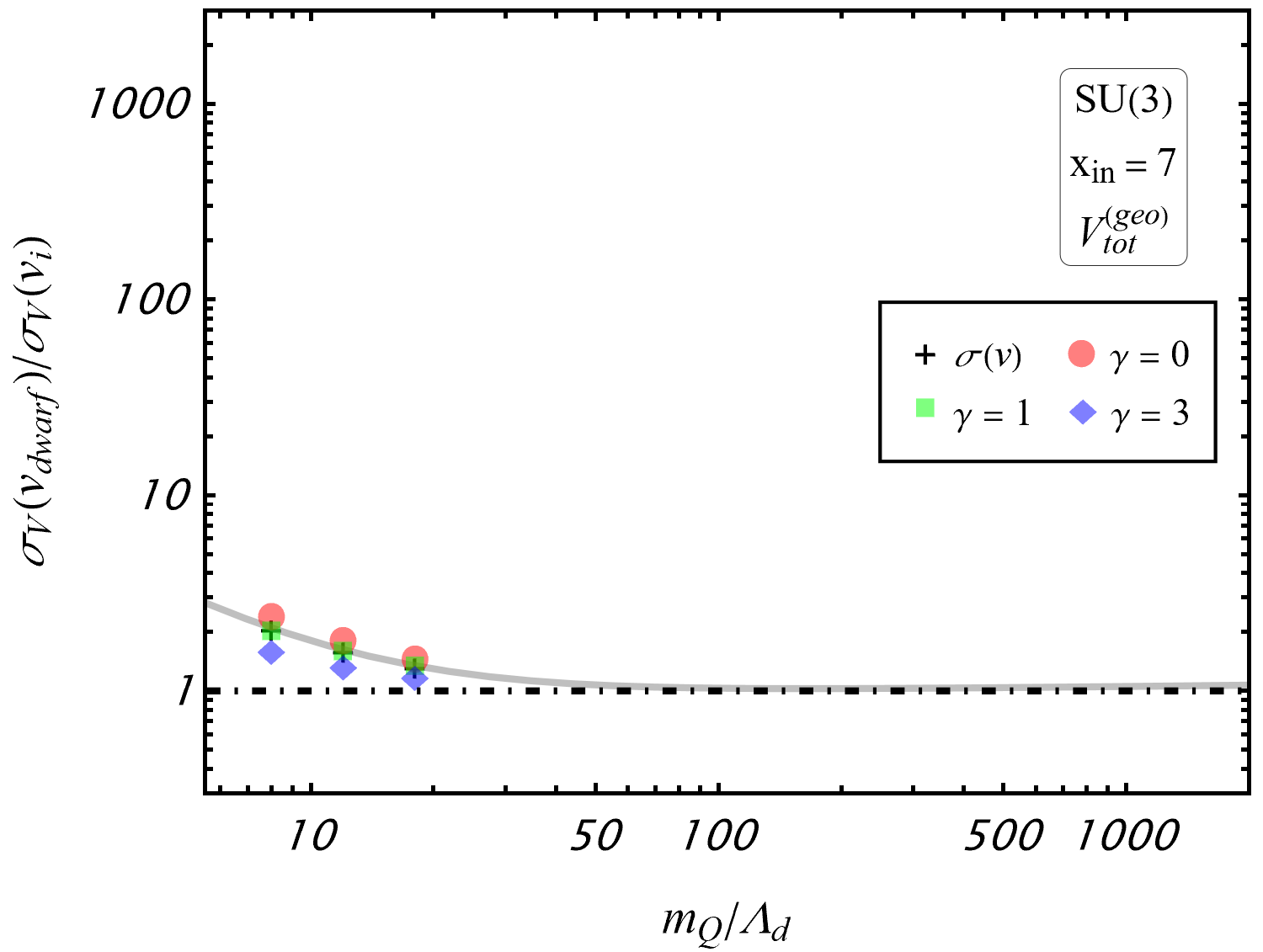} \includegraphics[height=5.25cm]{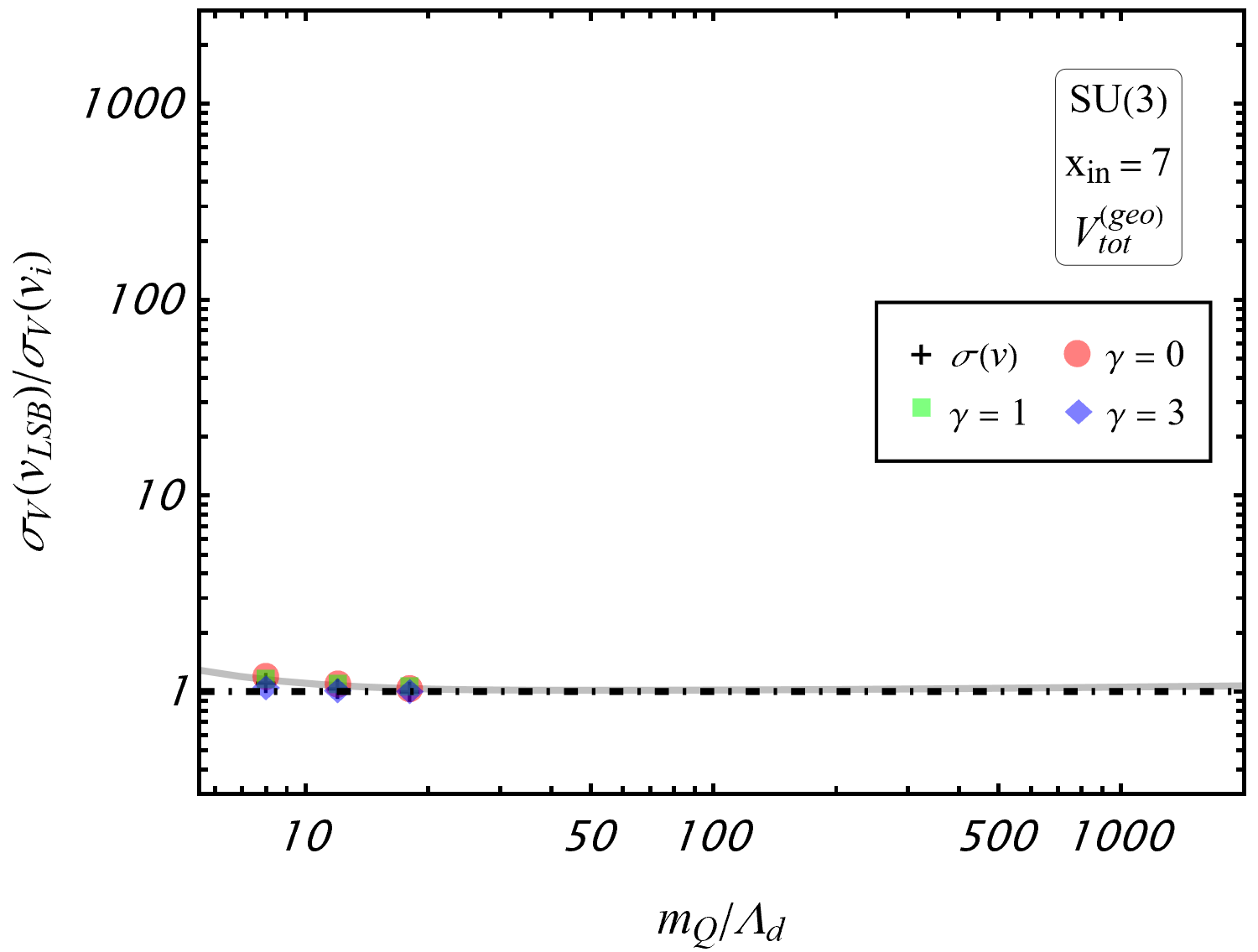}

             \includegraphics[height=5.25cm]{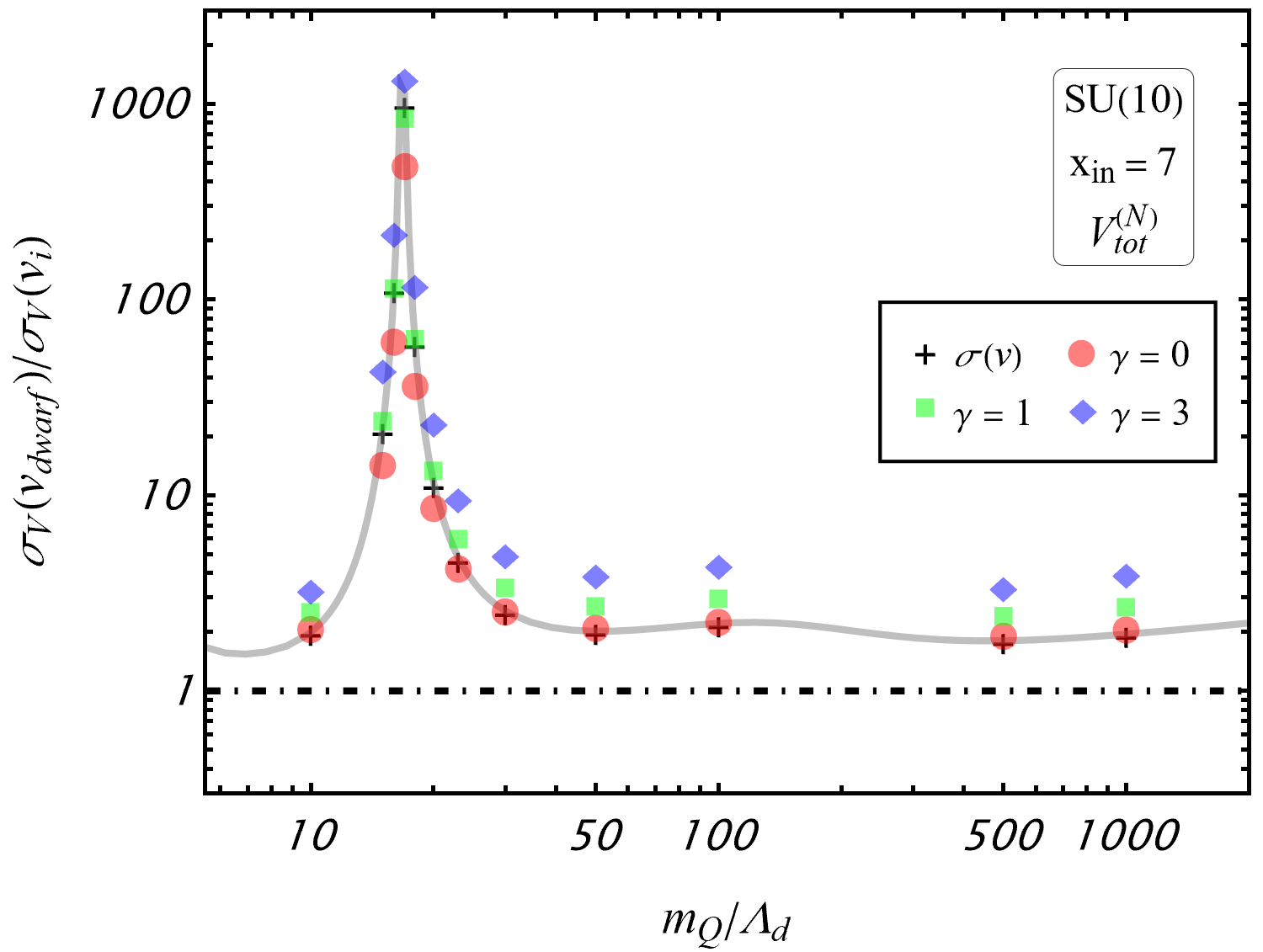} \includegraphics[height=5.25cm]{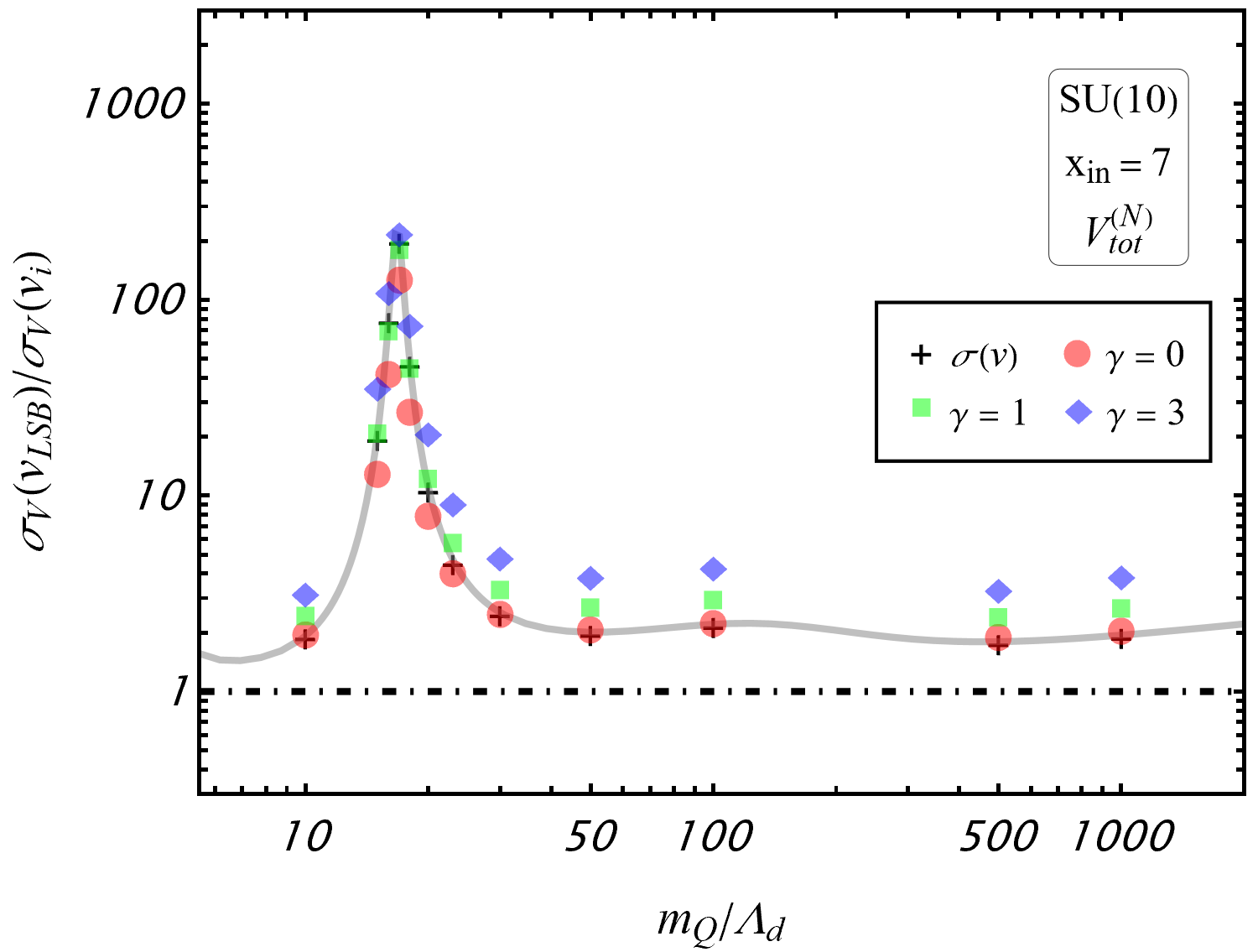}
     \caption{  Ratio between the self-interaction cross section $\sigma_V(v_i)$ at dwarfs (left panels, with $v_i=v_{\rm dwarf}=30$~km/s)  and low-surface-brightness galaxies (right panels, with $v_i=v_{\rm LSB}=100$~km/s), and that at cluster scales, $\sigma_V(v_{\rm cluster})$ with $v_{\rm cluster}=1500$~km/s, as a function of the hierarchy parameter $\xi=m_Q/\Lambda_d$.  We have used $V_{\rm tot}$ with $x_{\rm in}=7$ and $C$ fixed to reproduce $\sigma_{\rm geo}$ for $N=2,3$ while the repulsion used for $\SU(10)$ is the large $N$ result. Superimposed are the velocity-averaged enhancement factors $\overline{\sigma}_V(v_i)_\gamma/\,\overline{\sigma}_V(v_{\rm cluster})_\gamma$ for the selected points listed in \cref{tab:benchmark_choices}. Results are shown for $\gamma=0$ (red circles), $\gamma=1$ (green squares), and $\gamma=3$ (blue diamonds), corresponding respectively to the unweighted, rate-weighted, and energy-weighted velocity moments. The black $+$ markers denote the direct ratio $\sigma_V(v_i)/\sigma_V(v_{\rm cluster})$ for the same points, allowing a direct comparison with the velocity-averaged quantities. The dot-dashed horizontal black line indicates a ratio of unity, expected for velocity-independent self-interactions. Top, middle, and bottom panels correspond to $N=2$, $3$, and $10$, respectively.   }
    \label{fig:cross_section_velocity_ratio7_averages}
\end{figure}

For the remainder, we adopt the representative values
$x_{\rm in}=7$ and the effective repulsion determined by matching the geometrical cross section---see \cref{eq:geo_cross_seciton}.
These choices provide an intermediate benchmark while keeping the presentation sufficiently compact.
The resulting velocity-ratio scan is shown in \cref{fig:cross_section_velocity_ratio7_averages}, where we also include the ratio of cross sections at LSB-scale velocities, $v_{\rm LSB}=100$~km/s, over $\sigma_V$ at $v_{\rm cluster}=1500$~km/s.

The simplified scan confirms the conclusions drawn from the full parameter variation.
In particular, a pronounced velocity dependence occurs only for
$\xi\ll10^3$.
For larger hierarchies, the attractive interaction becomes increasingly suppressed, leading to nearly velocity-independent cross sections.
The exception is the $\SU(10)$ case with $x_{\rm in} \lesssim10$, where the relatively large chromoelectric polarizability enhances the induced interaction sufficiently to preserve some velocity dependence even for $\xi\gg10^3$.

Motivated by the different structures visible in
\cref{fig:cross_section_velocity_ratio7_averages}, we select the values
of $\xi$ listed in \cref{tab:benchmark_choices} to illustrate the
different classes of velocity evolution found in heavy quarks $SU(N)$ models. Set~1 corresponds to the mild enhancement at low $\xi\sim8$, present for all groups. The subsequent sets are only relevant for $\SU(10)$, as smaller $N$ values do not exhibit further distinct velocity behavior. Set~2 explores the resonance at $\xi\sim17$, Set~3 probes the intermediate enhancement immediately above the resonance at $\xi\sim25$, while Set~4 corresponds to the mild enhancement that persists at very large hierarchy parameters, $\xi\gtrsim100$.

\begin{table}[t] \centering \begin{tabular}{c|c|c|c}   & $\SU(2)$ & $\SU(3)$ & $\SU(10)$ \\ \hline Set 1 & $8,\;12,\;20$ &$8,\;12,\;18$  & $10$  \\ Set 2 &  &   & $15,\;16,\;17,\;18$ \\ Set 3 &   &   & $20,\;23,\;30$\\ Set 4 &   &   &  $50,\;100,\;500,\;1000 $ \end{tabular} \caption{Selected values for the hierarchy parameter $\xi=m_Q/\Lambda_d$ used to illustrate the different classes of velocity dependence displayed in \cref{fig:vel_evolution_sigmaVSvelocity}. Note that we fix $x_{\rm in}=7$ and take $V_{\rm tot}^{(\rm geo)}$ for $N=2,3$ while for $N=10$ we take $V_{\rm tot}^{(N)}$. } \label{tab:benchmark_choices} \end{table}

To quantify these points more realistically, \cref{fig:cross_section_velocity_ratio7_averages} also presents velocity-averaged enhancement factors,
\[
\dfrac{\overline\sigma_V(v)_\gamma }{\overline\sigma_V(v_{\rm cluster})_\gamma }=\frac{\langle\sigma_V(v)\,v^\gamma\rangle/\langle v^\gamma\rangle}
     {\langle\sigma_V(v_{\rm cluster})\,v_{\rm cluster}^\gamma\rangle/\langle v_{\rm cluster}^\gamma\rangle},
\]
for both
$\gamma=1$
(rate-weighted scattering)
and
$\gamma=3$
(energy-transfer weighting), as well as $\gamma=0$ (averaged cross section).
The qualitative conclusions remain unchanged under either averaging prescription.
Large enhancements at dwarf and LSB galaxy velocities coexist naturally with cluster-scale cross sections that remain suppressed.
The selected points span a wide range of enhancement factors, reflecting the diverse velocity dependences generated by the interplay between the intermediate-range induced-dipole interaction, the non-perturbative Yukawa tail, and the short-distance repulsive core.

\begin{figure}
             \centering \includegraphics[width=0.33\linewidth]{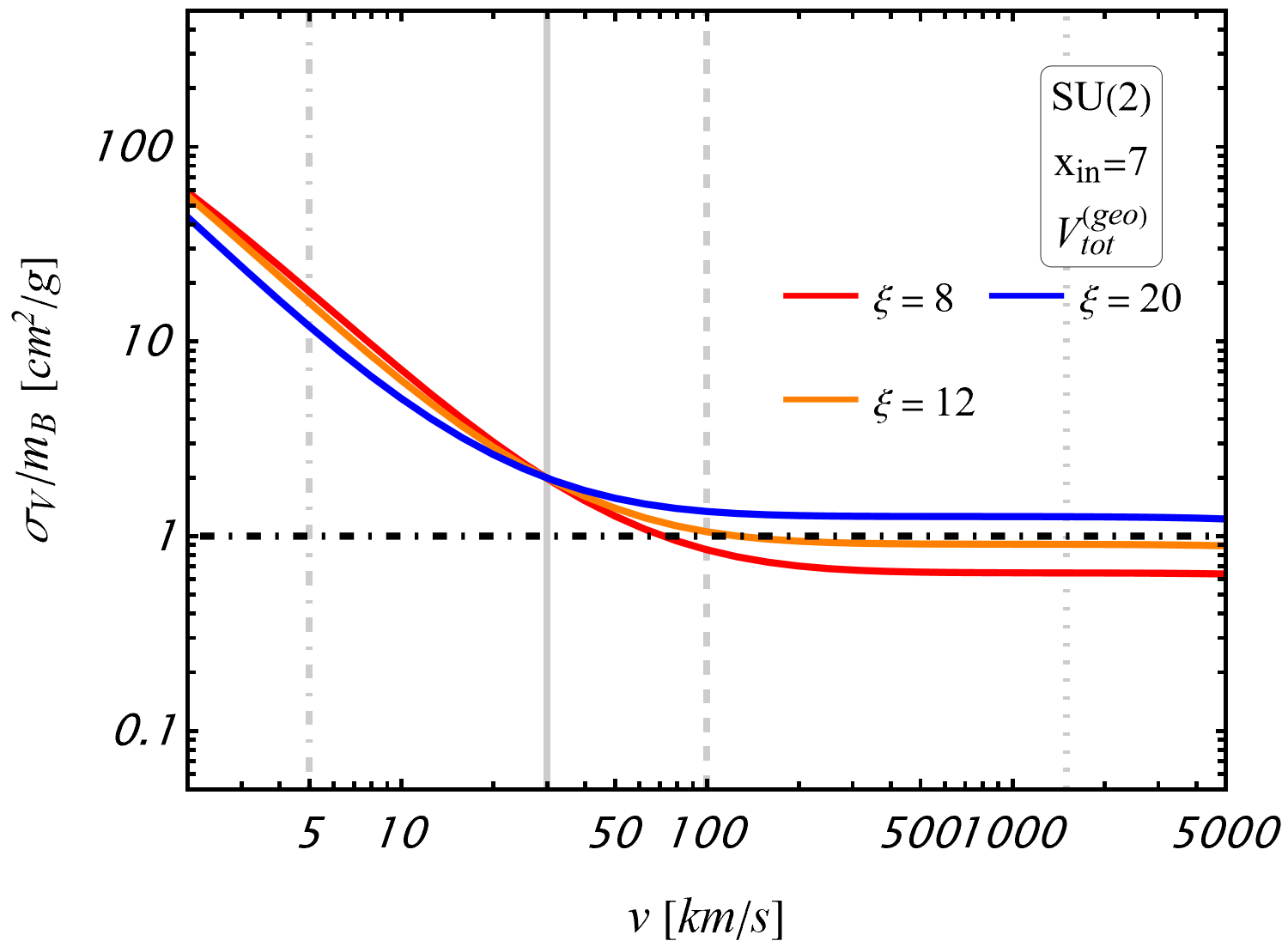}\includegraphics[width=0.33\linewidth]{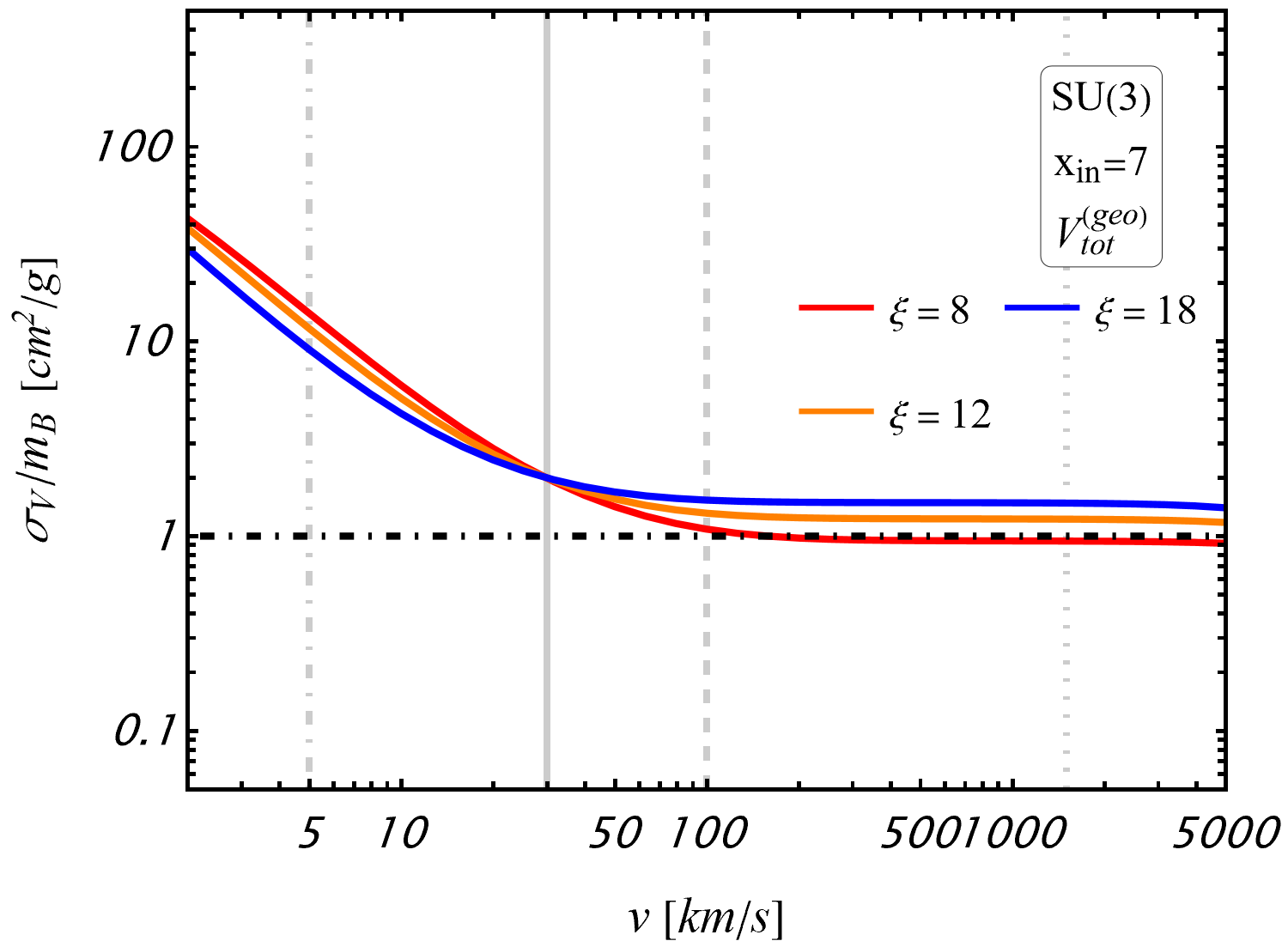}\includegraphics[width=0.33\linewidth]{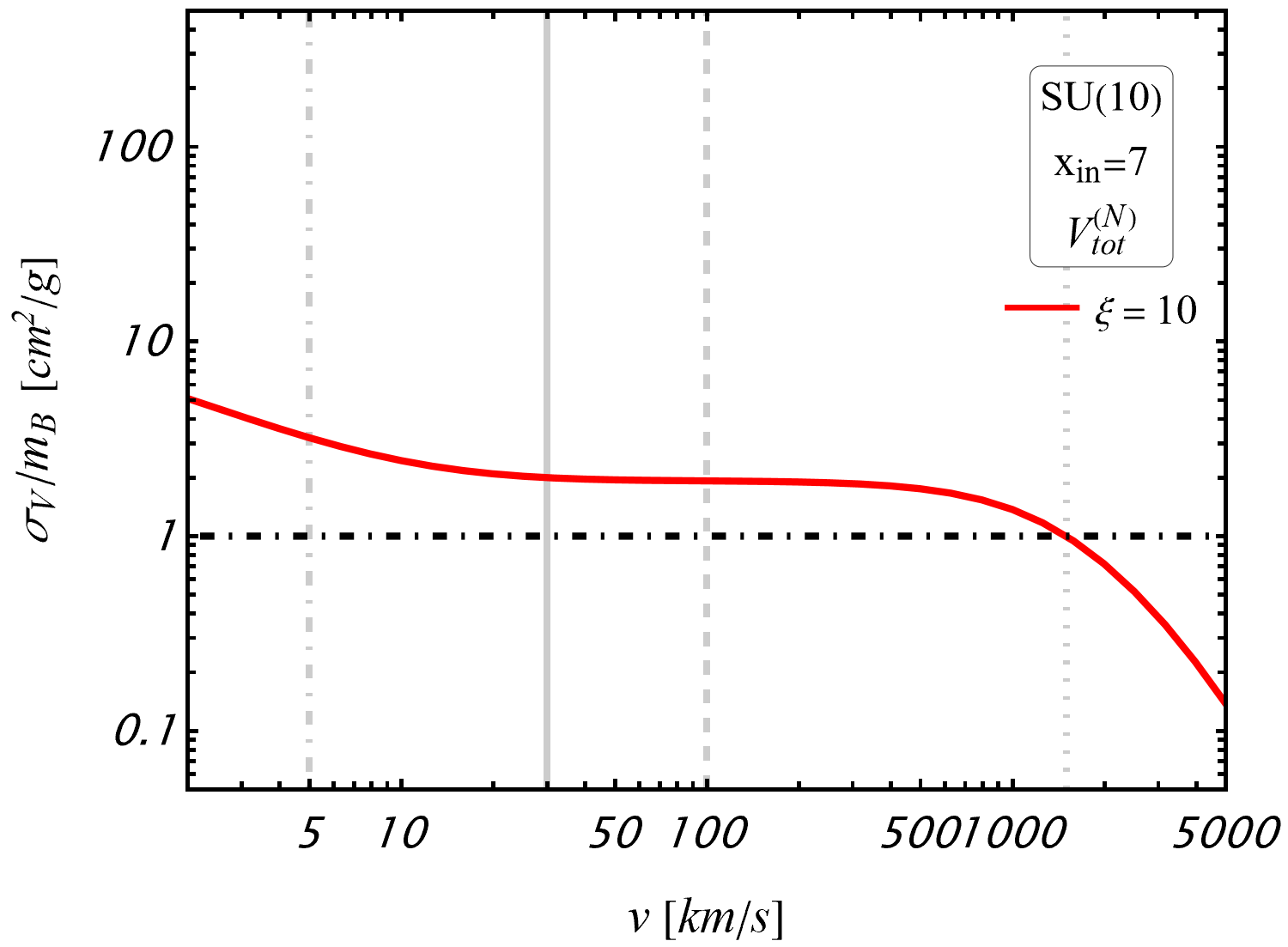}
             \includegraphics[width=0.33\linewidth]{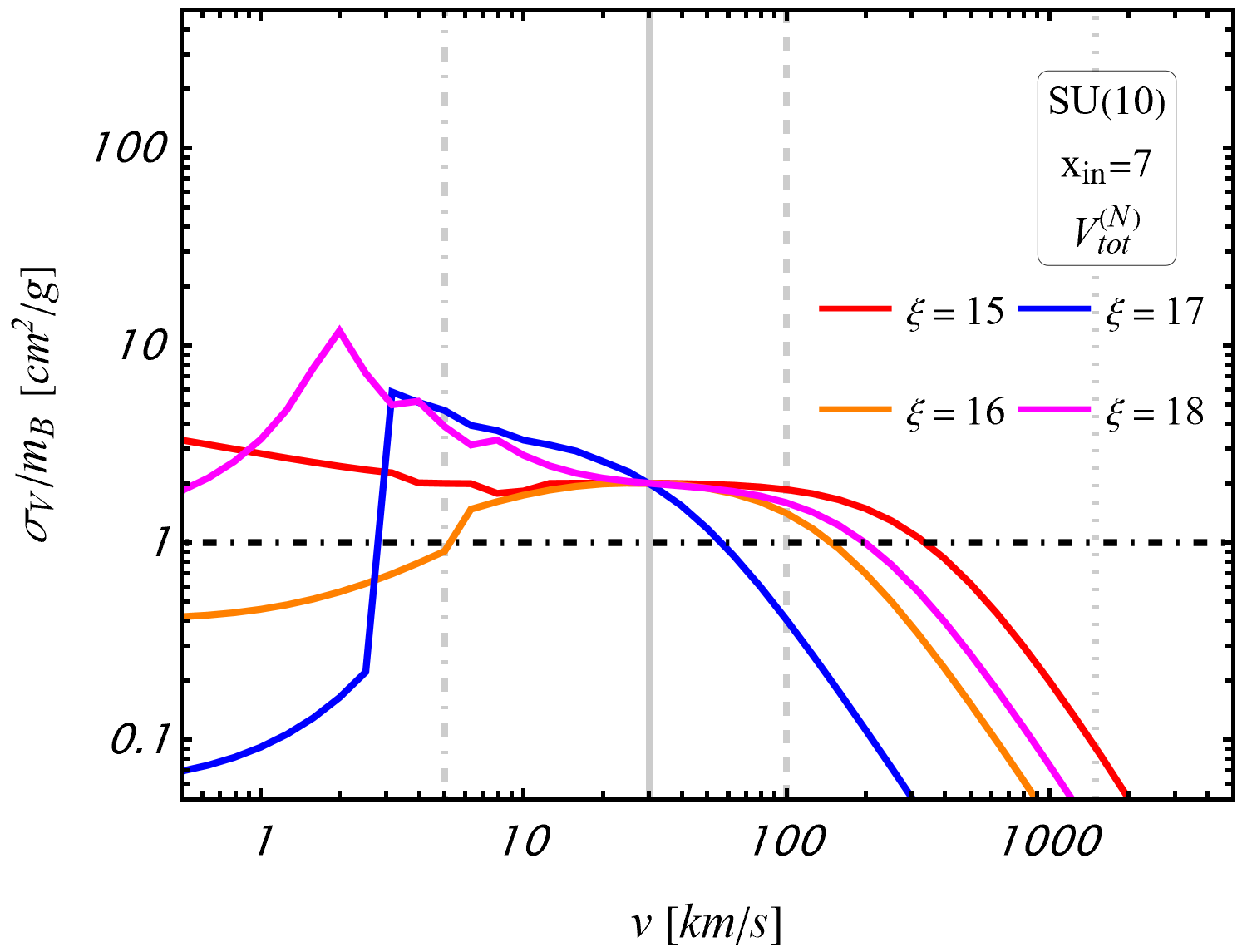} \includegraphics[width=0.33\linewidth]{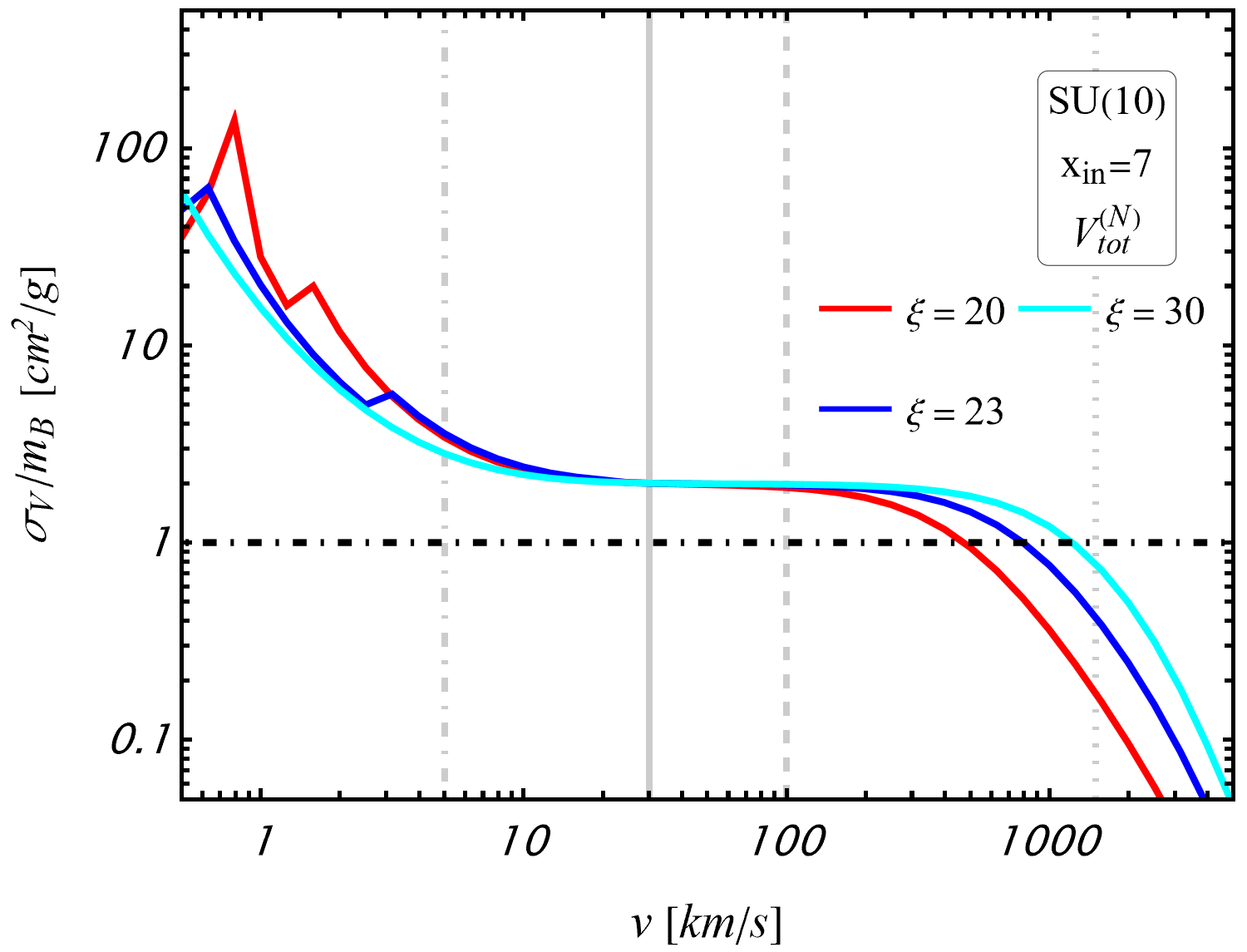}\includegraphics[width=0.33\linewidth]{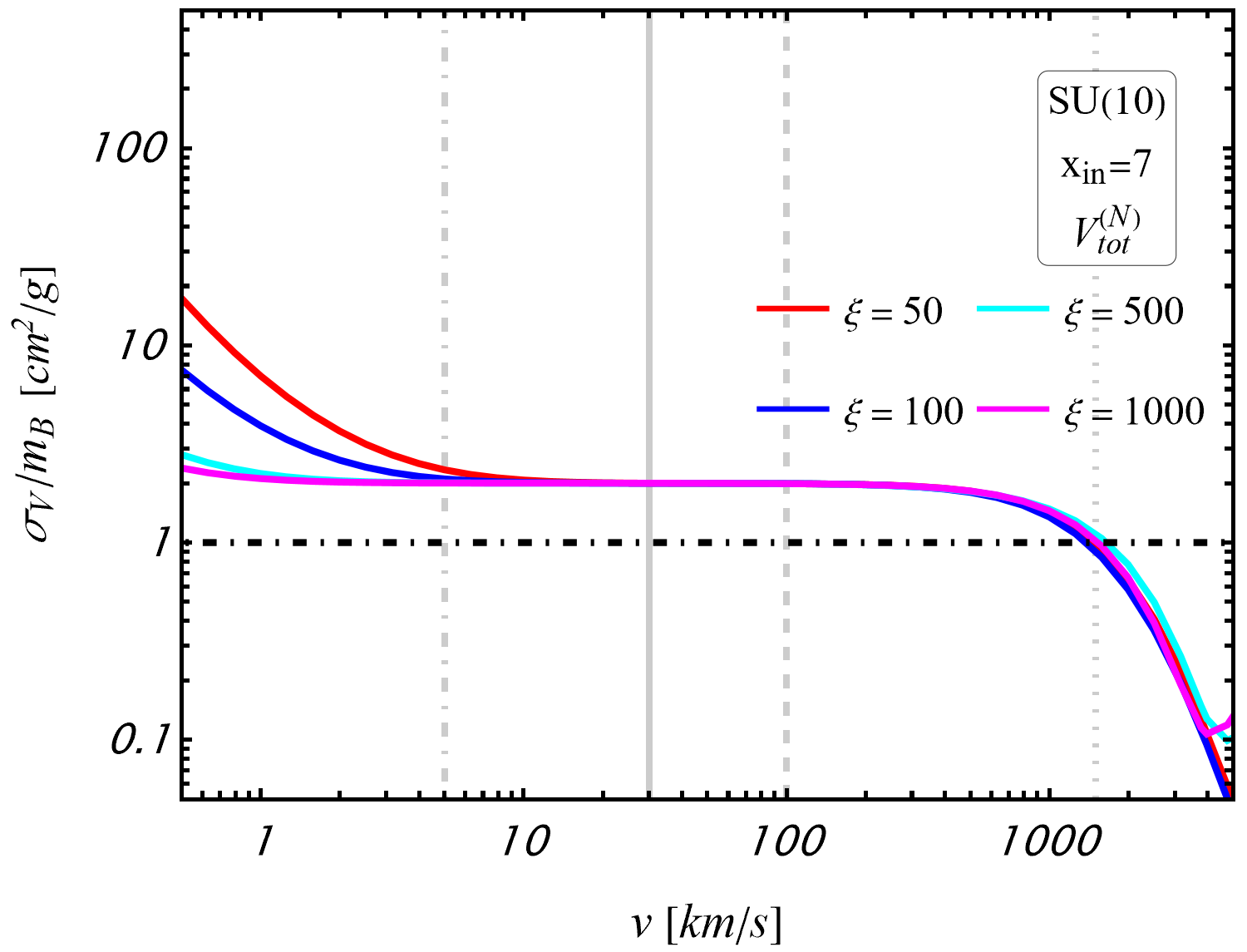}
\caption{Velocity evolution of $\sigma_V(v)/m_B$ for the selected values of $\xi=m_Q/\Lambda_d$ listed in \cref{tab:benchmark_choices}. The baryon mass $m_B$ is arbitrarily fixed in order to present $\sigma_V(v_{\rm dwarf})/m_B=2\;\mathrm{cm}^2/\mathrm{g}$. However, note that the velocity evolution is fully independent of $m_B$. Gray vertical dot-dashed, solid, dashed, and dotted lines correspond to $v_{\rm collapse}$, $v_{\rm dwarf}$, $v_{\rm LSB}$, and $v_{\rm cluster}$, respectively. The black horizontal dot-dashed line indicates the approximate cluster-scale bound, $\sigma_V/m_B=1\;\mathrm{cm}^2/\mathrm{g}$. Top: left, center, and right correspond respectively to $N=2,3,10$ for Set 1. Bottom: left, center, and right correspond respectively to Sets 2,3,4 for $\SU(10)$.  }
    \label{fig:vel_evolution_sigmaVSvelocity}
\end{figure}

The origin of these different enhancement factors becomes clear in
\cref{fig:vel_evolution_sigmaVSvelocity},
which displays the complete velocity dependence for each point.
The dark-baryon mass is chosen only to normalize the cross section to
$2~{\rm cm}^2/{\rm g}$
at  dwarf velocities $v_{\rm dwarf}=30$~km/s;
the shapes of the curves are entirely independent of this choice.
Several qualitatively distinct behaviors emerge.
Some points exhibit narrow resonant enhancements concentrated at low velocities, while others display broad plateaus extending from dwarf to galaxy scales.
For larger values of $\xi$, the velocity dependence becomes progressively smoother as the interaction is increasingly dominated by the perturbative induced-dipole force. However, in these cases, some resonant enhancement may also appear above cluster velocities. Despite these differences, all classes contain regions capable of simultaneously producing large self-interactions at small scales and relatively suppressed interactions at cluster scales.

These results demonstrate that heavy dark baryons in the large-$N$
regime can generate a rich variety of velocity-dependent
self-interaction cross sections.
In the following, we determine the overall dark-matter mass scale required to reproduce the phenomenologically preferred values,
$\sigma_V/m_B\sim1-10~{\rm cm}^2/{\rm g}$,
and identify the corresponding regions of the $(m_Q,\Lambda_d)$ parameter space.

\subsection{Mass scales}

We now translate the velocity-dependent cross sections into the corresponding dark-baryon mass scales. Although the velocity-averaged cross section is the quantity most directly related to astrophysical observables, computing it over the entire parameter space is numerically expensive since it requires solving the scattering problem over the full velocity range for every value of $\xi$ and $N$. Our analysis in the previous subsection showed that the velocity
averaging generally preserves the main features of the single-velocity
results and that only the large-$N$ regime produces relevant velocity
dependence. Since the interaction approaches an approximately
$N$-independent form at relatively large $N\sim 10$, we use $\SU(10)$ as a representative
finite-$N$ realization of this regime and evaluate the cross section at
representative dwarf, galaxy, and cluster velocities to efficiently
identify the corresponding phenomenologically relevant parameter space.

\begin{figure}
              \centering
             \includegraphics[width=0.5\linewidth]{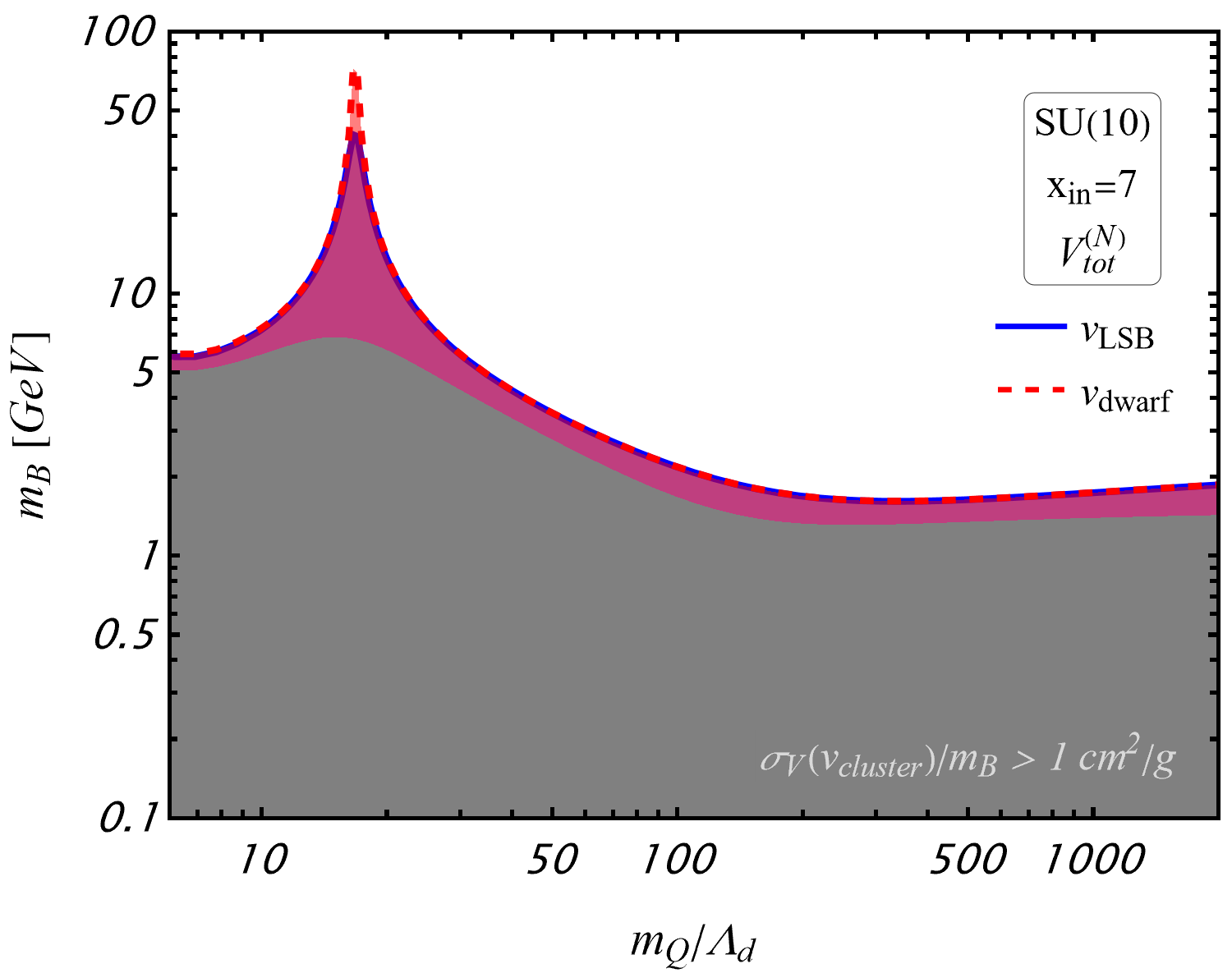}
    \caption{The maximal dark-baryon mass compatible with a viscosity self-interaction cross section of $1~{\rm cm}^2/{\rm g}$ at dwarf (dashed red) and galaxy (solid blue) velocities as a function of the hierarchy parameter $\xi=m_Q/\Lambda_d$ for $\SU(10)$. Results are shown for the effective potential $V_{\rm tot}$ with the repulsive-core strength given by the large $N$ result of Ref.~\cite{Adhikari:2013dfa} with $x_{\rm in}=7$. The shaded regions indicate baryon masses yielding $\sigma_V/m_B\ge1~{\rm cm}^2/{\rm g}$ at clusters (gray), galaxies (blue), and dwarfs (red). }
    \label{fig:minimal_mass}
\end{figure}

For fixed gauge group $N$, the velocity dependence varies only with the hierarchy parameter
$\xi=m_Q/\Lambda_d$, whereas the overall normalization of the cross section is determined by the baryon mass,
\begin{equation}
    \frac{\sigma_V}{m_B}\propto m_B^{-3}.
\end{equation}
Consequently, for each value of $\xi$ one can directly determine the largest baryon mass capable of producing
$\sigma_V/m_B\ge1~{\rm cm}^2/{\rm g}$
at dwarf or galaxy velocities, as well as the smallest baryon mass compatible with the cluster constraint
$\sigma_V/m_B\le1~{\rm cm}^2/{\rm g}$.
The resulting allowed mass ranges are shown in
\cref{fig:minimal_mass}.

Despite the large variations observed in the velocity dependence, the preferred dark-baryon masses occupy a comparatively narrow range because of the steep scaling
$\sigma_V/m_B\propto m_B^{-3}$.
The preferred dark-baryon masses are found between a
few hundred MeV and a few tens of GeV, remarkably close to the masses
of the proton and neutron that constitute ordinary visible matter. This result already hints at a possible common origin for the visible and DM abundances.

\begin{figure}
             \centering\includegraphics[width=0.6\linewidth]{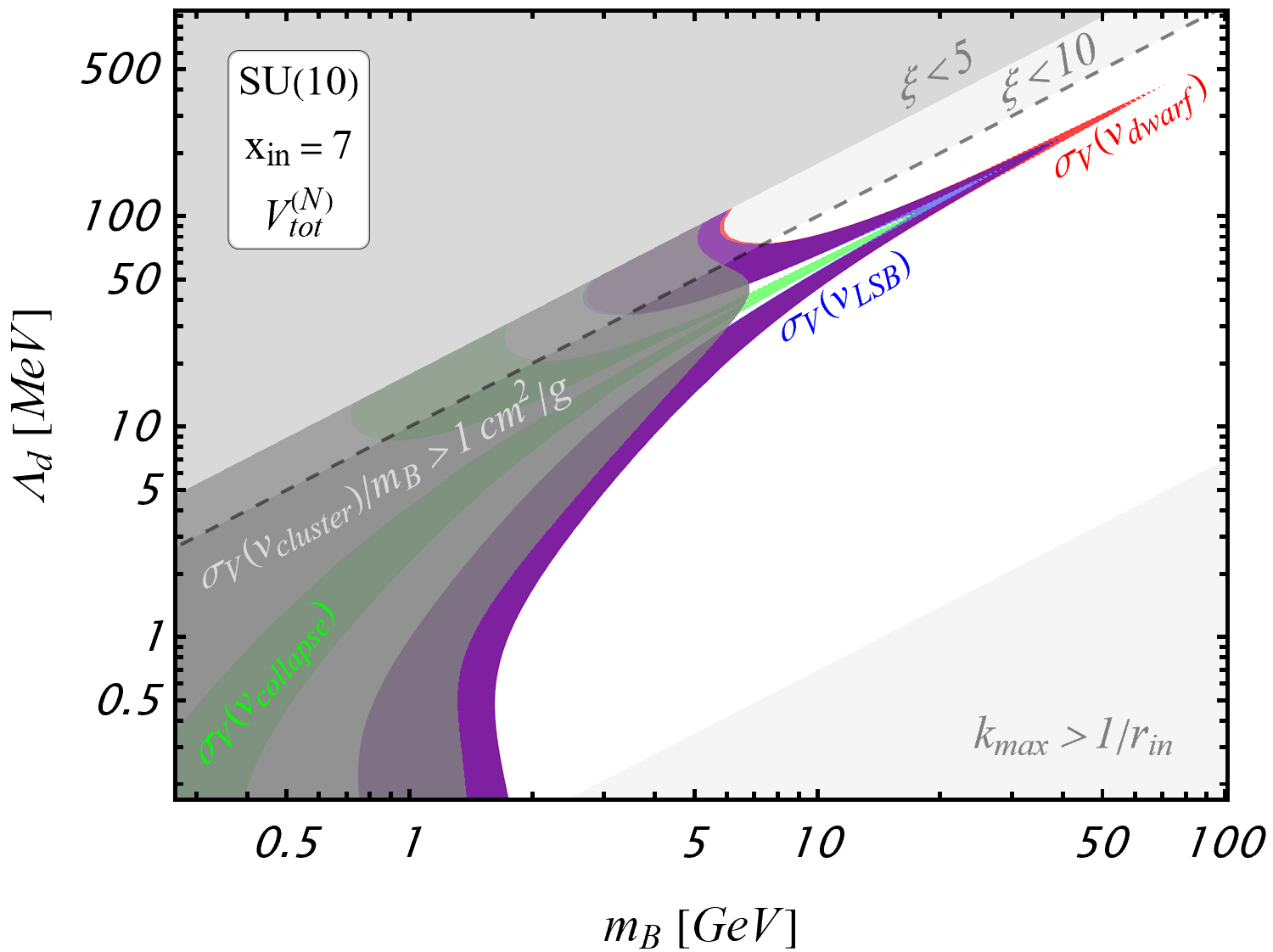}
\caption{Regions in the $(m_B,\Lambda_d)$ plane where the viscosity self-interaction cross section satisfies $1\;\mathrm{cm}^2/\mathrm{g}\leq\sigma_V/m_B\leq10\;\mathrm{cm}^2/\mathrm{g}$, evaluated at $v_{\rm dwarf}=30~\mathrm{km/s}$ (red) and $v_{\rm LSB}=100~\mathrm{km/s}$ (blue), using $V_{\rm tot}^{(N)}$ with $r_{\rm in}=7a_B$ for $\SU(10)$. Dark gray regions are excluded by the requirement $\sigma_V/m_B<1\;\mathrm{cm}^2/\mathrm{g}$ at $v_{\rm cluster}=1500~\mathrm{km/s}$. Light gray regions indicate where our theoretical approximations are expected to break down. In the upper-left region, the hierarchy $m_B/\Lambda_d$ becomes too small for our baryonic description to remain reliable, as confinement and relativistic effects become important. In the lower-right region, the minimum de Broglie wavelength (corresponding to $v_{\rm cluster}$) becomes smaller than the phenomenological regularization scale $r_{\rm in}$. Green regions satisfy $75\leq\sigma_V/m_B\leq1000~\mathrm{cm}^2/\mathrm{g}$ at $v_{\rm collapse}=5~\mathrm{km/s}$, where gravothermal core collapse may explain the JVAS B1938+666 system~\cite{Vegetti:2026mmx,Zhang:2026gur}.}
    \label{fig:plane_mqXlambda}
\end{figure}

The connection with QCD becomes even more apparent in the
$(m_B,\Lambda_d)$
plane shown in
\cref{fig:plane_mqXlambda}.
Besides predicting baryon masses close to the QCD scale, the preferred confinement scales are also found to lie near, or within roughly one order of magnitude below,
$\Lambda_{\rm QCD}$.
This simultaneous preference for QCD-like baryon and confinement scales is a striking outcome of the present framework.

Such scales may arise in ADM scenarios, where a dark baryon asymmetry comparable to the ordinary baryon asymmetry automatically yields the observed DM abundance when the dark baryon mass is of order the proton mass. Motivated by this coincidence, we discuss the implications for ADM in \cref{sec:asymmetric_DM}.

\subsection{Gravothermal core collapse}

Having established that heavy dark baryons can exhibit a rich velocity dependence capable of addressing the small-scale structure tensions of $\Lambda$CDM, we now investigate whether the same framework can also accommodate evidence for apparently core-collapsed subhalos. In particular, we consider the recently reported second perturber of the strong gravitational lens system JVAS B1938+666~\cite{Powell:2025rmj,Vegetti:2026mmx,Zhang:2026gur}. The perturber has a total mass of roughly $10^6M_\odot$ within a region of size $\sim100$~pc, together with an unresolved and very dense inner component at $\lesssim10$~pc embedded in a more extended envelope of size $\sim100$--$150$~pc. This combination of a compact central component and an extended envelope can be interpreted as the signature of a SIDM halo in the late stages of gravothermal core collapse~\cite{Lynden-Bell:1968awc,Balberg:2002ue,Turner:2020vlf,Outmezguine:2022bhq}. The corresponding analyses infer self-interaction cross sections of order $\sigma/m\sim100$--$800$~cm$^2$/g at characteristic velocities $v\sim5$~km/s~\cite{Vegetti:2026mmx,Zhang:2026gur}. More generally, gravothermal evolution has been proposed as a possible explanation for the large diversity of circular-velocity profiles observed among cuspy ultra-faint and cored bright satellites with similar characteristic dwarf-scale velocities~\cite{Zavala:2019sjk,Correa:2022dey,Correa:2020qam,Roberts:2024uyw}. In particular, Ref.~\cite{Correa:2020qam} finds that relatively large self-interactions, $\sigma/m\sim30$--$100~\mathrm{cm}^2/\mathrm{g}$, at dwarf-galaxy velocities can reproduce this diversity at dwarf-scales once gravothermal evolution is taken into account. Similarly, at the slightly higher velocities, Ref.~\cite{Roberts:2024uyw} has shown the diversity found in LSB galaxies could be explained by $\sigma/m\sim20$--$40~\mathrm{cm}^2/\mathrm{g}$.

As in the previous analyses, we first examine ratios of the self-interaction cross section evaluated at different characteristic velocities. In this case, however, we specifically search for regions of parameter space in which the cross section increases dramatically between cluster velocities and
$v_{\rm collapse}=5~\mathrm{km/s}$. Such requirements are necessary in order to guarantee  sufficient suppression at cluster velocities to satisfy cluster constraints.

\begin{figure}
\centering
\includegraphics[width=0.333\linewidth]{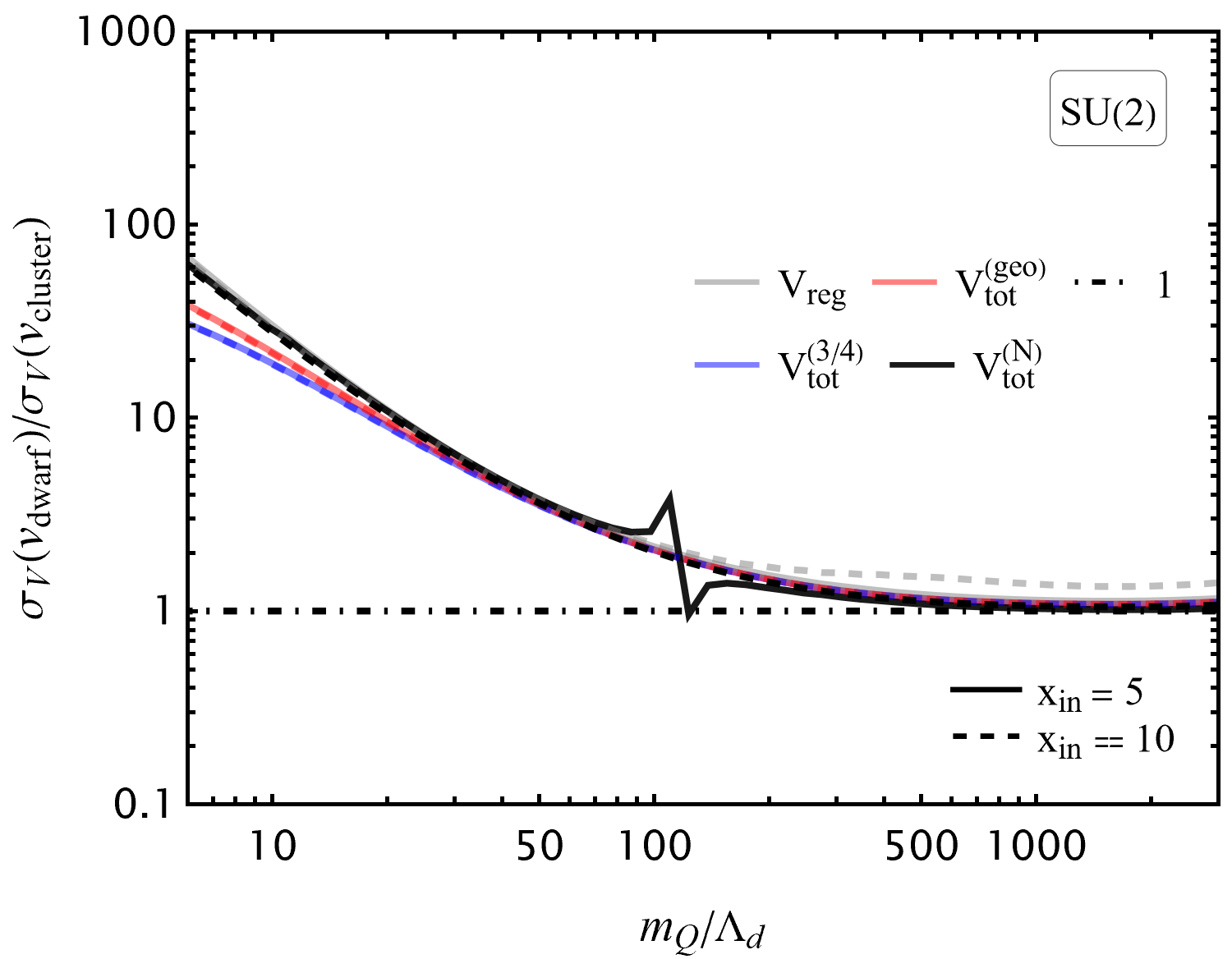}\includegraphics[width=0.333\linewidth]{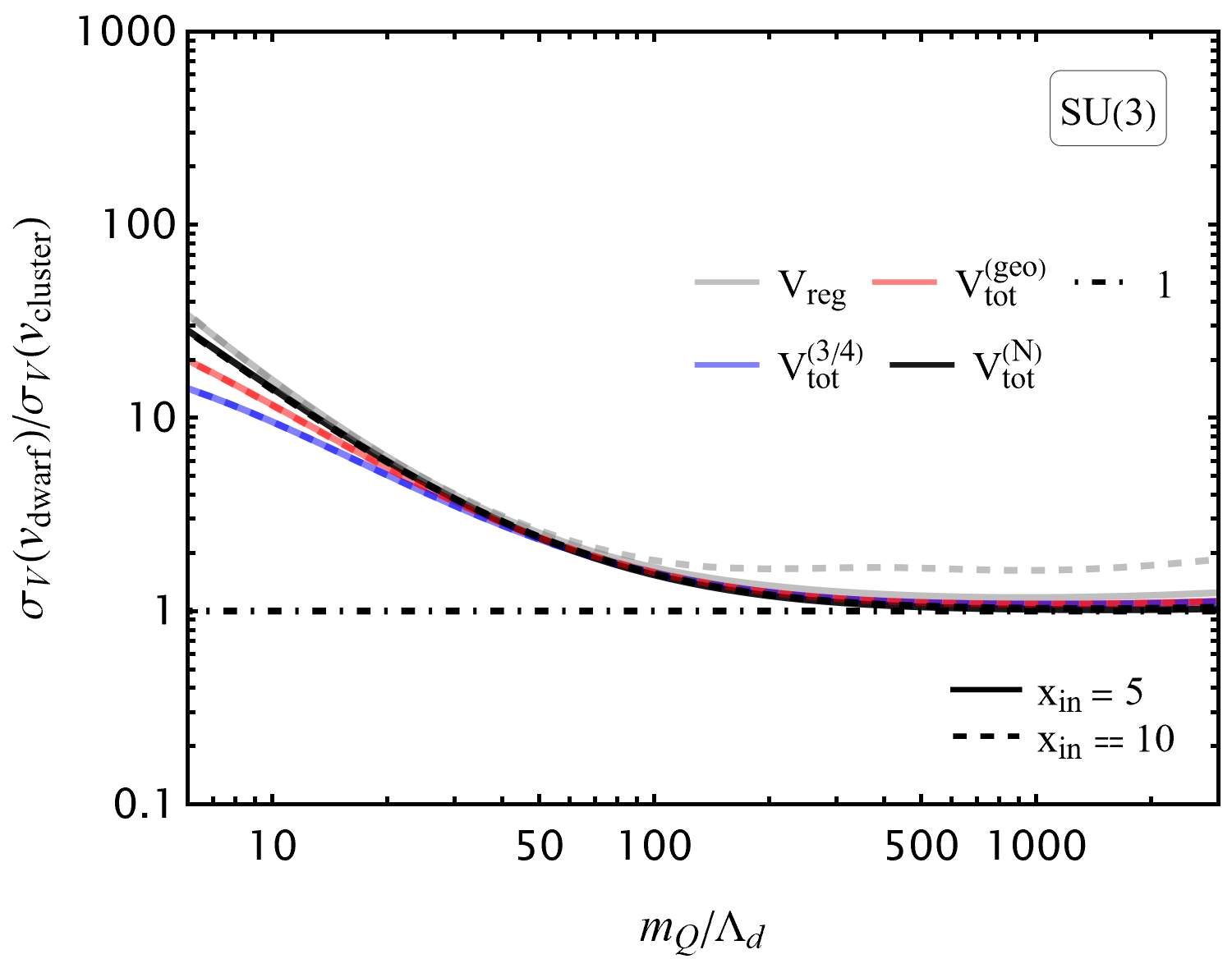}\includegraphics[width=0.333\linewidth]{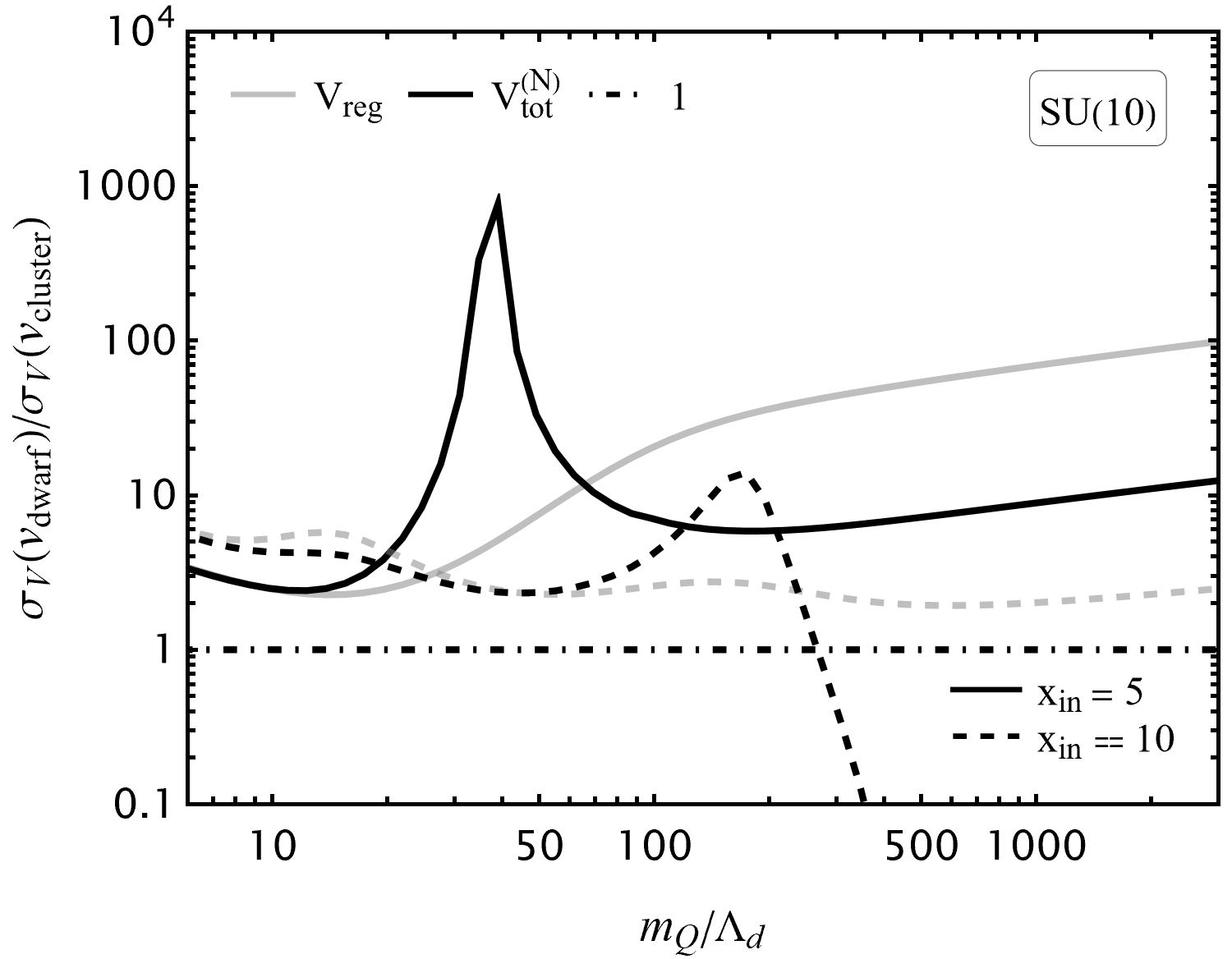}
\caption{Ratio between the self-interaction cross section at $v_{\rm collapse}=5$~km/s and the cross section evaluated at cluster velocities $v_{\rm cluster}=1500$~km/s, as a function of the hierarchy parameter $\xi=m_Q/\Lambda_d$. The gray, red, blue and black curves correspond to the potentials $V_{\rm reg}$, $V_{\rm tot}^{(\rm geo)}$, $V_{\rm tot}^{(3/4)}$ and $V_{\rm tot}^{(N)}$, respectively, using solid (dashed) lines for $x_{\rm in}=5$ ($x_{\rm in}=10$). The horizontal black line indicates a ratio of unity. The left, center, and right panels correspond to $N=2$, $3$, and $10$, respectively.}
\label{fig:cross_section_velocity_ratio_core_collapse}

\end{figure}

The resulting ratios are shown in
\cref{fig:cross_section_velocity_ratio_core_collapse}. For $N=2$ and
$3$, the enhancement at low $\xi$ is relatively mild, reaching its maximum only in the region $\xi\lesssim10$, where the
heavy-baryon description is approaching the limit of its regime of
validity. These
enhancements are therefore insufficient to reach the very large
cross sections required for the gravothermal-collapse scenario.
For $\SU(10)$, on the other hand, large enhancements between cluster
and collapse velocities occur in  a relatively narrow region
associated with a peak in $\xi$.

\begin{figure}
\centering
\includegraphics[width=0.33\linewidth]{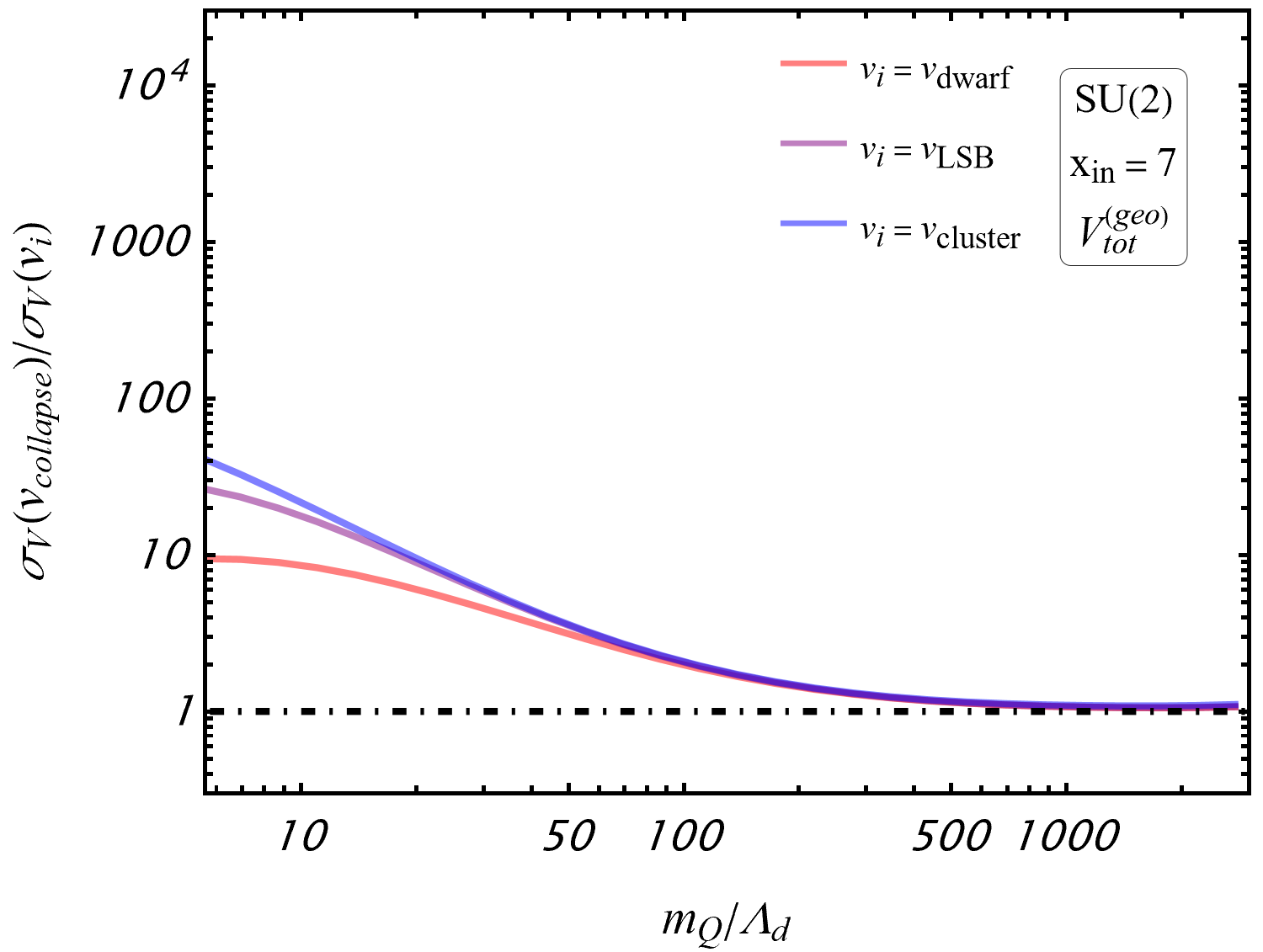}\includegraphics[width=0.33\linewidth]{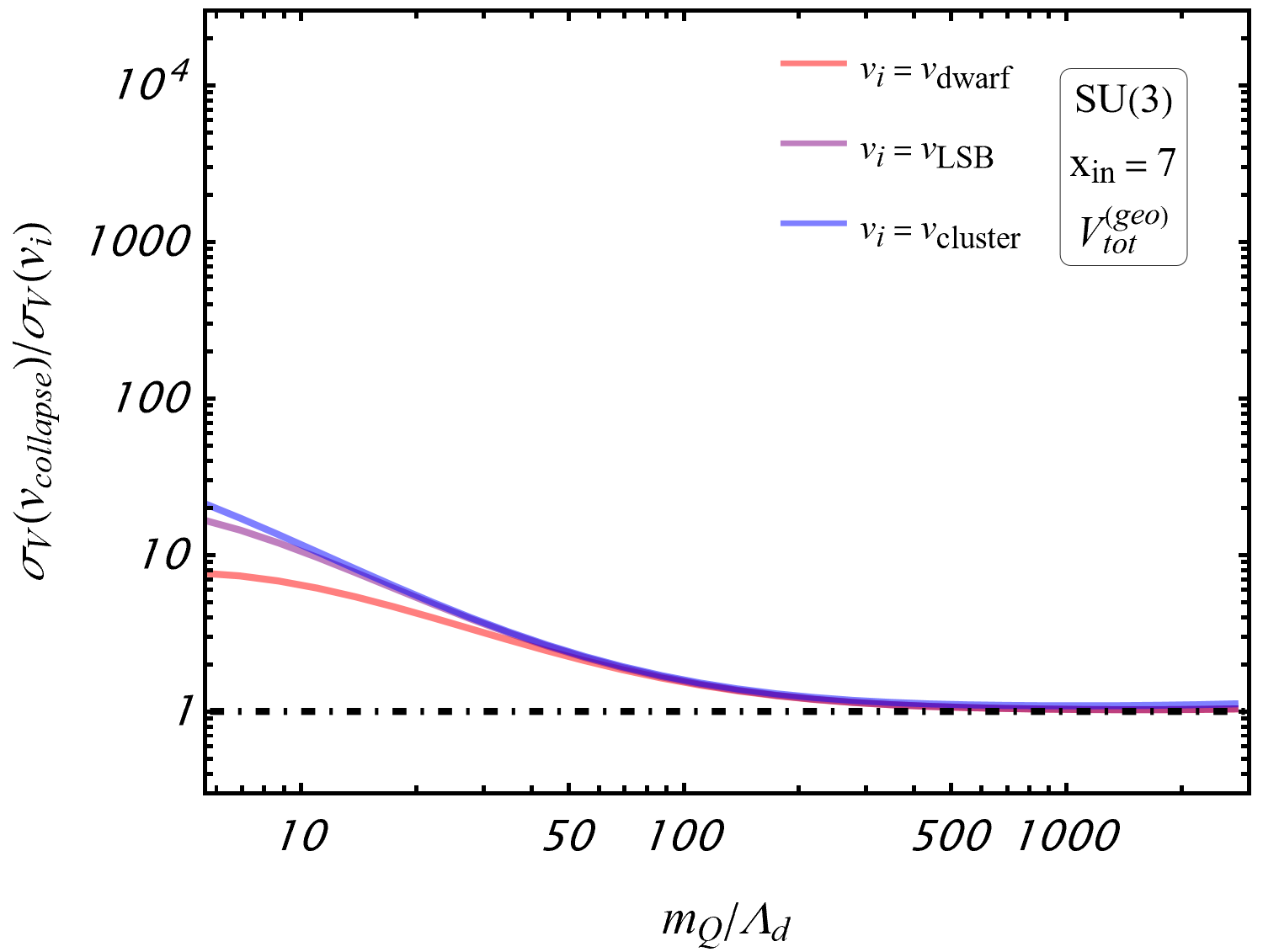}\includegraphics[width=0.33\linewidth]{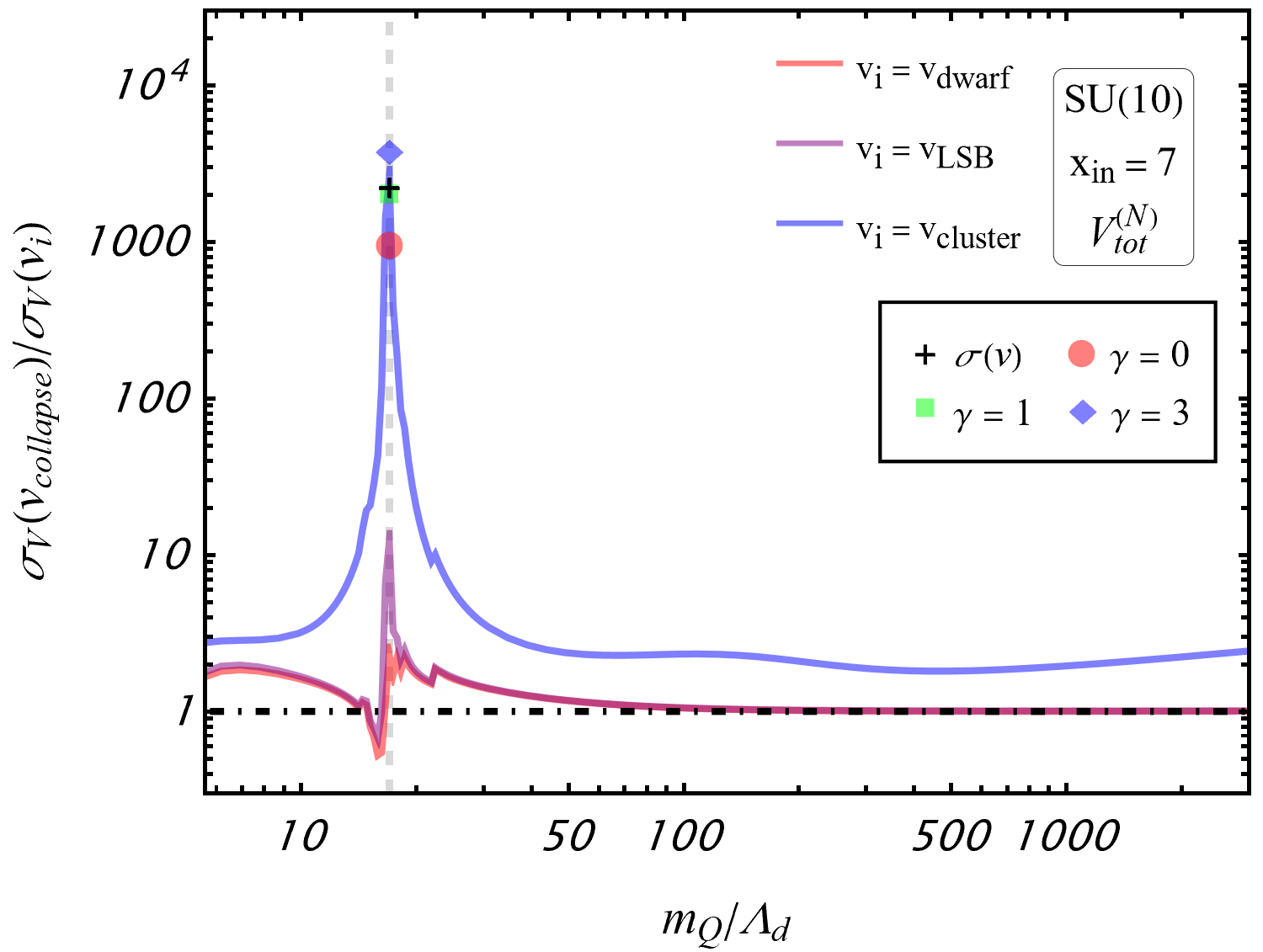}
\caption{Ratio between the self-interaction cross section at $v_{\rm collapse}=5$~km/s and the cross section at higher velocities $v_i$, computed using  $V_{\rm tot}^{(\rm geo)}$ for $N=2,3$ and $V_{\rm tot}^{(N)}$ for $N=10$, both  with $x_{\rm in}=7$. The red, purple, and blue curves correspond to $v_i$ equal to $v_{\rm dwarf}=30$~km/s, $v_{\rm LSB}=100$~km/s, and $v_{\rm cluster}=1500$~km/s, respectively. The horizontal dot-dashed black line indicates a ratio of unity.  Similarly to \cref{fig:cross_section_velocity_ratio7_averages}, we superimpose the velocity-averaged enhancement factors  $\overline{\sigma}_V(v_{\rm collapse})_\gamma/\,\overline{\sigma}_V(v_{\rm cluster})_\gamma$ using the benchmark point $\xi=17$, for $\gamma=0$ (red circle), $\gamma=1$ (green square), and $\gamma=3$ (blue diamond), corresponding respectively to the unweighted, rate-weighted, and energy-weighted velocity moments. The black $+$ marker denotes the direct ratio   $\sigma_V(v_{\rm \rm collapse})/\sigma_V(v_{\rm cluster})$ for comparison. The left, center, and right panels correspond to $N=2$, $3$, and $10$, respectively.}
\label{fig:cross_section_velocity_ratio7_core_colapse}
\end{figure}

The dependence on the short-distance suppression scale is very pronounced. Increasing the suppression to $x_{\rm in}\gtrsim9$ substantially reduces the attractive contribution at short distances and can eliminate the strong resonant enhancement necessary for gravothermal collapse, causing the cross section to approach a geometric-like behavior. Since $x_{\rm in}\lesssim9$ is motivated by our treatment of the short-distance baryon interaction (see \cref{sec:regularizing_potential}), these results provide a strong indication that, in the large-$N$
regime, extremely large self-interactions at very low velocities are
a plausible consequence of the dark-baryon potential rather than an
artifact of a particular choice of phenomenological parameters. For the representative choice $V_{\rm tot}^{(\rm geo)}$ for $N=2,3$ and $V_{\rm tot}^{(N)}$ for $N=10$ both with $x_{\rm in}=7$, the same behavior is displayed more clearly in \cref{fig:cross_section_velocity_ratio7_core_colapse}.

Having identified a $\xi$-region with the required velocity hierarchy, we next determine the corresponding dark baryon and confinement scales. The resulting parameter space is shown in \cref{fig:plane_mqXlambda}, where the green regions satisfy $$
75~\mathrm{cm}^2/\mathrm{g}
<
\sigma_V(v_{\rm collapse})/m_B
<
1000~\mathrm{cm}^2/\mathrm{g},$$
corresponding to an approximate uncertainty of 25\% below and above the range inferred for the collapsing cross section in Refs.~\cite{Vegetti:2026mmx,Zhang:2026gur}. As emphasized in Ref.~\cite{Vegetti:2026mmx}, values larger than this range could also be compatible with the properties of the second perturber, so we do not regard the upper edge of this interval as a strict physical limit.

For $\SU(10)$, the preferred dark baryon masses and confinement scales
once again lie within roughly one order of magnitude of the proton mass
and $\Lambda_{\rm QCD}$. This coincidence therefore persists even when the requirement of extremely large self-interactions at $v\sim5$~km/s is added to the small-scale and cluster constraints. Moreover, we find regions in which all of the phenomenological requirements can be simultaneously satisfied: self-interactions in the preferred range at dwarf and LSB scales,
$
\sigma_V/m_B\sim1\text{--}10~{\rm cm}^2/{\rm g}$,
compatibility with cluster constraints, $
\sigma_V/m_B\lesssim1~\mathrm{cm}^2/\mathrm{g}$,
and an extremely large cross section at $v_{\rm collapse}=5$~km/s,
$
\sigma_V/m_B\sim100\text{--}800~\mathrm{cm}^2/\mathrm{g}$,
of the magnitude required to reproduce the inferred properties of the second perturber in JVAS B1938+666. We emphasize that regions satisfying these requirements are found as long as
$x_{\rm in}\lesssim9$, although their precise location in parameter
space depends on the exact value of the regularization scale.

To illustrate the corresponding velocity evolution, we select a
benchmark point within the resonant enhancement region visible in
\cref{fig:cross_section_velocity_ratio7_core_colapse}. Specifically,
we choose $\xi=17$ and present its corresponding velocity-dependent
cross section in \cref{fig:vel_evolution_sigmaVSvelocity}.

\Cref{fig:cross_section_velocity_ratio7_core_colapse} also presents the velocity-averaged enhancement factors $\overline{\sigma}_V(v_{  \rm collapse})_\gamma/\,\overline{\sigma}_V(v_{\rm cluster})_\gamma$ with $
\gamma=0,1,3$ for our benchmark point. These quantities provide a more appropriate comparison between characteristic halo velocities than the cross section evaluated at a single velocity, since DM particles in halos follow a velocity distribution. They confirm that the benchmark model can produce extremely large self-interactions at $v_{\rm collapse}$ while remaining well below the observational bounds at cluster velocities.

The behavior at dwarf and LSB velocities requires more care. Once gravothermal evolution is taken into account, relatively large self-interaction cross sections at dwarf scales can remain compatible with observations and may even be favored in some analyses. In particular, Ref.~\cite{Correa:2020qam} finds that values substantially larger than those inferred from analyses based solely on core formation can be compatible with the observed diversity of dwarf-galaxy density profiles. We therefore use the gravothermal-collapse analysis of Ref.~\cite{Correa:2020qam} as the primary dwarf-scale comparison in this section, rather than interpreting the commonly quoted $\sigma/m\sim1$--$10$~cm$^2$/g range from core-formation analyses~\cite{Kaplinghat:2015aga} as a strict requirement.

\begin{figure}
\centering
\includegraphics[width=0.975\linewidth]{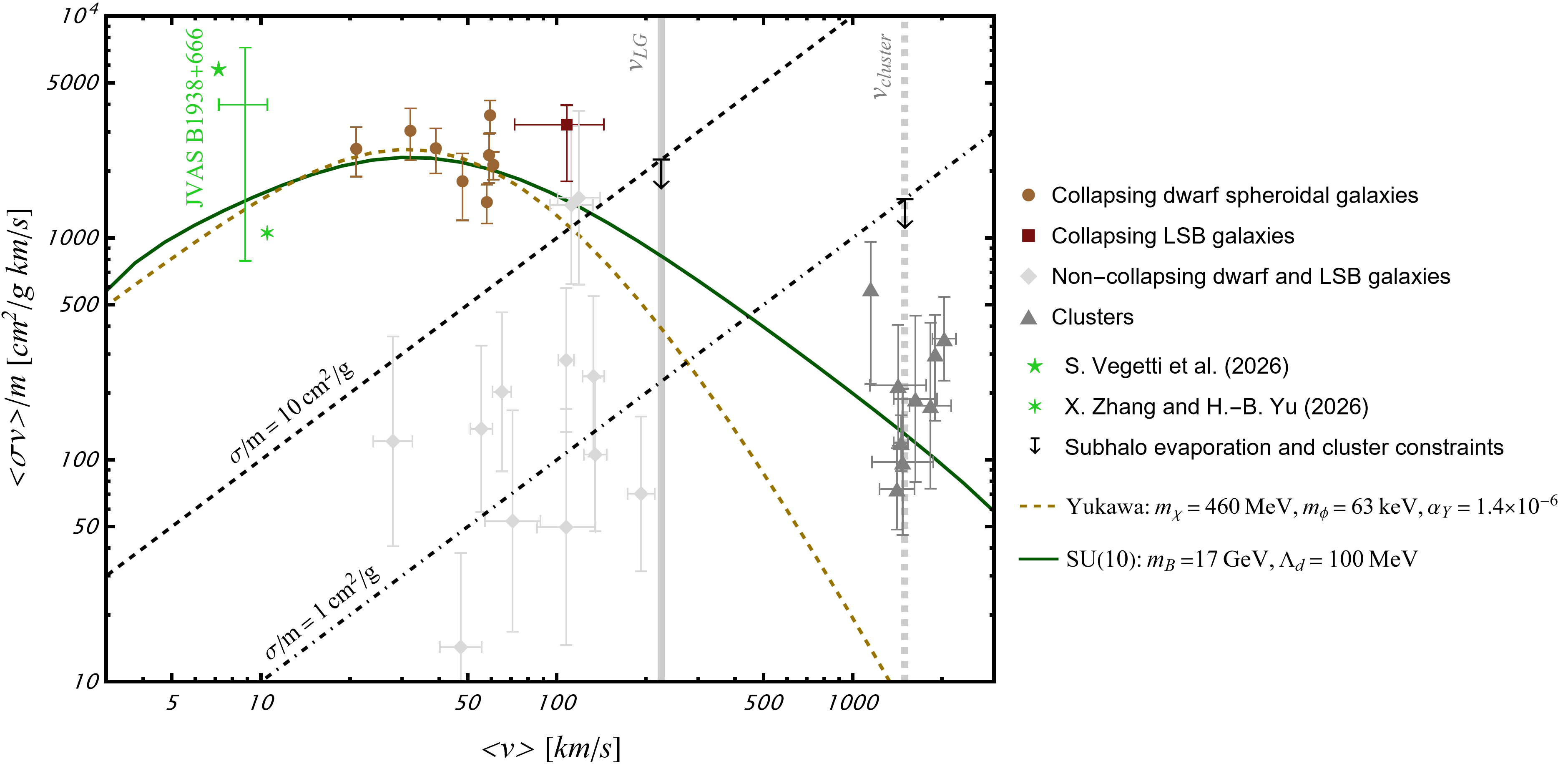}
\caption{Velocity-weighted  self-interaction cross section per unit dark matter mass as a function of the mean collision velocity. The gray data points correspond to an astrophysical compilation of cluster cross sections from Refs.~\cite{Kaplinghat:2015aga,Sagunski:2020spe,Andrade:2020lqq}, while the light-gray data points correspond to dwarf and LSB galaxies from Ref.~\cite{Kaplinghat:2015aga}. Both data sets are obtained under the assumption that gravothermal core collapse is absent. The brown points correspond to dwarf spheroidal galaxies data from Ref.~\cite{Correa:2020qam} and the dark red point to LSB galaxies from Ref.~\cite{Roberts:2024uyw}, where gravothermal collapse is included in the analyses. The green points and error bars indicate the approximate range of self-interaction cross sections inferred to explain the second perturber of JVAS B1938+666~\cite{Vegetti:2026mmx,Zhang:2026gur}. The dark green curve shows the velocity-averaged predictions of our benchmark model, where the dark baryon mass is chosen to provide the best fit to the gravothermally evolved dwarf-spheroidal data of Ref.~\cite{Correa:2020qam}. The dark yellow dashed curve shows the corresponding best-fit Yukawa model in the Born regime, included for comparison. The upper limit at Local Group velocities ($v_{\rm LG}\simeq225$~km/s) is taken from Ref.~\cite{Correa:2020qam} and corresponds to $\sigma/m<10$~cm$^2$/g, as inferred from the absence of significant subhalo evaporation~\cite{Vogelsberger:2012ku,Rocha:2012jg,Zavala:2012us}. We also indicate the commonly adopted cluster bound, $\sigma/m<1$~cm$^2$/g at $v_{\rm cluster}=1500$~km/s~\cite{Balan:2024cmq}.
}
\label{fig:final_plot}
\end{figure}

At the same time, our model exhibits a substantial decrease in the cross section between dwarf/LSB and cluster velocities. Consequently, they can satisfy the cluster constraints implying no gravothermal core collapse at cluster scales such that the inferred cross sections from cluster core-formation analyses still apply. This behavior is precisely what motivates the comparison with the full set of astrophysical determinations. We therefore compare the velocity-averaged predictions of our benchmark models with the observational compilation shown in \cref{fig:final_plot}, which combines the results of Refs.~\cite{Kaplinghat:2015aga,Correa:2020qam,Roberts:2024uyw,Andrade:2020lqq,Sagunski:2020spe,Vegetti:2026mmx,Zhang:2026gur,Rocha:2012jg,Zavala:2012us,Vogelsberger:2012ku}.\footnote{We exclude the Ursa Minor data point from Ref.~\cite{Correa:2020qam} in our compilation because its quoted uncertainties appear to be strongly underestimated. Notably, Ref.~\cite{Correa:2020qam} also excludes Ursa Minor from their Yukawa fit.} For our benchmark choice $\xi=17$, the baryon mass is fixed by reducing the $\chi^2$ of  the corresponding (velocity-averaged) fit with the gravothermal core collapse data for dwarf spheroidal galaxies of Ref.~\cite{Correa:2020qam}.

For comparison, we also perform a fit using a standard Yukawa potential in the Born regime. We use the transfer cross section given in Ref.~\cite{Tulin:2013teo},
\begin{equation}
\sigma_T(v)=
\dfrac{8\pi\alpha_Y^2}{m_\chi^2v^4}
\left[
\ln\left(1+\frac{m_\chi^2v^2}{m_\phi^2}\right)
-
\frac{m_\chi^2v^2}{m_\phi^2+m_\chi^2v^2}
\right],
\end{equation}
where $m_\chi$ is the DM mass, $m_\phi$ the mediator mass, and $\alpha_Y$ the coupling.  We minimize $\chi^2$ with respect to the two independent parameter combinations: $\alpha_Y^2/m_\chi^3$ and $m_\phi/m_\chi$. Additionally, we fix the third parameter by requiring the Born regime validity $\alpha_Y m_\chi \ll m_\phi$, specifically $\alpha_Y m_\chi=0.01 m_\phi$. For the resulting best-fit point, we then compute the velocity-averaged cross section and allow $\alpha_Y$ to be rescaled to minimize the corresponding $\chi^2$ with a velocity-averaged fit. This rescaling changes the best-fit value of $\alpha_Y$ by less than 5\%.

A fully consistent treatment would instead minimize the $\chi^2$ of the velocity-averaged cross section directly with respect to the independent Yukawa parameters. Such a fit is numerically more expensive because the velocity average of the Yukawa cross section does not generally admit a closed analytical form. Since the purpose of the Yukawa model here is to provide a comparison with the normalization-only fit performed for the heavy dark baryon benchmark, this approximate treatment is sufficient.

In \cref{tab:fit_results}, we present the total $\chi^2$ and corresponding $p$-values for the fits to the dwarf-spheroidal data of Ref.~\cite{Correa:2020qam}, together with the $\chi^2$ and $p$-values obtained when the same models are compared with the cluster data.\footnote{Since we select the $\xi$ benchmark that exhibits the desired decreasing velocity dependence of the cross section, we take both the Yukawa and confining heavy-quark models to have two fitted parameters when computing the $p$-values. A complete statistical analysis would require a detailed scan over $\xi$ and a subsequent minimization of $\chi^2$, which is computationally costly.} The latter data are not used to determine the model parameters and therefore provide an independent test of the predicted velocity dependence.

\begin{table}[t]
\centering
\begin{tabular}{c|c c|c c|}
\multirow{2}{*}{Model} &
\multicolumn{2}{c|}{Dwarf spheroidal galaxies} &
\multicolumn{2}{c|}{Clusters} \\
& $\chi^2$ & $p$-value & $\chi^2$ & $p$-value \\ \hline
Yukawa
& 12.9 & 0.045 & 31 & 0.00036 \\
$\SU(10)$
& 13.2 & 0.040 & 9.6 & 0.38 \\
\end{tabular}
\caption{Goodness of fit for the different models and benchmark points. The parameters are determined from the dwarf-spheroidal data of Ref.~\cite{Correa:2020qam}, while the cluster data~\cite{Kaplinghat:2015aga,Sagunski:2020spe,Andrade:2020lqq} provide an independent test of the resulting velocity dependence. Note that all fits  safely satisfy galaxy and cluster constraints and are expected to explain the second perturber of the strong gravitational lens system JVAS B1938+666~\cite{Powell:2025rmj,Vegetti:2026mmx,Zhang:2026gur}.}
\label{tab:fit_results}
\end{table}

Several conclusions can be drawn from \cref{fig:final_plot} and
\cref{tab:fit_results}. First, the heavy-dark-baryon benchmark can
reproduce the preferred self-interaction strengths over a broad range
of astrophysical systems within current observational uncertainties.
Second, the same velocity dependence that produces the very large cross
section required at $v\sim5$~km/s can strongly suppress the interaction
at large velocities, avoiding current constraints. The $\SU(10)$
benchmark provides a substantially better description of the cluster
data than the Yukawa benchmark---despite the cluster data not being
included in the fit---while giving a comparable fit to the
dwarf-spheroidal data. The Yukawa model can nevertheless accommodate
the inferred cross section of the second perturber, illustrating that
the distinguishing feature is the velocity dependence rather than the
overall normalization alone.

Future more precise determinations of self-interaction cross sections
across a wide range of velocities will therefore be crucial for
distinguishing between the different functional forms of the
interaction and, ultimately, for probing the microscopic nature of the
dark sector.

Finally, we again find that the dark baryon masses and confinement scales favored by the combined astrophysical requirements lie strikingly close to the corresponding QCD scales. The persistence of this coincidence, now including the requirement of extremely large self-interactions at very low velocities, further motivates an asymmetric heavy-dark-baryon interpretation. In such a scenario, a dark baryon asymmetry comparable to the visible baryon asymmetry can therefore account for the observed DM abundance while the QCD-like dark confinement scale and GeV-scale dark baryon mass are independently selected by the self-interaction phenomenology. We explore this possibility in the following section.

\section{Asymmetric heavy dark baryons}
\label{sec:asymmetric_DM}

Up to this point, we have implicitly assumed that dark baryons account for the entire DM relic abundance. We now discuss the cosmological production mechanisms capable of realizing this scenario while remaining consistent with the self-interaction requirements established in \cref{sec:results}. Interestingly, the parameter space selected by late-time halo dynamics also has important implications for the early-Universe evolution of the dark sector. The same parameters that determine the self-interaction cross section, namely the hierarchy $m_Q/\Lambda_d$ and the overall mass scale, also control the annihilation efficiency of the heavy dark quarks and the resulting abundances of dark baryons and dark glueballs.

Throughout the viable self-interacting region identified in \cref{sec:results}, the dark quark mass typically lies at the GeV scale, while the confinement scale is close to that of QCD, $\Lambda_d\sim100~\mathrm{MeV}$, up to order-one variations depending on the gauge group $N$, the short-distance regularization of the baryon--baryon potential, and the astrophysical observables considered. It is remarkable that these scales are selected entirely by late-time halo dynamics rather than by imposing the observed DM relic abundance. This naturally points toward an ADM interpretation, in which the relic abundance is determined by a primordial dark baryon asymmetry rather than thermal freeze-out~\cite{Nussinov:1985xr,Barr:1990ca,Kaplan:2009ag,Petraki:2013wwa,Zurek:2013wia}.

Above the confinement transition, the relevant degrees of freedom are the heavy vector-like dark quarks $Q$ and the dark gluons $g$. The symmetric $Q\bar Q$ population efficiently annihilates before confinement, predominantly into dark gluons. Once confinement occurs, any remaining quarks and antiquarks become incorporated into dark hadrons, after which baryon--antibaryon annihilations further suppress the symmetric baryonic component.\footnote{Mesons do not carry a conserved dark baryon number and therefore can decay into glueballs, with the exception of vector mesons in $\SU(2)$, which can carry $\SU(2)_B$. However, their abundance is suppressed by $Q\bar Q$ pair annihilation. }

An upper bound on the symmetric dark baryon relic abundance can already be obtained by considering only the perturbative annihilation process $Q\bar Q\rightarrow gg$ with  cross section~\cite{Gouttenoire:2023bqh}
\begin{equation}
   \langle \sigma_{\rm ann} v \rangle_{\bar{Q}Q} \sim \frac{\pi \alpha ^{2}_d (m_Q)}{m_Q^2}f_c(N)S(v,N)\,,
\end{equation}
where
\begin{equation}
	f_c(N)=\left(\frac{N^4-3N^2+2}{16N^3}\right)\,,
\end{equation}
while $S(v,N)$ is an $N$-dependent Sommerfeld enhancement factor which only leads to negligible corrections. The corresponding upper bound is
\begin{equation}
\label{eq:relic_pert}
\frac{\Omega_{\rm B}h^2}{0.12}
\lesssim
10^{-8}
\left[
\left(\dfrac{m_B}{\mathrm{GeV}}\right)
\ln(2\xi)
\right]^2
\zeta_Q^{-1},
\end{equation}
where $\zeta_Q\equiv T_{\rm SM}/T_d|_{T_d\simeq m_Q/40}$ denotes the SM--dark-sector temperature ratio at dark quark freeze-out.\footnote{The freeze-out temperature $T_d\simeq m_Q/40$ is obtained using the perturbative annihilation cross section. For $\xi=m_Q/\Lambda_d\lesssim40$, this approximation eventually breaks down because freeze-out occurs during the confinement transition.} Here and in the following, we assume that the energy density of the Universe is dominated by SM radiation.

For the parameter region relevant to self-interactions, namely $\xi\lesssim100$ and $m_B\lesssim50~\mathrm{GeV}$, the perturbative estimate already implies that the symmetric component contributes less than $0.1\%$ of the observed DM abundance for a dark sector with the same temperature as the SM ($\zeta_Q=1$). A colder dark sector suppresses the symmetric relic abundance even further.

An independent upper bound can be obtained by considering the non-perturbative baryon--antibaryon re-annihilation process studied in Ref.~\cite{Mitridate:2017oky}.  Assuming that the baryon abundance immediately after confinement, $Y_{B,0}$, is sufficiently large so that the $Y_{B,0}^{-1}$ term can be neglected when solving the Boltzmann equation---otherwise the symmetric abundance would already be negligible before re-annihilation---their result implies
\begin{equation}
\label{eq:relic_non_pert}
\frac{\Omega_{\rm B}h^2}{0.12}
\lesssim
10^{-10}
\left(\dfrac{m_B}{\mathrm{GeV}}\right)^2
\dfrac{\alpha_B}{\sqrt{N}}\xi\,
\zeta_g^{-1}
<
10^{-10}
\left(\dfrac{m_B}{\mathrm{GeV}}\right)^2
\xi\,\zeta_g^{-1},
\end{equation}
where $\zeta_g\equiv T_{\rm SM}/T_d|_{T_d\simeq\Lambda_d}$ denotes the SM--dark-sector temperature ratio at the confinement transition and we have used the baryon--antibaryon annihilation cross section provided in eq.~(3.11) of Ref.~\cite{Mitridate:2017oky}.\footnote{Note that there are stable rearrangement channels, such as $QQ$, which nevertheless yield negligible relic abundances~\cite{Mitridate:2017oky}. For $\SU(2)$, rearrangement into stable vector mesons is also possible~\cite{Francis:2018xjd}. However, as shown previously, the $\SU(2)$ model is irrelevant for SIDM. Therefore additional stable states  do not affect our conclusions.} In the last inequality we have used $\alpha_B<1$, which holds throughout the validity region of our analysis (see \cref{fig:couplingXratio}), together with $N>1$. Therefore, even if the dark baryon abundance immediately after confinement were initially large, the efficient rearrangement of baryons into mesons described in Ref.~\cite{Mitridate:2017oky} would remove essentially the entire symmetric component for the 10~GeV-scale baryon masses and $\xi\lesssim100$ relevant to this work. Only a hot dark sector could possibly leave a non-negligible symmetric relic abundance. Such scenarios, however, would lead to a Universe dominated by dark-sector entropy and make the overproduction of dark glueballs extremely difficult to avoid, as discussed below.

A fully quantitative determination of the symmetric dark baryon relic abundance would require solving the Boltzmann equations through the confinement transition, consistently including perturbative $Q\bar Q$ annihilation, bound-state formation~\cite{Binder:2023ckj} (including unitarity restoration~\cite{Flores:2026yay}), baryon--antibaryon re-annihilation~\cite{Mitridate:2017oky} (including phase-transition effects \cite{Gouttenoire:2023bqh}), and possible entropy injection from dark glueball decays~\cite{Gouttenoire:2023bqh}.\footnote{All of these effects are expected to further reduce the symmetric dark baryon abundance.} Nevertheless, the qualitative conclusion is already robust: throughout the self-interacting parameter space identified in \cref{sec:results}, a purely symmetric thermal history generically underproduces the observed DM abundance. This strongly motivates an asymmetric origin for the surviving dark baryons~\cite{Bottaro:2021snn}.

The model contains an accidental dark baryon number under which the lightest dark baryon is stable. Analogously to the SM, we may therefore assume that an initial dark baryon asymmetry, $\eta_D$, is generated in the early Universe. For the GeV-scale dark baryon masses favored by our self-interaction analysis, the required dark baryon asymmetry is therefore comparable to the visible baryon asymmetry. This is the usual ADM coincidence, but here it arises in a particularly suggestive way: the GeV scale is not chosen solely to reproduce $\Omega_{\rm DM}/\Omega_b$, but is instead selected independently by the requirement of large, velocity-dependent self-interactions (see also Ref.~\cite{Chung:2025gev}).

The microscopic origin of $\eta_D$ is, however, model dependent. It may arise from a common visible--dark asymmetry generated at high temperatures, from higher-dimensional transfer operators relating visible baryon or lepton number to dark baryon number, or from the out-of-equilibrium decay of heavy states carrying both visible and dark quantum numbers~\cite{Kaplan:2009ag,Petraki:2013wwa,Zurek:2013wia,Bottaro:2021snn}. In particular, dark grand-unification models that aim to explain the DM--baryon coincidence predict DM to be composed of dark baryons together with comparable visible and dark asymmetries, $\eta_D\sim\eta_B$, see Refs.~\cite{Murgui:2021eqi,Chung:2024dgu}. Since the present work focuses on the low-energy phenomenology of the confining dark sector, we do not specify a complete ultraviolet mechanism for generating the asymmetry.

The asymmetric interpretation is particularly attractive for models with large self-interactions. In a generic symmetric scenario, the same dark-sector dynamics responsible for enhancing elastic scattering can also lead to efficient annihilation and observable indirect-detection signals. In an asymmetric scenario, by contrast, the dominant late-time process is elastic baryon--baryon scattering, which governs halo dynamics, rather than baryon--antibaryon annihilation. If a small residual antibaryon population survives, however, $B\bar B$ annihilations may still affect the thermal history or generate indirect signals~\cite{Baldes:2017gzw}.\footnote{The importance of these effects also depends on the portal through which dark mesons or glueballs decay into the visible sector and is therefore model dependent.} Nevertheless, the estimates of \cref{eq:relic_pert,eq:relic_non_pert} indicate that, throughout the parameter space favored by self-interactions, the symmetric component is expected to be negligible, making observable indirect signals unlikely.

A crucial cosmological issue is the relic abundance of dark glueballs. Since dark gluons are massless in the deconfined phase ($T\gtrsim\Lambda_d$), they are relativistic  and behave as dark radiation. Consequently, they retain the large number density characteristic of relativistic species. Once the dark-sector temperature falls below the confinement scale, $T_d\sim\Lambda_d$, these gluons hadronize into a spectrum of glueballs, the lightest scalar state having mass $m_{\rm DG}\simeq6.4\,\Lambda_d$. As a result, the large relativistic gluon abundance is converted into a potentially enormous relic abundance of non-relativistic glueballs~\cite{Carenza:2022pjd,Carenza:2023eua},
\begin{equation}
\label{eq:rhoDG}
\frac{\Omega_{\rm DG}h^2}{0.12}\simeq
10^6
\left(\frac{\Lambda_d}{100\,\mathrm{MeV}}\right)
\zeta_g^{-3}\,.
\end{equation}  For the region favored by self-interactions, $\Lambda_d\sim10$--$100~\mathrm{MeV}$, the corresponding glueball masses lie in the range of tens to hundreds of MeV. Therefore, if glueballs are stable or sufficiently long-lived, their relic abundance must be strongly suppressed. This typically requires a substantially colder dark sector; for $100~\mathrm{MeV}$, a glueball fraction below roughly $1\%$ implies $\zeta_g\gtrsim5\cdot 10^2$, corresponding to $T_d\lesssim 2\cdot 10^{-3}\,T_{\rm SM}$.

Alternatively, because glueballs are not protected by the accidental dark baryon number,   they do not need to be stable.  If suitable portal interactions are present, glueball decays can deplete the glueball abundance while reheating the visible sector~\cite{Forestell:2016qhc}.\footnote{Note that $C$-preserving portals lead to the stability of the lightest $C$-odd glueball for $N\geq 3$~\cite{McKeen:2024trt}.} Such scenarios must nevertheless satisfy the usual constraints from Big Bang nucleosynthesis, the cosmic microwave background, and dark radiation~\cite{Juknevich:2009ji,Forestell:2017wov,Carenza:2022pjd,Carenza:2023eua,McKeen:2024trt}.

The confinement transition could also generate gravitational waves.
With no light quarks and $m_Q\gg T_c$, the thermal gauge dynamics
approaches pure $SU(N)$ Yang--Mills theory, whose transition is
first order for $N\geq3$ \cite{YMtransition}.
For our parameters, $T_c=\mathcal{O}(\Lambda_d)\sim10$--$100\,\mathrm{MeV}$, a transition at this scale could produce nanohertz gravitational
waves, provided that its duration and energy release are suitable.
Assuming a phase-transition origin for the entire background and
relativistic bubble walls, the corrected NANOGrav 15-year analysis
\cite{NANOGravNP,NANOGravNPErratum} favours
$T_*=8.8$--$50\,\mathrm{MeV}$,
$\alpha_*=0.66$--$5.0$, and $\beta/H_*\simeq7$--$30$
for sound waves, or $T_*=79$--$349\,\mathrm{MeV}$,
$\alpha_*>1.7$, and $\beta/H_*\lesssim6.7$
for bubble collisions, at the $68\%$ credible level.
Here $\alpha_*$ measures the transition strength relative to
the radiation density, while $\beta/H_*$ is its inverse duration
in Hubble units. The fitted temperature refers to the cosmological
bath and need not coincide with the dark-sector temperature.
Lattice-informed effective models of pure Yang--Mills theory
predict only percent-level supercooling and
$\beta/H_*\sim10^4$ near $N=10$--$20$, with correspondingly
weak gravitational-wave signals \cite{YMsupercooling,YMGW}.
Large $N$ therefore does not guarantee a signal of the strength
observed by PTAs.

The cold history considered above imposes a further suppression.
Writing $r=T_d/T_{\rm SM}=\zeta_g^{-1}$ at confinement,
the stable-glueball scenario requires $r=\mathcal{O}(10^{-3})$.
The dark-to-visible radiation-density ratio is parametrically
$\rho_{\rm rad,d}/\rho_{\rm rad,SM}
\sim2(N^2-1)r^4/g_{*,\rm SM}$.
Without substantial supercooling, the energy released by the
transition is likewise a small fraction of the total cosmic
energy density. For $N=10$, $T_d\sim100\,\mathrm{MeV}$,
and $r=10^{-3}$, one has $T_{\rm SM}\sim100\,\mathrm{GeV}$
and a radiation-density ratio of order $10^{-12}$.
Assuming radiation domination and no subsequent entropy injection,
the sound-wave peak today is approximately
$20\,\mu\mathrm{Hz}\,(\beta/H_*)/v_w$, where $v_w$ is the
bubble-wall velocity, well above the PTA band for the transition
rates considered here. A warmer dark sector avoids this energy
suppression, but requires a viable mechanism for glueball decay
and a consistent treatment of reheating, which is beyond the scope of this work.

\section{Conclusions}\label{sec:conclusion}

In this work, we have studied self-interacting dark matter in the form of heavy baryons arising from a confining $\SU(N)$ dark sector. Unlike scenarios in which self-interactions are mediated by an additional elementary light particle, the interaction here emerges from the composite nature of the dark matter itself. The long-distance interaction between these composite states is generated by their induced chromo-electric dipole moments, providing a natural mechanism for sizable and strongly velocity-dependent self-interactions. We derived the resulting baryon-baryon potential and identified the London, Casimir--Polder, and glueball-mediated regimes, while parametrizing the short-distance region where the long-distance description breaks down.

By solving the scattering problem beyond the Born approximation, we
found that the phenomenology depends strongly on the number of colors.
For $N=2$ and $3$, the attractive interaction remains too weak to
overcome the short-distance repulsive core, and no phenomenologically
relevant velocity dependence is obtained within our minimal treatment.
Mild enhancements at low $\xi$ occur only near the boundary of validity
of the heavy-baryon description and are insufficient to reach the
self-interaction strengths required by small-scale structure. In contrast, in the large-$N$ regime the attractive interaction remains
competitive with the repulsive core for $N\sim10$, and the resulting
cross sections exhibit an approximately $N$-independent qualitative behavior over
the range $N\sim10$--$100$. For much larger $N$, the Pauli repulsion
eventually becomes dominant, leading to a geometrical-like scattering
behavior. Taking $\SU(10)$ as a representative finite-$N$
realization of the large-$N$ regime, we find that the self-interaction
strength can vary substantially across the range of velocities relevant
for dwarf galaxies, low-surface-brightness galaxies, and galaxy
clusters.

We further investigated whether this velocity dependence can accommodate gravothermal core collapse, focusing on the second perturber of the strong gravitational lens system JVAS B1938+666~\cite{Powell:2025rmj,Vegetti:2026mmx,Zhang:2026gur}. In the large-$N$ regime, we find regions of parameter space in which the self-interaction cross section rises by orders of magnitude towards $v\sim5~\mathrm{km/s}$, reaching $\mathcal{O}(100)~\mathrm{cm}^2/\mathrm{g}$, while remaining consistent with the self-interaction phenomenology at larger velocities. Thus, the same microscopic interaction can simultaneously accommodate the self-interactions inferred from dwarf spheroidal galaxies and galaxy clusters and the much larger cross sections potentially associated with the second perturber of JVAS B1938+666. For the $\SU(10)$ benchmark considered here, the resulting velocity dependence also provides a substantially better description of the cluster data than the Yukawa benchmark, while giving a comparable fit to the dwarf-spheroidal data. The Yukawa interaction can nevertheless accommodate the inferred cross section of the second perturber, illustrating that the velocity dependence, rather than the overall normalization alone, can provide a means of distinguishing different microscopic interactions.

Remarkably, the astrophysical requirements select dark baryon masses and confinement scales close to the ordinary baryon and QCD scales, respectively. In the large-$N$ regime, the viable regions typically contain GeV-scale dark baryons and $\Lambda_d\sim\Lambda_{\rm QCD}$. These scales are not imposed by requiring the correct relic abundance, but instead emerge from the late-time self-interaction phenomenology. The same parameter region has important cosmological implications. Efficient annihilation of the symmetric heavy-quark population, followed by baryon-antibaryon re-annihilation after confinement, strongly suppresses the symmetric dark baryon abundance. This naturally motivates an asymmetric origin for the surviving dark matter. For GeV-scale dark baryons, a dark baryon asymmetry comparable to the visible baryon asymmetry can account for the observed abundance, providing a realization of the asymmetric-dark-matter coincidence. A complete cosmological realization must, however, specify the origin of the asymmetry and address the potentially large abundance of dark glueballs, for example through a colder dark sector or suitable glueball decays.

Taken together, our results show that the phenomenology is strongly
dependent on the number of colors: while small $N$ does not yield
phenomenologically relevant self-interactions within the heavy-baryon
regime, the large-$N$ regime provides a viable realization with strong
velocity dependence and an approximately $N$-independent behavior for
$N\sim10$. For $N\gg10$, the Pauli repulsion becomes increasingly
dominant, leading to a geometrical-like scattering behavior. For
intermediate $N=4$--$8$, the present approximations do not yet allow a
definitive conclusion. Improved baryon wavefunctions and
nonperturbative calculations will be needed to determine how the
large-$N$ behavior is approached.

{Future observations of dwarf and low-surface-brightness galaxies, galaxy clusters, and strongly lensed subhalos will be crucial for determining whether the velocity dependence suggested by current data is realized in nature. In particular, confirming or excluding gravothermal evolution in systems such as JVAS B1938+666 could discriminate between different microscopic models of dark matter. Beyond these astrophysical probes, the first-order confinement transition in our dark sector offers a complementary cosmological probe through potentially observable nanohertz gravitational waves, although a detectable signal is not guaranteed in the minimal model and is strongly suppressed in the cold stable-glueball history. Developing a complete model of the confining sector that accounts for the origin of the dark asymmetry and the cosmological fate of dark glueballs is a promising direction for future work.}

\section*{Acknowledgements}
We are deeply indebted to Giacomo Landini for participating in the early stages of this work and for many useful discussions. We thank Bryan Zaldivar for kindly providing access to some of his numerical codes.  We are also grateful to  Yi Chung, Avirup Ghosh, Pablo Figueroa, Camilo Garcia-Cely, Benjamin Grinstein, Manoj Kaplinghat, Juan Miguel Nieves, Antonio Pich, Anthony Thomas, and Hai-Bo Yu for useful discussions. We acknowledge the use of ChatGPT and Claude in cross-checking the calculations and the writing. GG was supported by the Doctoral School ``Karlsruhe School of Elementary and Astroparticle Physics: Science and Technology (KSETA)” through the GSSP program of the German Academic Exchange Service (DAAD).

This work is partially supported by the \emph{Australian Research Council} through the ARC Centre of Excellence for Dark Matter Particle Physics (CE200100008), the Spanish \emph{Agencia Estatal de Investigación} MICINN/AEI (10.13039/501100011033) grant PID2023-148162NB-C21, the \emph{Generalitat Valenciana} through the GenT Excellence
Program (CIESGT2024-007), and the \emph{Severo Ochoa} project MCIU/AEI CEX2023-001292-S. This work has received funding from the European Union’s Horizon Europe research and
innovation programme under the Marie Skłodowska-Curie Staff Exchange grant agreement
No 101086085 – ASYMMETRY.

\appendix

\bibliographystyle{JHEP}
\bibliography{biblio.bib}	\vspace{3 cm}

\section{Color factor for the dipole operator}
\label{app:color-contraction}

We follow Ref.~\cite{Peskin:1979va} to compute the color factor entering the baryonic dipole operator. The color wave function of an $SU(N)$ baryon is the completely antisymmetric singlet state~\cite{Brambilla:2009cd},
\begin{equation}
    |\mathcal B\rangle
    =
    \frac{\epsilon_{i_1 i_2\cdots i_N}}{\sqrt{N!}}
    |i_1 i_2 \cdots i_N\rangle \,.
\end{equation}
The derivation proceeds in complete analogy with Eqs.~(3.1)--(3.2) and (5.5)--(5.6) of Ref.~\cite{Peskin:1979va}. We focus on the simplest diagrams, namely those without connected gluonic vertices and with only direct gluon insertions on the Wilson loop (see Fig.~5 of Ref.~\cite{Peskin:1979va}).

The contribution of the diagram of Fig.~5 of Ref.~\cite{Peskin:1979va} can be decomposed into configurations in which both gluons attach to the same quark line and configurations in which they attach to different quark lines.\footnote{For compactness, we denote
$A^0(\mathbf x_{q_i},t)$ simply by $A_{q_i}$ and
$A^0(\mathbf x_{q_i},t-\tau)$ by $A'_{q_i}$, where
$A^{0,a}$ is the time component of the gluonic field with color index $a$, $\mathbf x_{q_i}$ is the position of quark $i$, and  $t-\tau$ and $t$ are the interaction times of the first and second gluons, respectively.}
Denoting this amplitude by $\mathcal M_5^{(\mathcal B)}$, one finds
\begin{multline}
\mathcal M_5^{(\mathcal B)}
\propto
\frac{1}{N!}
\sum_{q=1}^N
\epsilon_{j_1\cdots i_q\cdots j_N}
[(\tau^a)_{i_qk}A_{q_q}]
[(\tau^b)_{k j_q}A'_{q_q}]
\epsilon_{j_1\cdots j_q\cdots j_N}
\\
+
\frac{1}{N!}
\sum_{\substack{p,q=1\\p\neq q}}^N
\epsilon_{j_1\cdots i_p\cdots i_q\cdots j_N}
[(\tau^a)_{i_pj_p}A_{q_p}]
[(\tau^b)_{i_qj_q}A'_{q_q}]
\epsilon_{j_1\cdots j_p\cdots j_q\cdots j_N}\,.
\label{eq:baryon_fig5}
\end{multline}
Schematically,
\begin{equation}
\mathcal M_5^{(\mathcal B)}
\propto
\frac{1}{N!}\,
\epsilon
\left(
\tau^a A_{q_1}
+\cdots+
\tau^a A_{q_N}
\right)
\left(
\tau^b A'_{q_1}
+\cdots+
\tau^b A'_{q_N}
\right)
\epsilon \,.
\end{equation}

 As an example, consider the configuration where the first gluon interacts with quark $1$ and the second with quark $N$. One obtains
\begin{multline}
\frac{1}{N!}
\epsilon_{m j_2\cdots j_{N-1} n}
[(\tau^a)_{m j_1}A_{q_1}]
[(\tau^b)_{n j_N}A'_{q_N}]
\epsilon_{j_1\cdots j_N}
\\
=
\frac{1}{N(N-1)}
\left(
{\rm tr}[\tau^a]\,
{\rm tr}[\tau^b]\,
A_{q_1}A'_{q_N}
-
{\rm tr}[\tau^a\tau^b]\,
A_{q_1}A'_{q_N}
\right)
\\
=
-\frac{\delta^{ab}}
{2N(N-1)}
A_{q_1}A'_{q_N}\,,
\end{multline}
where we used the Levi-Civita contraction $\epsilon_{m j_2\cdots j_{N-1} n} \epsilon_{j_1\cdots j_N}=(N-2)!(\delta_{m j_1}\delta_{n j_N}-\delta_{m j_N}\delta_{n j_1})$ as well as the properties of the $\SU(N)$ generators,
${\rm tr}[\tau^a]=0$ and
${\rm tr}[\tau^a\tau^b]=\delta^{ab}/2$.

More generally, whenever the two gluons are attached to the same quark line, all color indices of the Levi-Civita tensor except one are contracted, yielding
\begin{equation}
\frac{1}{N!}(N-1)!
\frac{\delta^{ab}}{2}
=
\frac{\delta^{ab}}{2N}\,.
\end{equation}
Conversely, when the two gluons are attached to different quark lines, two indices remain uncontracted and one finds
\begin{equation}
-\frac{1}{N!}(N-2)!
\frac{\delta^{ab}}{2}
=
-\frac{\delta^{ab}}
{2N(N-1)}\,,
\end{equation}
where the relative minus sign follows directly from the antisymmetry of the Levi-Civita tensor.

For comparison, the mesonic amplitude reads (see Eq.~(3.1) of Ref.~\cite{Peskin:1979va})
\begin{multline}
\mathcal M_5^{(\mathcal M)}
\propto\frac{1}{N} \left(
\delta_{i \bar j}
[(\tau^a)_{i k}A_{q}][
(\tau^b)_{k j}A'_{q}]\delta_{j \bar j}   +  \delta_{j \bar i}
[(-\tau^a)_{k \bar i}A_{\bar q}][
(-\tau^b)_{\bar j k}A'_{\bar q}]\delta_{j \bar j}    \right)                 \\+\frac{1}{N} \left(\delta_{ i \bar i}
[(-\tau^a)_{\bar j \bar i}A_{\bar q}][
(\tau^b)_{i j}A'_{q}]\delta_{j \bar j}  +\delta_{ i \bar i}
[(\tau^a)_{i j}A_{ q}][
(-\tau^b)_{\bar j \bar i}A'_{\bar q}]\delta_{j \bar j}  \right) \,,
\label{eq:meson_fig5}
\end{multline}
since antiquarks transform in the anti-fundamental representation,
\begin{equation}
(\tau^a_{\rm anti})_{ij}
=
-(\tau^a_{\rm fund})_{ji}\,.
\end{equation}
Schematically,
\begin{equation}
\mathcal M_5^{(\mathcal M)}
\propto
\frac{1}{N}
\delta
\left(
\tau^a A_q
-
\tau^a A_{\bar q}
\right)
\left(
\tau^b A'_q
-
\tau^b A'_{\bar q}
\right)
\delta \,.
\end{equation}

The relative factor of $(N-1)$ between the equal- and different-quark contractions allows the baryonic color structure to be reorganized into pairwise differences of gauge fields, exactly as in the mesonic derivation of Ref.~\cite{Peskin:1979va}. For a given quark, say $q_1$, one may rewrite its diagonal contribution as
\begin{equation}
A_{q_1}A'_{q_1}
=
\frac{1}{N-1}
\sum_{i=2}^{N}
\left(
A_{q_1}-A_{q_i}
\right)
\left(
A'_{q_1}-A'_{q_i}
\right)
+\cdots ,
\end{equation}
where the ellipsis denotes mixed terms involving quark $1$ and the remaining quarks as well as diagonal terms for the other quarks divided by $N-1$. Expanding the sum in the right-hand side reproduces the diagonal contribution $A_{q_1}A'_{q_1}$ together with off-diagonal terms such as
\begin{equation}
-\frac{1}{N-1}
A_{q_1}A'_{q_i}\,,
\qquad
-\frac{1}{N-1}
A_{q_i}A'_{q_1}\,,
\end{equation}
which occur with precisely the sign and relative normalization dictated by the color contractions.

Repeating this construction for all quarks yields
\begin{equation}
\mathcal M_5^{(\mathcal B)}
\propto\frac{1}{2N(N-1)}
\sum_{i<j}
(A_{q_i}-A_{q_j})
(A'_{q_i}-A'_{q_j})\,,
\end{equation}
which, in the dipole $a_B\rightarrow0$  and retarded $\tau \to0 $ limits,\footnote{The non-retarded regime introduces derivatives with respect to time which are fully incorporated in \cref{app:highern}, e.g. see \cref{eq:operator_definition_bound}.}  becomes
\begin{equation}
\frac{1}{2N(N-1)}
\sum_{i<j}
(A_{q_i}-A_{q_j})
(A'_{q_i}-A'_{q_j})
\;\rightarrow\;
\frac{1}{2N(N-1)}
\sum_{i<j}
(\mathbf r_{ij}\!\cdot\!\boldsymbol\nabla A)^2 \,.
\end{equation}
The interaction therefore depends only on relative coordinates and admits a dipole interpretation.

If the baryonic wave function is approximated by
\begin{equation}
\psi
\propto
\exp
\left(
-\dfrac{1}{a_B}\sum_{i<j}r_{ij}
\right)\,,
\end{equation}
as in Ref.~\cite{Mitridate:2017oky}, all quark pairs contribute equally to the matrix elements. Under this assumption, one may replace the sum over pairs by the contribution of a representative pair,
\begin{equation}
\frac{1}{2N(N-1)}
\sum_{i<j}
(\mathbf r_{ij}\!\cdot\!\boldsymbol\nabla A)^2
\rightarrow
\frac{1}{4}
(\mathbf r\!\cdot\!\boldsymbol\nabla A)^2 \,,
\end{equation}
where $\mathbf r$ denotes the separation of an arbitrary quark pair. For mesons, one instead finds
\begin{equation}
\frac{1}{2N}
(\mathbf r\!\cdot\!\boldsymbol\nabla A)^2\,,
\end{equation}
showing an additional large-$N$ suppression. Baryons do not exhibit this suppression because the number of contributing quark pairs grows as $N(N-1)/2$ exactly canceling the large $N$ suppression.

Finally, gauge invariance guarantees that the same color coefficient accompanies the remaining diagrams in the multipole expansion, ensuring that the final operator can be assembled into a gauge-invariant combination of the gauge field $A^\mu$ order by order in the $SU(N)$ coupling. More precisely, the $\boldsymbol\nabla A^a$ term becomes the chromo-electric field $\boldsymbol{E}^a$.

Therefore, under the assumption of equal contributions from all quark pairs, the color coefficient appearing in the polarizability calculation of Ref.~\cite{Peskin:1979va} is modified according to
\begin{equation}
\frac{16\pi}{3N^2}
\Bigg|_{\rm meson,\;large\;N}
\longrightarrow
\frac{8\pi}
{3C_N\times2}
\Bigg|_{\rm baryon}\,,
\end{equation}
where we have written the result in terms of the exact Casimir $C_N=(N^2-1)/(2N)$ whereas Ref.~\cite{Peskin:1979va} works in the large-$N$ limit, $C_N\rightarrow N/2$. In particular, for $\SU(2)$ the baryonic and mesonic color factors coincide, as expected since an $\SU(2)$ baryon is simply a two-body state with the same interaction strength as a meson.

\section{The London to Casimir-Polder transition}
\label{app:highern}

In this appendix we derive the transition between the retarded Casimir-Polder potential, which scales as $r^{-7}$, and the non-retarded London potential, which scales as $r^{-6}$. We show that the full transition can be accurately approximated by matching these two asymptotic regimes at an appropriate distance scale $r_{\rm ret}$, providing a simple description suitable for the self-interaction calculations of this work.

\subsection{Bound-state dominated regime}

We first consider the bound-state dominated regime, which is expected to provide an excellent approximation at large $N$. In this limit, assuming that all quark pairs contribute equally, the operator sum in \cref{eq:operator_definition} simplifies to
\begin{equation}
\label{eq:operator_definition_bound}
\sum_{p} c_p \mathcal{O}_p(t,r)
=
-
\frac{2\pi \alpha_d}{3\,\delta\epsilon_B}
\langle \phi | r^i |2P_a \rangle
\langle 2P_a |r^i | \phi \rangle
\sum_{n=2 \;({\rm even})}^{\infty}
\frac{1}{(\delta\epsilon_B)^{n-2}}
\,
{\rm tr}
\!\left[
E^i
(-iD^0)^{n-2}
E^i
\right]\,,
\end{equation}
where we have retained only the contribution from the lowest adjoint excitation
$1S_s\rightarrow2P_a$, such that
\[
H_a+\epsilon_B
\rightarrow
\epsilon_B-\epsilon_{B,a}^{(2P)}
\equiv
\delta\epsilon_B\,.
\]
The dependence on the operator order is therefore entirely encoded in the number of temporal derivatives and the corresponding powers of $(\delta\epsilon_B)^{-1}$. In our quark-diquark (two-body) approximation,
\begin{equation}
\delta
=
\frac34
+
\frac{1}{2C_N}
-
\frac{1}{4C_N^2}
\overset{N\rightarrow\infty}{\longrightarrow}
0.75\,.
\end{equation}

Passing to frequency space through a Fourier transform, we obtain~\cite{Peskin:1979va}
\begin{multline}
\int d\epsilon\,
e^{it\epsilon}
\sum_p c_p \mathcal O_p(\epsilon,r)
=
-
d_2^{(2P)}a_B^3
\int d\epsilon\,
e^{it\epsilon}
{\rm tr}
\!\left[
E^2
\right]
\sum_{n=2 \;({\rm even})}^{\infty}
\left(
\frac{i\epsilon}{\delta\epsilon_B}
\right)^{n-2}
\\
=
-
\frac{d_2^{(2P)}}{r}
a_B^3
\int d\lambda\,
e^{it\lambda/r}
{\rm tr}
\!\left[
E^2
\right]
\frac{\Lambda^2}
{\Lambda^2+\lambda^2}\,,
\end{multline}
where
\begin{equation}
d_2^{(2P)}a_B^3
=
\frac{2\pi\alpha_d}
{3\,\delta\epsilon_B}
\langle \phi | r^i |2P_a \rangle
\langle 2P_a |r^i | \phi \rangle\,,
\end{equation}
and in the second line we introduced the dimensionless variables $
\epsilon={\lambda}/{r}$ and $
\Lambda=\delta\epsilon_B r$.

Using this result, the perturbative evaluation of the vacuum expectation value of the
${\rm tr}[E^2]$ follows directly from the analysis of Ref.~\cite{Peskin:1979va}, yielding
\begin{equation}\label{eq:general_pertu_potential}
V(r)
=
-\frac{N^2-1}{64\pi^2}
\left(
d_2^{(2P)}a_B^3
\right)^2
\frac{f(\Lambda)}{r^7}\,,
\end{equation}
with
\begin{equation}
f(\Lambda)
=
\frac{4}{\pi}
\int_0^\infty
d\lambda
\left(
\frac{\Lambda^2}
{\Lambda^2+\lambda^2}
\right)^2
e^{-2\lambda}
\left(
3+6\lambda+5\lambda^2+2\lambda^3+\lambda^4
\right)\,.
\end{equation} Since we assume the validity of the multipole expansion when writing \cref{eq:operator_definition}, \cref{eq:general_pertu_potential}  is consistent for separations well outside the baryonic core and within the perturbative regime,
\[
a_B \ll r\ll \Lambda_d^{-1} \,.
\]

The asymptotic limits of the integral are straightforward to evaluate. For
$\Lambda\rightarrow\infty$, corresponding to
$r\gg(\delta\epsilon_B)^{-1}$, we obtain
\begin{equation}
V_\infty(r)
=
-(N^2-1)
\frac{23}{64\pi^3}
\left(
d_2^{(2P)}a_B^3
\right)^2
\frac{1}{r^7}\,,
\end{equation}
which precisely reproduces the Casimir-Polder potential found in the heavy-quarkonium literature~\cite{Fujii:1999xn}.  Here, we used the analytical expression for the integral
\begin{equation}
\int_0^\infty
d\lambda\,
e^{-2\lambda}\,\lambda^n\,= \frac{n!}{2^{n+1}}
\end{equation}
to obtain $f(\infty)=23/\pi$.

In the opposite limit,
$\Lambda\rightarrow0$, corresponding to
$r\ll(\delta\epsilon_B)^{-1}$, we find
\begin{equation}
V_0(r,\delta)
=
-(N^2-1)
\frac{3}{64\pi^2}
\left(
d_2^{(2P)}a_B^3
\right)^2
\frac{\delta\,\epsilon_B}{r^6}\,,
\end{equation}
which is the expected London dispersion force between two induced non-permanent dipoles when retardation effects become negligible.

The full interpolation between the two regimes is encoded in the function
$f(\Lambda)$. For practical purposes, however, it is sufficient to approximate the potential by the two asymptotic expressions matched continuously at
\begin{equation}
r_{\rm ret}
=
\frac{23}{3\pi}
\frac{1}{\delta\,\epsilon_B}\,.
\end{equation}

\begin{figure}[t]
\centering
\includegraphics[width=0.475\linewidth]{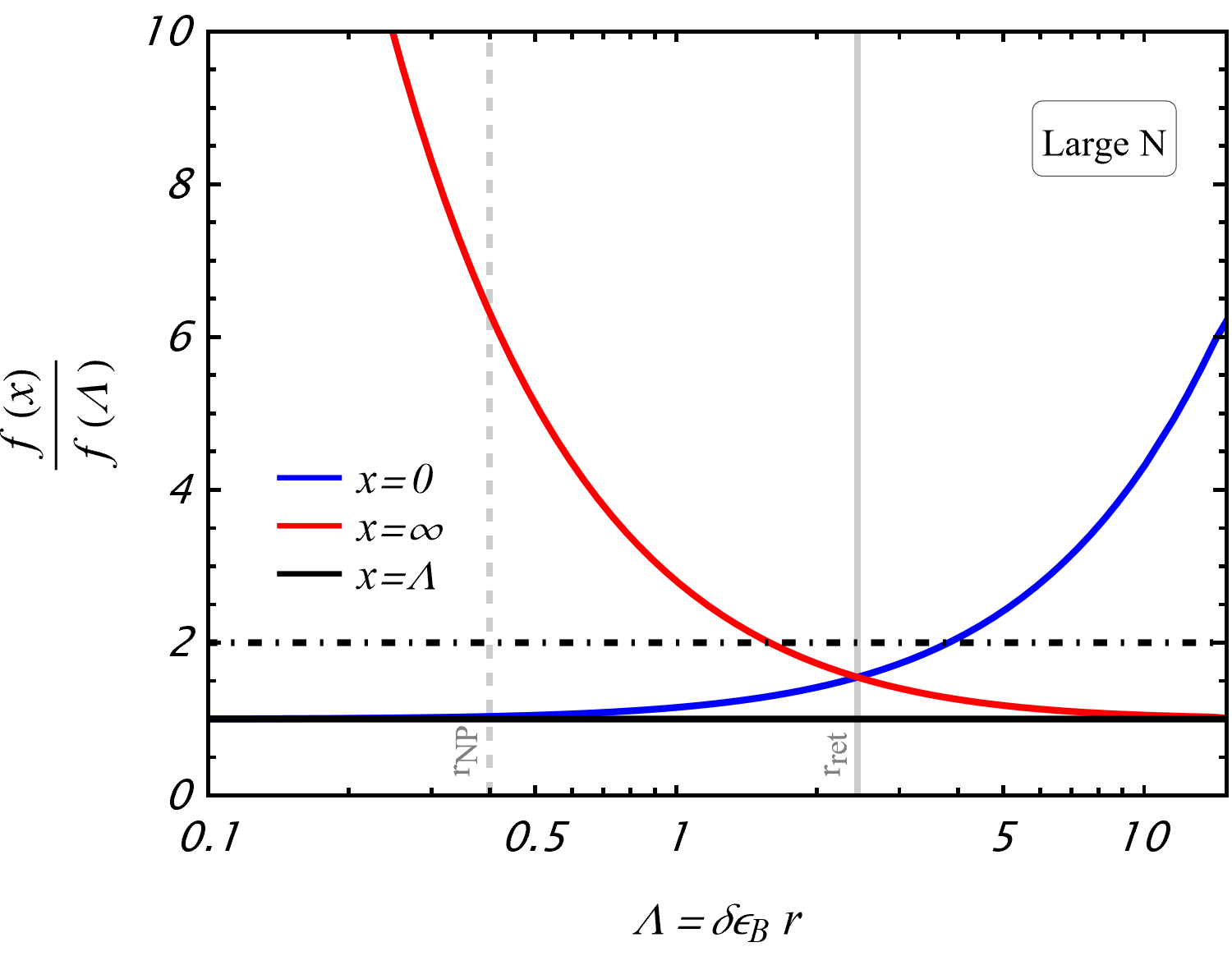} \hspace{0.1cm}
\includegraphics[width=0.475\linewidth]{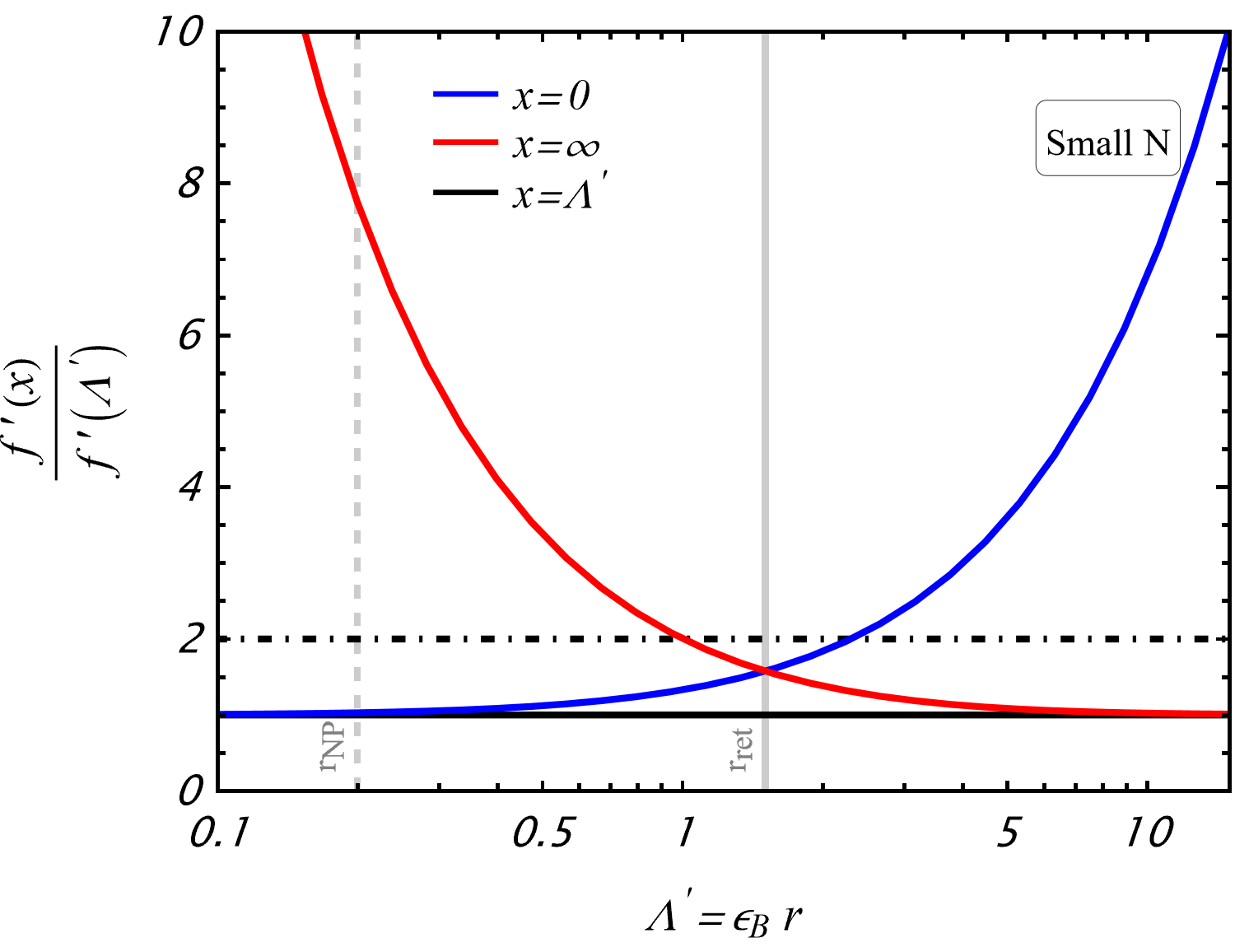}
\caption{ Transition between the London and Casimir-Polder potentials. The blue curves show the ratio of the $r^{-6}$ London dispersive approximation to the full potential, while the red curves show the ratio of the $r^{-7}$ Casimir-Polder approximation to the full potential. The black horizontal lines correspond to ratios of $1$ (solid) and $2$ (dot-dashed). Vertical gray lines stand for   the non-perturbative truncation $r_{\rm NP}$ (dashed) and for the $r^{-6}\leftrightarrow r^{-7}$ matching point $r_{\rm ret}$ (solid). The left panel corresponds to the bound-state dominated transition $1S_s\rightarrow2P_a$ (valid for large $N$), whereas the right panel shows transitions dominated by free adjoint quarks (valid for small $N$).}
\label{fig:V7V6comparison}
\end{figure}

The left panel of \cref{fig:V7V6comparison} compares this approximation with the full numerical evaluation of $f(\Lambda)$ and shows that the agreement is excellent. Neglecting the higher-order operators with $n>2$ introduces an error of one order of magnitude only for $
r\lesssim {0.3}/(\delta\epsilon_B)
$. Given our inability to reliably compute the corresponding $n>2$ non-perturbative contributions, we truncate the $n=2$ non-perturbative result at
\begin{equation}
r_{\rm NP}
=
\frac{0.3}{\delta\epsilon_B}\,,
\end{equation}
for large $N$, corresponding to a slightly optimistic estimate.

\paragraph{Including the full bound-state tower: }

The previous analysis can be straightforwardly generalized to include the complete tower of adjoint bound states $kP$, with integer $k\ge2$, using the fact that its energy difference $\delta_k$ is bounded both from below and from above as $1>\delta_k \geq \delta$.

The inclusion of all bound states leads to (using $\Lambda'\equiv\epsilon_B r$) \begin{equation}
\sum_{k=2}^\infty
d_2^{(kP)}
\sum_{n=2\;({\rm even})}^\infty
\left(
\frac{i\epsilon}
{\delta_k\epsilon_B}
\right)^{n-2}
=
\sum_{k=2}^\infty
d_2^{(kP)}
\frac{\delta_k^2\Lambda'^2}
{\delta_k^2\Lambda'^2+\lambda^2}
=
\sum_{k=2}^\infty
d_2^{(kP)}
g(\delta_k)\,,
\end{equation}
where
\begin{equation}
d_2^{(kP)}a_B^3
=
\frac{2\pi\alpha_d}
{3\,\delta_k\epsilon_B}
\langle \phi | r^i |kP_a \rangle
\langle kP_a |r^i | \phi \rangle \,.
\end{equation}

In the limit $\Lambda' \to \infty$, the function $g(x)$ converges to $g(\delta_k)\to 1$ for any $k$ such that the sum simply converges to the $V_{\infty}$ potential as before with  the substitution of $d_2^{(2P)}$  by
\begin{equation}
d_2^{(2P)}\to d_2^{(\rm bounded)}
=
\sum_{k=2}^\infty
d_2^{(kP)}\,.
\end{equation}

For finite $\Lambda'$, noting that the function $g(x)$ is monotonically increasing for $x>0$,
\begin{equation}
g(\delta)
\leq
g(\delta_k)
<
g(1)\,,
\end{equation}
one can deduce bounds
\begin{equation}
g(\delta)
d_2^{(\rm bounded)}
<
\sum_{k=2}^\infty
d_2^{(kP)}g(\delta_k)
<
g(1)
d_2^{(\rm bounded)}\,.
\end{equation}
Applying these inequalities to the potential for $\Lambda' \to 0$ yields
\begin{equation}
|V_0(r,\delta)|
<
|V_0(r)|
<
|V_0(r,1)|\,.
\end{equation}

These bounds are particularly useful in the large-$N$ regime, such as for $\SU(10)$, where the chromo-electric polarizability is strongly dominated by bound-state contributions in close analogy with the hydrogen atom. Moreover, since
$\delta\simeq3/4$ for large $N$, the interval between the two bounds is relatively narrow. In our numerical calculations we therefore adopt the midpoint approximation
\begin{equation}
V_0(r)
\simeq
V_0(r,7/8)\,.
\end{equation}

\subsection{Continuum-dominated regime}

The situation is considerably more involved when the chromo-electric polarizability is dominated by continuum states, since the operator sum cannot be resummed as straightforwardly as in the bound-state case. We therefore focus on the limit in which the continuum quarks can be approximated as free particles, which provides a good description for the small values of $N$ of interest here, particularly $N=2$ and $N=3$.

In this regime (using $d_n$ from Ref.~\cite{Peskin:1979va}; see \cref{eq:free_dn}),
\begin{equation}
\label{eq:operator_definition_free}
\sum_p c_p \mathcal O_p(t,r)
=
-
\frac{256}{C_N}
\sqrt{\pi}
a_B^3
\sum_{n=2\;({\rm even})}^{\infty}
\frac{\Gamma\!\left(n+\frac52\right)}
{\Gamma(n+5)}
\frac{1}{\epsilon_B^{n-2}}
\,
{\rm tr}
\!\left[
E^i
(-iD^0)^{n-2}
E^i
\right]\,.
\end{equation}
Proceeding as above, one obtains
\begin{equation}
V(r)
=
-\frac{N^2-1}{64\pi^2}
\left(
d_2^{(\rm free)}a_B^3
\right)^2
\frac{f'(\Lambda')}{r^7}\,,
\end{equation}
with
\begin{equation}
f'(\Lambda')
=
\frac{2^{18}3^2}
{49\pi^2}
\int_0^\infty
d\lambda\,
S(\lambda,\Lambda')^2
e^{-2\lambda}
\left(
3+6\lambda+5\lambda^2+2\lambda^3+\lambda^4
\right)\,,
\end{equation}
where
\begin{equation}
S(\lambda,\Lambda')
=
\sum_{n=2\;({\rm even})}^{\infty}
\frac{\Gamma\!\left(n+\frac52\right)}
{\Gamma(n+5)}
\left(
\frac{i\lambda}{\Lambda'}
\right)^{n-2},
\qquad
\Lambda'=r\epsilon_B \,.
\end{equation}

To evaluate $S(\lambda,\Lambda')$, it is convenient to rewrite the series in a form amenable to resummation. Introducing
$n=2m+2$, one has
\begin{equation}
(i\lambda)^{n-2}
=
(-1)^m\lambda^{2m}\,.
\end{equation}
Using the beta-function representation,
\begin{equation}
\frac{\Gamma\left(n+\frac52\right)}
{\Gamma(n+5)}
=
\frac{1}{\Gamma\left(\frac52\right)}
\int_0^1
dt\,
t^{n+\frac32}
(1-t)^{\frac32}\,,
\end{equation}
the sum becomes
\begin{align}
S(\lambda,\Lambda')
&=
\frac{1}{\Gamma\left(\frac52\right)}
\int_0^1
dt\,
t^{7/2}
(1-t)^{3/2}
\sum_{m=0}^\infty
\left(
-\frac{\lambda^2 t^2}{\Lambda'^2}
\right)^m \,.
\end{align}
The geometric series can be resummed (and analytically continued for $|\lambda t/\Lambda'|>1$),
\begin{equation}
\sum_{m=0}^{\infty}
\left(
-\frac{\lambda^2 t^2}{\Lambda'^2}
\right)^m
=
\frac{\Lambda'^2}
{\Lambda'^2+\lambda^2 t^2}\,,
\end{equation}
leading to
\begin{equation}
S(\lambda,\Lambda')
=
\frac{\Lambda'^2}
{\Gamma\left(\frac52\right)}
\int_0^1
dt\,
\frac{
t^{7/2}(1-t)^{3/2}
}
{
\Lambda'^2+\lambda^2 t^2
}\,.
\end{equation}
The large-$\Lambda'$ limit is straightforward,
\begin{equation}
S(\lambda,\infty)
=
\frac{7}{768}
\sqrt{\pi}\,,
\end{equation}
which, once substituted into $f'(\Lambda')$ and integrated over $\lambda$, reproduces
\[
f'=\frac{23}{\pi}\,,
\]
as expected from retaining only the leading $n=2$ operator. For $\Lambda'\to 0$, a numerical evaluation yields \begin{equation}
    f'(r)
\approx
4.86\,
\epsilon_B r \, ,
\end{equation}
which precisely reproduces the London scaling.

The full $V_0 \to V_\infty$ transition is numerically computed and presented in the right panel of \cref{fig:V7V6comparison}. The figure also shows that neglecting the operators with $n>2$ introduces an order-of-magnitude error only for $
r\lesssim {0.15}/{\epsilon_B}$.  Given the absence of a reliable non-perturbative treatment of these higher-order contributions, we truncate the $n=2$ non-perturbative result at
\begin{equation}
r_{\rm NP}
=
\frac{0.2}{\epsilon_B}\,,
\end{equation}
for $N=2,3$, corresponding to a slightly optimistic estimate.

\section{Chromo-electric polarizability}
\label{app:dipole-transition-integration}

The chromo-electric polarizability of heavy hadrons originates from second-order transitions induced by the chromo-electric dipole operator. Since the dipole operator carries one unit of angular momentum and changes the color representation of the quark pair, it connects the singlet $\ell$-state to intermediate color-adjoint $(\ell+\Delta \ell)$-states with $\Delta \ell =\pm 1$.
Physically, the hadron absorbs a chromo-electric gluon, propagates as an excited color-adjoint configuration, and subsequently re-emits a gluon to return to the singlet state. We will be interested only in singlet hadrons in the ground state $1S$.

For mesons, the chromo-electric polarizability is given by~\cite{Fujii:1999xn,Brambilla:2015rqa}
\begin{equation}
\alpha_{\mathcal M}
=
\frac{d_{2}^{(\mathcal M)} a_M^3}{4\pi\alpha_d}
=
\frac{1}{3N}
\left\langle
\phi
\left|
r^i
\frac{1}{H_a+\epsilon}
r^i
\right|
\phi
\right\rangle \,,
\end{equation}
where $H_a$ denotes the effective Hamiltonian governing the intermediate adjoint state and $\phi$ is the singlet ground-state wave function, which can be approximated by a Coulombic wave function in the heavy-quark limit.

In \cref{app:color-contraction} we derived the corresponding expression for baryons,
\begin{equation}
\alpha_{\mathcal B}
=
\frac{d_2 a_B^3}{4\pi\alpha_d}
=
\frac{1}{3N(N-1)}
\sum_{i<j}^N
\left\langle
\phi
\left|
r_{ij}^k
\frac{1}{H_a+\epsilon_B}
r_{ij}^k
\right|
\phi
\right\rangle
=
\frac16
\left\langle
\phi
\left|
r^k
\frac{1}{H_a+\epsilon_B}
r^k
\right|
\phi
\right\rangle \,,
\end{equation}
where in the last step we assumed that all quark pairs contribute equally, such that $r$ may be interpreted as the separation of an arbitrary quark pair. This assumption is expected for a spherically symmetric ground state and is realized explicitly for the large-$N$ Hartree wave functions of Ref.~\cite{Albertus:2015fca}, trivially for $N=2$, and for the $N=3,4,5,6$ approximate wave functions of Ref.~\cite{Mitridate:2017oky}.

At first sight one might expect that the baryonic polarizability can be obtained from the mesonic result simply by replacing $a_M\rightarrow a_B$. However, the intermediate Hamiltonian $H_a$ is qualitatively different in the two systems.\footnote{Strictly speaking, several adjoint configurations exist for baryons, implying that $H_a$ is a matrix in the space of adjoint states. In this work we approximate $H_a$ by an average effective Hamiltonian as described in the main text---see \cref{sec:2p1p1_baryon_properties}. A full matrix treatment for $\SU(3)$ was performed in Ref.~\cite{Dong:2022rwr}.} In particular, the baryonic system can contain intermediate adjoint bound states, which have no direct analogue in the mesonic calculation and give rise to important modifications of the chromo-electric polarizability.

Throughout this appendix we model the baryon as an effective quark-``diquark'' two-body system in order to obtain analytical expressions. As discussed in the main text, this approximation reproduces the baryonic parameters $a_B$ and $\epsilon_B$ within at most a factor of a few for $N=2,3,4$ as well as for large $N$. For $\SU(3)$ we additionally employ exact results available in the literature, while for large $N$ we make use of the moments of the baryon density distribution  derived in Ref.~\cite{Cohen:2011cw}.

To evaluate the polarizability, one can decompose $\alpha_{\mathcal{B}}$ into two parts by inserting the completeness relation~\cite{Fano:1968idi},
\begin{equation}
1=
\sum_\nu |\nu\rangle\langle \nu|
+
\int dk\, |k\rangle\langle k|\,,
\end{equation}
which includes both discrete adjoint bound states $|\nu\rangle$ and the continuum spectrum of unbound adjoint states $|k\rangle$. Since the initial baryon is in the $1S$ ground state, only intermediate states with orbital angular momentum $\ell=1$ contribute.

If the adjoint and singlet potentials were identical, one would recover the standard hydrogenic problem (adopting our two-body approximation), in which the $2P$ state provides the dominant contribution and the continuum accounts for only about $20\%$ of the total $1S$ electric polarizability~\cite{Mitroy:2010qkk}. The differences between the singlet and adjoint potentials in the present case, however, can substantially modify this picture.

To make the analogy with the electric polarizability of hydrogen explicit, inserting the completeness relation into the mesonic expression yields
\begin{equation}
\alpha_{\mathcal M}
=
\frac{1}{N}
\sum_{k'}
\frac{
\langle\phi|z|k'_a\rangle
\langle k'_a|z|\phi\rangle
}
{E_{k',a}+\epsilon_M}\,,
\end{equation}
 where spherical symmetry was used to replace
$r^i r^i \rightarrow 3z^2$ and the notation $k'$ collectively denotes both discrete and continuum states.  Additionally, the meson binding energy $\epsilon_M$ is equivalent to $\epsilon_B$ in our ``quark-diquark'' approximation after the substitution $\mu\to m_Q/2$.

For comparison, the electric dipole polarizability governing the quadratic Stark effect~\cite{https://doi.org/10.1002/andp.19143480702} in the hydrogen atom (with Bohr  radius $a_0$) is~\cite{fitzpatrick_stark_2016}
\begin{equation}
\alpha_P
=
8\pi\alpha_{\rm em}
\sum_{k'}
\frac{
\langle1S|z|k'\rangle
\langle k'|z|1S\rangle
}
{E_{k'}-E_{1S}}
=
18\pi a_0^3 \,,
\end{equation}
for the Coulomb potential
\begin{equation}
V(r)=-\frac{\alpha_{\rm em}}{r}\,.
\end{equation}
In the following we reproduce this result in the appropriate limit and generalize it to the case in which the intermediate adjoint states experience a potential different from that of the initial singlet configuration.

Before presenting the full analytical computation, it is useful to separately study the contributions from intermediate adjoint bound states and from the continuum of unbound states. This decomposition is physically important because the approximations adopted throughout this work rely on the relative size of these two contributions. In particular, for $N=2$ and $N=3$ we will employ the free-adjoint-only approximation, which is justified only if the continuum dominates the chromo-electric polarizability and is well described by free-adjoint quarks, whereas for large $N$ we will approximate the polarizability by its bound-state contribution, assuming that intermediate adjoint bound states provide the leading effect. The analytical method developed later in this appendix exploits closure and directly computes the full spectral sum, but it does not permit a decomposition into separate bound and unbound contributions. Therefore, estimating these two sectors independently remains essential for assessing the validity of the approximations used in our work.

\subsection{Bound-state contributions}

The contribution from the continuum of unbound quarks has already been computed analytically for mesons under the approximation that the intermediate adjoint quarks are free~\cite{Peskin:1979va}---and semi-analytically including the repulsive adjoint potential via the Coulombic wave functions in the continuum~\cite{Brambilla:2015rqa}. We revisit the free results in the next section, including perturbative corrections arising from weak adjoint potentials. Here we instead focus on transitions between singlet baryons and intermediate adjoint bound states.

The leading bound state contribution is expected to arise from the  excitation,
\begin{equation}
1S_{\rm singlet}\rightarrow \nu P_{\rm adjoint}\,,
\end{equation}
which possesses the smallest energy gap,
\begin{equation}
\Delta E_\nu
=
\epsilon_B
\left(
1-\frac{1}{(\nu\kappa)^2}
\right)\,,
\end{equation}
with
\begin{equation} \label{eq:kappa_definition}
\kappa\equiv \frac{C_N}{C_N-1} \,,
\end{equation}
which follows from our quark-diquark approximation, and $\nu=2$ corresponds to the lowest possible transition.

We therefore begin by evaluating explicitly the lowest
 transition, $1S\rightarrow2P$. Besides providing the dominant bound state contribution, this calculation illustrates the general structure of the dipole matrix elements that enter the chromo-electric polarizability. We then extend the result to arbitrary $1S\rightarrow \nu P$ transitions and finally numerically sum over the full tower of adjoint bound states.

 \subsubsection{Lightest bound-state transition: $1S\to2P$}

We consider an effective two-body system interacting through the Coulomb potential
\begin{equation}
V(r)
=
-\kappa_\pm\frac{\alpha_d}{r}\,,
\end{equation}
where
\begin{equation}
\kappa_\pm
=
C_N-\frac{1\pm1}{2}\,,
\end{equation}
with the plus (minus) sign corresponding to the color-adjoint (singlet) state. The associated Bohr radius is
\begin{equation}
a_{s/a}
=
\frac{1}{\kappa_\mp \mu\alpha_d}\,,
\end{equation}
where $\mu$ denotes the reduced mass. The normalized bound-state wave functions factorize as
\begin{equation}
\psi_{\nu \ell m}(r,\theta,\phi)
=
R_{\nu \ell}(r)Y_{\ell m}(\theta,\phi)\,.
\end{equation}
For the states relevant to the $1S\rightarrow2P$ transition one has
\begin{align}
\psi_{100}(r)
&=
\frac{1}{\sqrt{\pi a_0^3}}
e^{-r/a_0}\,,
\\
\psi_{21m}(r,\theta,\phi)
&=
\frac{1}{2\sqrt{6a_0^3}}
\frac{r}{a_0}
e^{-r/(2a_0)}
Y_{1m}(\theta,\phi)\,,
\end{align}
with $a_0$ standing for the Bohr radius of the corresponding baryon, $m=-1,0,1$ and
\begin{align}
Y_{1,-1}
&=
\sqrt{\frac{3}{8\pi}}
\sin\theta\,e^{-i\phi}\,,
&
Y_{1,0}
&=
\sqrt{\frac{3}{4\pi}}
\cos\theta\,,
&
Y_{1,1}
&=
-\sqrt{\frac{3}{8\pi}}
\sin\theta\,e^{i\phi}\,.
\end{align}

Since the dipole operator transforms as a vector under rotations, it can be decomposed in terms of the $Y_{1m}$ harmonics. Consequently, the orthogonality relation
\begin{equation}
\int d\Omega\,
Y_{\ell m}^*
Y_{\ell' m'}
=
\delta_{\ell\ell'}
\delta_{mm'}
\label{eq:orthogo_harmonics}
\end{equation}
immediately implies the selection rule
\begin{equation}
\Delta\ell=1\,,
\end{equation}
so that an initial $1S$ state couples only to $P$-wave intermediate states.

\subsection*{Matrix element}

The relevant dipole matrix element is
\begin{equation}
\langle
1S_s
|
r^k
|
\nu P_a,m
\rangle
=
\int d^3r\,
\psi_{100,s}^*(\mathbf r)\,
r^k\,
\psi_{\nu 1m,a}(\mathbf r)\,.
\end{equation}
Writing
\begin{equation}
r^k=r\hat r^k\,,
\end{equation}
with
\begin{equation}
\hat r^k
=
(\sin\theta\cos\phi,
\sin\theta\sin\phi,
\cos\theta)
=
\sqrt{\frac{2\pi}{3}}
\left(
Y_{1,-1}-Y_{1,1},
i(Y_{1,-1}+Y_{1,1}),
\sqrt2\,Y_{1,0}
\right)\,,
\end{equation}
the matrix element factorizes into radial and angular pieces,
\begin{equation}
\langle
1S
|
r^k
|
2P,m
\rangle
=
\mathcal R\,
\int d\Omega\,
\hat r^k
Y_{1m}\,,
\end{equation}
where (noting that $a_a/a_s=\kappa$)
\begin{equation}
\mathcal R
=
\int_0^\infty
dr\,
r^3
\psi_{100,s}(r)
R_{21,a}(r)
=
64
\sqrt{\frac{6}{\pi}}
a_s
\frac{\kappa^{5/2}}
{(2\kappa+1)^5}\,.
\end{equation}

\subsection*{Angular structure}

Defining
\begin{equation}
\mathcal A_m^k
=
\sqrt{\frac{3}{2\pi}}
\int d\Omega\,
\hat r^k
Y_{1m}\,,
\end{equation}
the angular integrals can be computed directly using Eq.~(\ref{eq:orthogo_harmonics}), yielding the values reported in \cref{tab:angular_results}.

\begin{table}[h]
    \centering
    \begin{tabular}{|c|c|c| c |c|c|c| c |c|c|c|}
         $k$& $m$ & $\mathcal{A}^k_m$ & & $k$& $m$ & $\mathcal{A}^k_m$ & & $k$& $m$ & $\mathcal{A}^k_m$\\
         1&$-1$&1& &2& $-1$ & $-i$ & & 3& $-1$ & 0\\
         1& 0 & 0 & & 2& 0 & 0 & &  3& 0 & $\sqrt{2} $\\
          1& 1 & $-1$ & & 2& 1 & $-i$ & &  3& 1 & 0
    \end{tabular}
    \caption{Results for the angular integration.  }
    \label{tab:angular_results}
\end{table}

Since the dipole operator appears quadratically in the polarizability, one first sums over the spatial index,
\begin{equation}
\langle
1S_s
|
r^k
|
2P_a,m
\rangle
\langle
2P_a,m
|
r^k
|
1S_s
\rangle
=
\frac{2\pi}{3}
\mathcal R^2
\sum_k
\mathcal A_m^k
\mathcal A_m^{k*}
=
\frac{4\pi}{3}
\mathcal R^2 \,,
\end{equation}
which is independent of the quantum number $m$. Summing over the three degenerate $m$ states therefore gives
\begin{equation}
\langle
1S_s
|
r^k
|
2P_a
\rangle
\langle
2P_a
|
r^k
|
1S_s
\rangle
=
\sum_{m,k}
\left|
\langle
1S
|
r^k
|
2P,m
\rangle
\right|^2
=
4\pi\mathcal R^2 \,.
\end{equation}
For the full $1S\rightarrow2P$ transition, one thus obtains
\begin{equation}
\langle
1S_s
|
r^k
|
2P_a
\rangle
\langle
2P_a
|
r^k
|
1S_s
\rangle
=
98304\,a_s^2
\frac{\kappa^5}
{(2\kappa+1)^{10}} \,.
\end{equation}In the large-$N$ limit $\kappa\rightarrow1$ gives
\begin{equation}
\langle1S|r^k|2P\rangle^2
\simeq
1.665\,a_s^2\,.
\end{equation}
For small values of $N$ the transition is significantly suppressed:
\begin{align}
N=3 &: \qquad 0.029\,a_s^2,
\\
N=4 &: \qquad 0.261\,a_s^2,
\\
N=5 &: \qquad 0.502\,a_s^2.
\end{align}
This suppression originates primarily from the normalization of the radial wave functions, which disfavors the larger adjoint bound states encountered at small $N$. For $N=2$, the present treatment breaks down since the interaction between quarks in the adjoint channel becomes repulsive and no adjoint bound state exists.

\subsubsection{General bound-state results}

The previous calculation can be generalized to arbitrary $1S\rightarrow \nu P$ transitions. The hydrogenic radial wave function is
\begin{equation}
\tilde R_{\nu \ell}(r)
=
\mathcal N_{\nu \ell}
(2r_\nu)^\ell
e^{-r_\nu}
L_{\nu-\ell-1}^{(2\ell+1)}(2r_\nu)\,,
\end{equation}
where
\begin{equation}
r_\nu
=
\frac{r}{\nu a_0}\,,
\end{equation}
and
\begin{equation}
\mathcal N_{\nu \ell}
=
\frac{2}{\nu^2}
\sqrt{
\frac{(\nu-\ell-1)!}
{[(\nu+\ell)!]^3a_0^3}
}\,.
\end{equation}
The generalized Laguerre polynomials are
\begin{equation}
L_s^{(r)}(x)
=
\sum_{q=0}^{s}
(-1)^q
\frac{[(s+r)!]^2}
{(s-q)!(r+q)!}
\frac{x^q}{q!}\,.
\end{equation}
The dipole matrix element for a transition into an arbitrary $\nu P$ bound state becomes
\begin{multline}
\sum_{m,k}
\left|
\langle
1S
|
r^k
|
\nu P,m
\rangle
\right|^2
=
4\pi
\left(
\int_0^\infty
dr\,
r^3
\psi_{100,s}
R_{\nu 1,a}
\right)^2
\\
=
256\,a_B^2
\nu^7
\kappa^5
(2\kappa-1)^2
(\nu^2-1)
\frac{(\nu\kappa-1)^{2\nu-6}}
{(1+\nu\kappa)^{2\nu+6}}\,,
\end{multline}
which reproduces the previous expression for $\nu=2$. Since all intermediate states have $\ell=1$, the angular analysis remains unchanged.

The bound-state contribution to the chromo-electric coefficient $d_2$ is therefore
\begin{equation}
d_2^{(\rm bound)}
=
\frac{1024\pi}{3C_N}
\sum_{\nu>1}
\nu^9
\kappa^7
(2\kappa-1)^2
(\nu^2-1)
\frac{(\nu\kappa-1)^{2\nu-7}}
{(1+\nu\kappa)^{2\nu+7}} \,.
\end{equation}

More generally,
\begin{multline}
d_n^{(\rm bound)}
=
\frac{4\pi}
{3a_B^2C_N}
\sum_{\nu>1}
\frac{
\langle1S_s|r^k|\nu P_a\rangle
\langle \nu P_a|r^k|1S_s\rangle
}
{(\Delta E_\nu/\epsilon_B)^{n-1}}
\\
=
\frac{1024\pi}{3C_N}
\sum_{\nu>1}
\nu^{5+2n}
\kappa^{3+2n}
(2\kappa-1)^2
(\nu^2-1)
\frac{(\nu\kappa-1)^{2\nu-n-5}}
{(1+\nu\kappa)^{2\nu+n+5}} \,.
\end{multline}

Since $\kappa>1$, the lowest $2P$ state becomes increasingly dominant as $n$ grows. For $N=3$, including all excited bound states increases $d_n$ by approximately $71\%$ for $n=2$, by $58\%$ for $n=20$, and by less than $9\%$ for $n=200$. For $n\gtrsim421$, the correction is below $1\%$. Consequently, the asymptotic large-$n$ tail of the operator expansion is accurately described by the $2P$ contribution alone.

 \subsection{Continuum contribution}

The contribution from the continuum of adjoint states can, in principle, be computed using the exact Coulomb scattering wave functions. However, the resulting matrix elements involve non-trivial integrals that generally have to be evaluated numerically, as done in Ref.~\cite{Brambilla:2015rqa}. Since our goal is to obtain analytical expressions for the coefficients $d_n$, we instead follow Ref.~\cite{Peskin:1979va} and approximate the intermediate adjoint quarks as free particles.

This approximation is expected to be reliable whenever the adjoint potential is sufficiently weak, namely for small values of $N$. In particular, it provides a good description of the cases of primary interest in this work, $N=2$ and $N=3$, for which the continuum contribution constitutes the dominant component of the chromo-electric polarizability. In this limit one finds~\cite{Peskin:1979va}
\begin{equation}\label{eq:free_dn}
    d_n^{\rm free}
    =
    \dfrac{256}{C_N}\sqrt{\pi }
    \frac{\Gamma \left(n+\frac{5}{2}\right)}
         {\Gamma (n+5)}\,.
\end{equation}

For $n=2$, one can systematically include the effect of a weak adjoint potential by following an approach analogous to Ref.~\cite{Peskin:1979va}. More specifically, one works in momentum space using the free wave functions and expands the Green's function as a series of potential insertions between free propagators. This procedure yields
\begin{equation}
    d_{2}^{\rm pert}
    =
    d_{2}^{\rm free}
    \left(
    1-0.238\kappa_{\rm eff}
    +0.057 \kappa_{\rm eff}^2
    -\mathcal{O}(\kappa_{\rm eff}^3)
    \right)
    \approx
    \frac{d_{2}^{\rm free}}
         {1+0.238\kappa_{\rm eff}}\,,
\end{equation}
where $    \kappa_{\rm eff}=-(C_N-1)$ for baryons. Since $|\kappa_{\rm eff}|$ grows with $N$, the weak adjoint potential approximation rapidly deteriorates and should only be regarded as quantitatively reliable for $N\lesssim5$.

\subsection{Analytical evaluation of the full spectral sum}

The separate analyses above provide information on the relative importance of intermediate bound and unbound states and justify the approximations adopted for different values of $N$. We now turn to an analytical method that computes the complete chromo-electric polarizability by summing over the entire adjoint spectrum through closure---following closely the method introduced by Dalgarno and Lewis in Ref.~\cite{10.1098/rspa.1955.0246}. The advantage of this approach is that it yields a compact expression for the total $d_2$ without requiring the explicit construction of all intermediate states. Its limitation, however, is that the resulting expression contains only the full spectral sum and does not allow one to disentangle the individual contributions from bound and continuum states. Moreover, it is not easily generalized for $d_n$ with $n>2$. Therefore, the separate computations of the previous subsections are indispensable for interpreting the physical origin of the final result and for assessing the validity of the approximations used in the current work.

The singlet and adjoint wavefunctions satisfy
\begin{align}
    \frac{1}{2\mu}\nabla^2\psi_s +(E_s-V_s)\psi_s&=0\,,
    \\
    \frac{1}{2\mu}\nabla^2\psi_a +(E_a-V_a)\psi_a&=0 \,.
\end{align}
Combining the two equations, one obtains
\begin{equation}
(E_s-E_a)
\int d^3x\,f(\mathbf x)\psi_a^\ast\psi_s
=
\frac{1}{2\mu}
\int d^3x\,f(\mathbf x)
\left(
\psi_s\nabla^2\psi_a^\ast
-
\psi_a^\ast\nabla^2\psi_s
+
2\mu\delta V\psi_a^\ast\psi_s
\right)\,,
\end{equation}
where $\delta V\equiv V_s-V_a$ and $f$ is an arbitrary function.

After integrating by parts and neglecting the boundary terms,
\begin{equation}
(E_s-E_a)\langle a|f|s\rangle
=
\frac{1}{2\mu}
\int d^3x\,
\psi_a^\ast\psi_s
\left(
\nabla^2f
+
2\nabla f\cdot\nabla\ln\psi_s
+
2\mu\delta V\,f
\right)\,.
\end{equation}
If a vector function $f^k$ satisfies
\begin{equation}\label{eq:dalgarno_eq}
\nabla^2f^k
+
2\nabla f^k\cdot\nabla(\ln\psi_s)
+
2\mu\delta V\,f^k
=
r^k\,,
\end{equation}
then, using closure and the fact that for the ground state $|s\rangle$ we have
\(
\langle s|r^k|s\rangle=0
\),
the full dipole polarizability becomes\footnote{We note once more that our $V_a$ denotes the averaged adjoint potential. The sum over intermediate adjoint states therefore amounts to an overall multiplicity factor $(N-1)$ relative to a single adjoint channel. At the same time, the total binding energy scales as $E_{a/s}\simeq N\epsilon_{B,a/s}$. Thus, when forming the ratio of the intermediate-state contribution to the corresponding energy difference, these overall factors of $N$ cancel. This justifies treating the intermediate state as a single average adjoint channel and energies as  $E_{a/s}\simeq \epsilon_{B,a/s}$.}
\begin{equation}
d_2=
\frac{4\pi\alpha_d}{a_B^3}
\frac16
\sum_{a}
\frac{
\langle s|r^k|a\rangle
\langle a|r^k|s\rangle
}
{E_{a}-E_{s}}
=
-\frac{4\pi}{3a_B^4}
\frac{1}{C_N}
\langle s|f^kr^k|s\rangle \,.
\end{equation}

In the limit $\delta V=0$, corresponding to identical singlet and intermediate potentials, one should recover the hydrogen result,
\begin{equation}
C_Nd_2=\frac{\alpha_P}{4a_0^3}
=\frac{9\pi}{2}\,,
\end{equation}
providing a useful consistency check.

\subsubsection{Solving for $f^k$}

For the Coulomb potentials relevant to our problem,
\(
\delta V=-\alpha_d/r
\),
and using the $1S$ wavefunction,
\cref{eq:dalgarno_eq}
becomes
\begin{equation}
\nabla^2f^k
-\frac{2}{a_B}
(\hat r\!\cdot\!\nabla)f^k
-\frac{2}{C_N}\frac{1}{a_Br}f^k
=
r^k \,.
\end{equation}
Rotational invariance implies
\begin{equation}
f^k(\mathbf r)=\hat r^kF(r)\,,
\qquad
\hat r^k=\frac{r^k}{r}\,,
\end{equation}
which reduces the problem to
\begin{equation}
F''
+
\left(
\frac{2}{r}
-\frac{2}{a_B}
\right)F'
-
\frac{2}{r^2}F
-
\frac{2}{C_N}\frac{1}{a_Br}F
=
r \,.
\end{equation}
Introducing
\begin{equation}
x=\frac{r}{a_B}\,,
\qquad
G(x)=\frac{F(r)}{a_B^3}\,,
\end{equation}
one finds
\begin{equation}\label{eq:G_equation}
G''
+
\left(
\frac{2}{x}-2
\right)G'
-
\frac{2}{x^2}G
-
\frac{2\epsilon_N}{x}G
=
x \,,
\end{equation}
where
\begin{equation}
\epsilon_N\equiv \frac{1}{C_N}\,.
\end{equation}

For large $N$, $\epsilon_N\sim1/N$ and the last term becomes negligible, reducing the problem to the standard hydrogen atom.

\paragraph{Hydrogen limit (\(\epsilon_N=0\)).}

Equation~\eqref{eq:G_equation} admits the polynomial solution
\begin{equation}
G(x)
=
-\frac14x^2
-\frac12x\,,
\end{equation}
which implies
\begin{equation}
f^k(\mathbf r)
=
-\left(
\frac{a_B}{4}r
+
\frac{a_B^2}{2}
\right)r^k \,.
\end{equation}
Consequently,
\begin{equation}
\langle s|f^kr^k|s\rangle
=
-\frac{27}{8}a_B^4\,,
\end{equation}
and therefore
\begin{equation}
C_Nd_2=\frac{9\pi}{2}\,,
\end{equation}
reproducing exactly the well-known hydrogen polarizability.

\paragraph{General case (\(\epsilon_N\neq0\)).}

Rather than solving \cref{eq:G_equation} explicitly, it is sufficient to introduce the moments
\begin{equation}
M_p\equiv
\int_0^\infty dx\,
x^p e^{-2x}G(x)\,,
\end{equation}
for which
\begin{equation}
\langle s|f^kr^k|s\rangle
=
4a_B^4M_3 \,.
\end{equation}

Multiplying \cref{eq:G_equation} by
\(x^q e^{-2x}\),
integrating over
\(x\in[0,\infty)\),
and repeatedly integrating by parts generates recursion relations among the moments. The first two moments satisfy
\begin{align}
&M_2=
-\frac{3}{8(1+\epsilon_N)}\,,
\\
&4M_2-2(2+\epsilon_N)M_3
=
\frac{15}{8}\,,
\end{align}
which yield
\begin{equation}
M_3
=
-\frac{3(9+5\epsilon_N)}
{16(1+\epsilon_N)(2+\epsilon_N)} \,.
\end{equation}

Finally,
\begin{equation}
\langle s|f^kr^k|s\rangle
=
-\frac{3(9+5\epsilon_N)}
{4(1+\epsilon_N)(2+\epsilon_N)}
a_B^4 \,,
\end{equation} which leads to the $d_2$ result reported in the main text.\footnote{The closure method can, in principle, be generalized to higher moments $d_n$. Following Ref.~\cite{10.1111/j.1365-2966.2005.08859.x}, one may define
\(
f(\lambda)=e^{-\lambda H_a}F e^{\lambda H_s}
\)
such that
\(
\langle a|d^n f/d\lambda^n|s\rangle|_{\lambda=0}
=(E_s-E_{a})^n\langle a|F|s\rangle
\).
 This relation suggests a possible recursive construction of higher-order coefficients $d_n$. In practice, however, the resulting hierarchy of differential equations rapidly becomes cumbersome, and we therefore restrict ourselves to  the exact evaluation of $d_2$ only.}

For $\epsilon_N=0$, the hydrogen result
\(
|\langle s|f^kr^k|s\rangle|
=
27a_B^4/8
\)
is recovered. Moreover, for $\epsilon_N=1$, corresponding to free intermediate adjoint baryons, one finds
\begin{equation}
C_Nd_2=\frac{7\pi}{3}\,,
\end{equation}
which precisely reproduces the free-adjoint result obtained by neglecting the adjoint potential. The negative sign for our \(
\langle s|f^kr^k|s\rangle
\)  results originates solely from our convention of defining the binding energy as a positive quantity.

\subsubsection{Summary of contributions}

Comparing the full $d_2$ computed in previous section, we find that the free-adjoint approximation already captures most of the chromo-electric polarizability for small $N$. For $N=2$, we have
\begin{equation}
    C_Nd_2 \approx6.3\,,\qquad
    C_Nd_{2}^{\rm free} \approx7.3\,,\qquad
    C_Nd_{2}^{\rm pert}\approx 6.9\,,
\end{equation}
while for $N=3$,
\begin{equation}
    C_Nd_2 \approx 8.3\,,\qquad
    C_Nd_{2}^{\rm free} \approx7.3\,,\qquad
    C_Nd_{2}^{\rm pert}\approx 8.0\,,\qquad
    C_Nd_{2}^{\rm free}+C_Nd_{2}^{\rm bound}\approx 7.5\,.
\end{equation}
The agreement is therefore remarkably good for the cases of primary interest in this work and justifies our use of the free-adjoint approximation whenever an analytical treatment of the continuum contribution is required.

However, for larger values of $N$, corrections to the free approximation and the contribution of intermediate bound states become increasingly important. For example, for $N=4$ one finds
\begin{align}
    C_Nd_2 \approx 9.4\,,&&
    C_Nd_{2}^{\rm free} \approx7.3\,,&&
    C_Nd_{2}^{\rm pert}\approx 9.3\,,&&
    C_Nd_{2}^{\rm free}+C_Nd_{2}^{\rm bound}\approx 9.1\,.
\end{align}
Note that, at large $n$, the continuum contribution (in the free approximation) decreases rapidly with increasing $n$, $d_{n}^{\rm free}
    \propto
     {\Gamma(n+5/2)/\Gamma(n+5)}
    \sim n^{-5/2}$ and becomes completely negligible compared with the bound-states contribution for $n\gtrsim50$.

Finally, the full spectral sum presented here allows us to demonstrate that the free-adjoint approximation provides an excellent description of the continuum contribution for $N=2$ and $N=3$, whereas, for large $N$, the chromoelectric polarizability is dominated by intermediate adjoint bound states---in direct analogy with the hydrogen atom. This behaviour also justifies our bound-state approximation adopted for $\SU(10)$.

\subsection{Comparison with previous results}

In this subsection we compare our analytical expressions with previous results available in the literature in order to assess the accuracy of our approximations. We first consider mesons, for which results are available for generic $N$, and subsequently discuss baryons, where only numerical $\SU(3)$ and large-$N$ calculations exist.

For mesons, treating the intermediate adjoint quarks as free particles gives~\cite{Peskin:1979va}
\begin{equation}
    \alpha_{\mathcal{M}}^{(\rm free)}
    =
    \frac{7}{6}\frac{1}{N C_N}
    \frac{a_M^3}{\alpha_d}\,,
\end{equation}
while including the repulsive adjoint potential yields the semi-analytical expression~\cite{Brambilla:2015rqa}
\begin{equation}
    \alpha_{\mathcal{M}}^{(\rm semi\mbox{-}analytic)}
    =
    \frac{1024}{7}
    I(\rho)\,
    \rho(\rho+2)^2\,
    \alpha_{\mathcal{M}}^{(\rm free)}\,,
\end{equation}
where $\rho=(N^2-1)^{-1}$ and
\begin{equation}
I(\rho)=
\int_{0}^{\infty} dp\,
p^{3}
\frac{\left(1+\dfrac{\rho^{2}}{p^{2}}\right)
\exp\!\left(\dfrac{4\rho}{p}\arctan p\right)}
{\left(\exp\!\left(\dfrac{2\pi\rho}{p}\right)-1\right)
(1+p^{2})^{7}} \, .
\end{equation}

For large $N$, the integral simplifies to
\begin{equation}
I
\xrightarrow[N\to\infty]{}
\frac{N^2}{2\pi}
\int_{0}^{\infty}dp\,
\frac{p^4}{(1+p^2)^7}
+\mathcal O(N^0)
=
\frac{7}{4096}N^2
+\mathcal O(N^0)\,,
\end{equation}
which reproduces the free result in the large-$N$ limit.

Our analytical expression for mesons, which can be obtained by setting
$\epsilon_N=N/(2C_N)$, is
\begin{equation}
\alpha_{\mathcal M}^{(\rm analytic)}
=\left(1+\frac{1}{14-21 N^2}+\frac{6}{7-14 N^2} \right) \alpha_{\mathcal M}^{(\rm free)}\,.
\end{equation}

For $N=3$, our result exactly reproduces the numerical value
$I=0.01143$ reported in Ref.~\cite{Brambilla:2015rqa}. We have also evaluated the integral numerically for $N=5$ and $N=10$ and again find perfect agreement with our analytical expression. Furthermore, both approaches converge to the free-particle result for $N\rightarrow\infty$. As expected, our method is therefore exact for two-body systems. To our knowledge, this constitutes the {\it first fully analytical expression for the chromo-electric polarizability of mesons for arbitrary $N$}, reproducing exactly the semi-analytical results of Ref.~\cite{Brambilla:2015rqa}.

We now turn to baryons, for which numerical results are available for $\SU(3)$~\cite{Dong:2022rwr} and semi-analytical large-$N$ estimates have been obtained in Ref.~\cite{Cohen:2011cw}. For $\SU(3)$, using $\epsilon_N=1/C_N$, our analytical result is
\begin{equation}
\alpha_{\mathcal B}^{(\rm analytic)}(N=3)
=
\frac{1}{3}
\frac{1}{C_N\alpha_d}
\frac{3(9+5\epsilon_N)}
{4(1+\epsilon_N)(2+\epsilon_N)}
a_B^3
\approx
\frac{0.707}
{\alpha_d^4m_Q^3}\,,
\end{equation}
while for large $N$ we obtain
\begin{equation}
\alpha_{\mathcal B}^{(\rm analytic)}(N\gg1)
\approx
\frac{18}{N^4}
\frac{1}{\alpha_d^4m_Q^3}\,.
\end{equation}
Ref.~\cite{Dong:2022rwr} instead finds for $\SU(3)$
\begin{equation}
\alpha_{\mathcal B}(N=3)
=
\frac{2.4}
{\alpha_d^4m_Q^3}\,,
\end{equation}
which differs from our estimate by approximately a factor of three. In the main text we therefore adopt the more precise $\SU(3)$ results of Ref.~\cite{Dong:2022rwr}, together with the baryon wave-function properties computed in Ref.~\cite{Jia:2006gw}.

\subsubsection{Large-$N$ limit}

At large $N$, intermediate bound states dominate the chromo-electric polarizability, while continuum contributions are expected to account for only about $20\%$ of the total, analogously to the hydrogen atom. Using
\begin{equation}
\sum_{\nu \geq 2}
\frac{
\langle 1S_s|r^k|\nu P_a\rangle
\langle \nu P_a|r^k|1S_s\rangle
}
{E_{a,\nu}-E_{s}}
=
-2\mu
\langle 1S_s|f^kr^k|1S_s\rangle \,,
\end{equation} where we neglect the contribution from the continuum,
one obtains the bounds
\begin{equation}
\frac{\langle 1S_s|r^2|1S_s\rangle}
{\epsilon_B}
\lesssim
\sum_{\nu \geq 2}
\frac{
\langle 1S_s|r^k|\nu P_a\rangle
\langle \nu P_a|r^k|1S_s\rangle
}
{E_{a,\nu}-E_{s}}
\lesssim
\frac{\langle 1S_s|r^2|1S_s\rangle}
{\frac34\epsilon_B}\,.
\end{equation}
Using our analytical result
\begin{equation}
\langle 1S_s|f^kr^k|1S_s\rangle
\xrightarrow[]{N\to\infty}
-\frac{27}{8}a_B^4\,,
\end{equation}   as well as $2\epsilon_B \mu= a_B^{-2}$,
we obtain
\begin{equation}
\frac{81}{32}a_B^2
\lesssim
\langle 1S_s|r^2|1S_s\rangle
\lesssim
\frac{27}{8}a_B^2\,,
\end{equation}
or equivalently,
\begin{equation}
\frac{10.1}
{(m_Q\alpha_dN)^2}
\lesssim
\langle 1S_s|r^2|1S_s\rangle
\lesssim
\frac{13.5}
{(m_Q\alpha_dN)^2}\,.
\end{equation}
A representative value is therefore
\begin{equation}
\langle 1S_s|r^2|1S_s\rangle
\approx
\frac{11.8}
{(m_Q\alpha_dN)^2}\,.
\end{equation}
This should be compared with the more accurate large-$N$ result of Ref.~\cite{Cohen:2011cw},
\begin{equation}
\langle 1S_s|r^2|1S_s\rangle
\approx
\frac{21.5}
{(m_Q\alpha_dN)^2}
=
\frac{21.5}
{(2m_Q\alpha_B)^2}\,,
\end{equation}
showing that our estimate differs by approximately a factor of two. We can therefore improve our large-$N$ estimate by adopting the more precise values
\begin{equation}
a_B\simeq
\frac{0.876}{\alpha_Bm_Q}\,,
\qquad
\epsilon_B\simeq
0.21\,\alpha_B^2m_Q\,,
\end{equation}
obtained in Ref.~\cite{Cohen:2011cw} along with their $\langle 1S_s|r^2|1S_s\rangle$. This gives
\begin{equation}
d_2
\approx
\frac{2}{N}
\frac{4\pi\alpha_B}{a_B^3}
\frac16
\frac{\langle 1S_s|r^2|1S_s\rangle}
{\epsilon_B}
\approx
\frac{1}{N}
\frac{\pi}{0.876^3}
\frac13
\frac{21.5}{0.21}
\approx
\frac{160}{N}\,.
\end{equation}
By comparison, our analytical estimate gives
\begin{equation}
d_2\simeq \frac{28}{N}\,,
\end{equation}
corresponding to a difference of approximately a factor of six. This discrepancy is driven by the cumulative errors in our estimates of
$a_B$, $\epsilon_B$, and $\langle r^2\rangle$, which are individually modest but add up constructively. Nevertheless, the resulting error in the baryonic polarizability itself is only of a factor of four, since our estimate overpredicts the binding energy by roughly a factor of two while underpredicting $\langle r^2\rangle$ by a similar amount.

\section{Spectral reconstruction of the retarded interaction potential}
\label{app:potentail_sum}

In this appendix, we evaluate the full  retarded ($n=2$) potential obtained from the spectral representation of \cref{eq:spectral_function_potential_def} and compare it with the simplified description adopted in the main text. Our goal is to verify that the potential can be accurately approximated by the sum of the non-perturbative Yukawa contribution generated by the lightest glueball and the perturbative Casimir--Polder interaction.

The spectral representation requires a matching scale $s_0$ separating the non-perturbative and perturbative regimes. Above this scale, the perturbative spectral density is given by~\cite{Fujii:1999xn}
\begin{equation} \label{eq:spectral_density_pertubative}
    \rho_{\rm pt}(m^2_\sigma)= \frac{23}{15}
    \frac{N^2-1}{4\pi^2}
    \left(
    \beta_d
    \frac{\alpha_d(m_\sigma)}{8\pi}
    \right)^2
    m_\sigma^4 \,.
\end{equation}
where the factor $23/15$ accounts for the contribution of the tensor $2^{++}$ two-gluon state.
  Below $s_0$, we approximate the non-perturbative spectral density by the contribution of the lightest glueball alone, as given in \cref{eq:spectral_density_glueball}. A more complete treatment would also include heavier scalar glueball resonances and multi-glueball states, which may generate additional structure in the spectral density near the matching region. Incorporating these contributions would require a detailed model of the full confined spectrum and is beyond the scope of the present work.

To study the resulting potential, we evaluate
\begin{equation}
    I(r)=
     \dfrac{1}{\Lambda_d^7}\left(
\frac{4\pi^2}{b}
\right)^2
    \int dm_\sigma^2\,
    \rho(m_\sigma^2)
    \frac{e^{-m_\sigma r}}{4\pi r}\,,
\end{equation}
and compare it with the sum of the normalized Yukawa and Casimir--Polder potentials. The latter can be found by simply neglecting the dark gauge coupling evolution in \cref{eq:spectral_density_pertubative}, $\alpha_d(m_\sigma) \to \alpha_d$, and integrating from $m_\sigma=\sqrt{s_0}$ to infinity while the former simply arises if you integrate $\rho_\theta(m_\sigma)$ over $m_\sigma=m_{\rm DG}$. Note that the fact that we start the perturbative integration at a non-zero $m_\sigma$ automatically turns the Casimir--Polder potential negligible for $r \gg 1/\sqrt{s_0}$.\footnote{Integrating from zero to infinity would lead to the standard Casimir-Polder potential without suppression.}

The matching scale $s_0$ can be determined through quark--hadron duality\footnote{Note that we remove the factor $23/15$ as we are only comparing the spectral densities of scalar quanta; however, including it has a negligible consequence as it only changes $\sqrt s_0$ by less than 10\%. }~\cite{Fujii:1999xn,Shifman:2000jv},
\begin{equation}
    \int_{m_{\rm DG}^2}^{s_0}
    dm_\sigma^2\,
    \rho_\theta(m_\sigma^2)
    =
    \int_{m_{\rm DG}^2}^{s_0}
    dm_\sigma^2\,
   \dfrac{15}{23} \rho_{\rm pt}(m_\sigma^2)\,.
\end{equation}
We obtain
\begin{equation}
    \sqrt{s_0}\sim 12\,\Lambda_d \,.
\end{equation}    More precisely, $\sqrt{s_0}/\Lambda_d\approx12.9,12.5,12.2$,   for $\SU(2)$, $\SU(3)$ and $\SU(10)$, respectively. However, the resulting potential is essentially unchanged if one instead chooses $\sqrt{s_0}\sim m_{\rm DG}$. This insensitivity originates from the hierarchy
\begin{equation}
    s_0>m_{\rm DG}^2\gg\Lambda_d^2\,,
\end{equation}
which implies that a possible strong running of $\alpha_d$, which happens near the scale $\Lambda_d$, is removed from the spectral integration.

\begin{figure}[t]
\centering \includegraphics[width=0.32\linewidth]{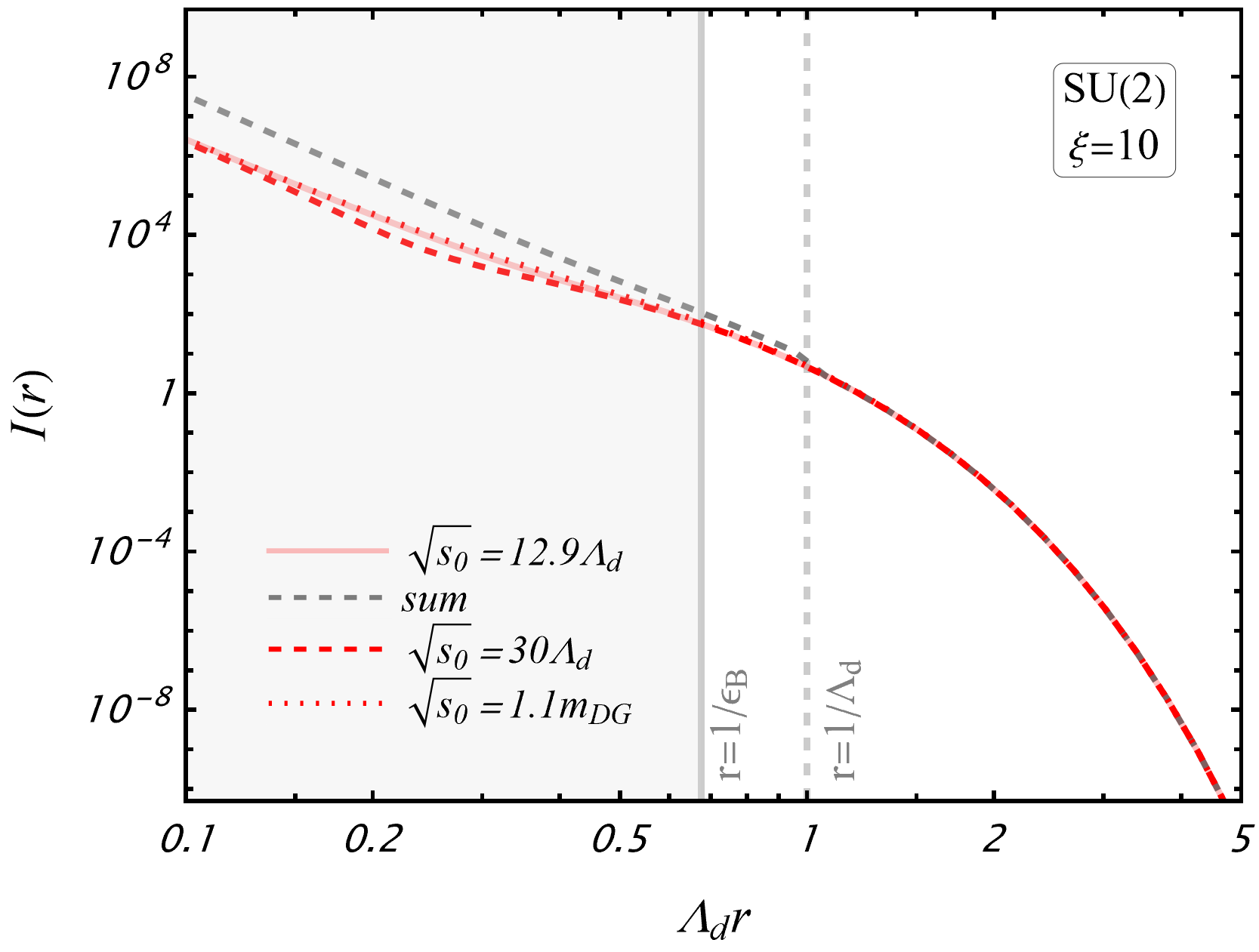}\includegraphics[width=0.32\linewidth]{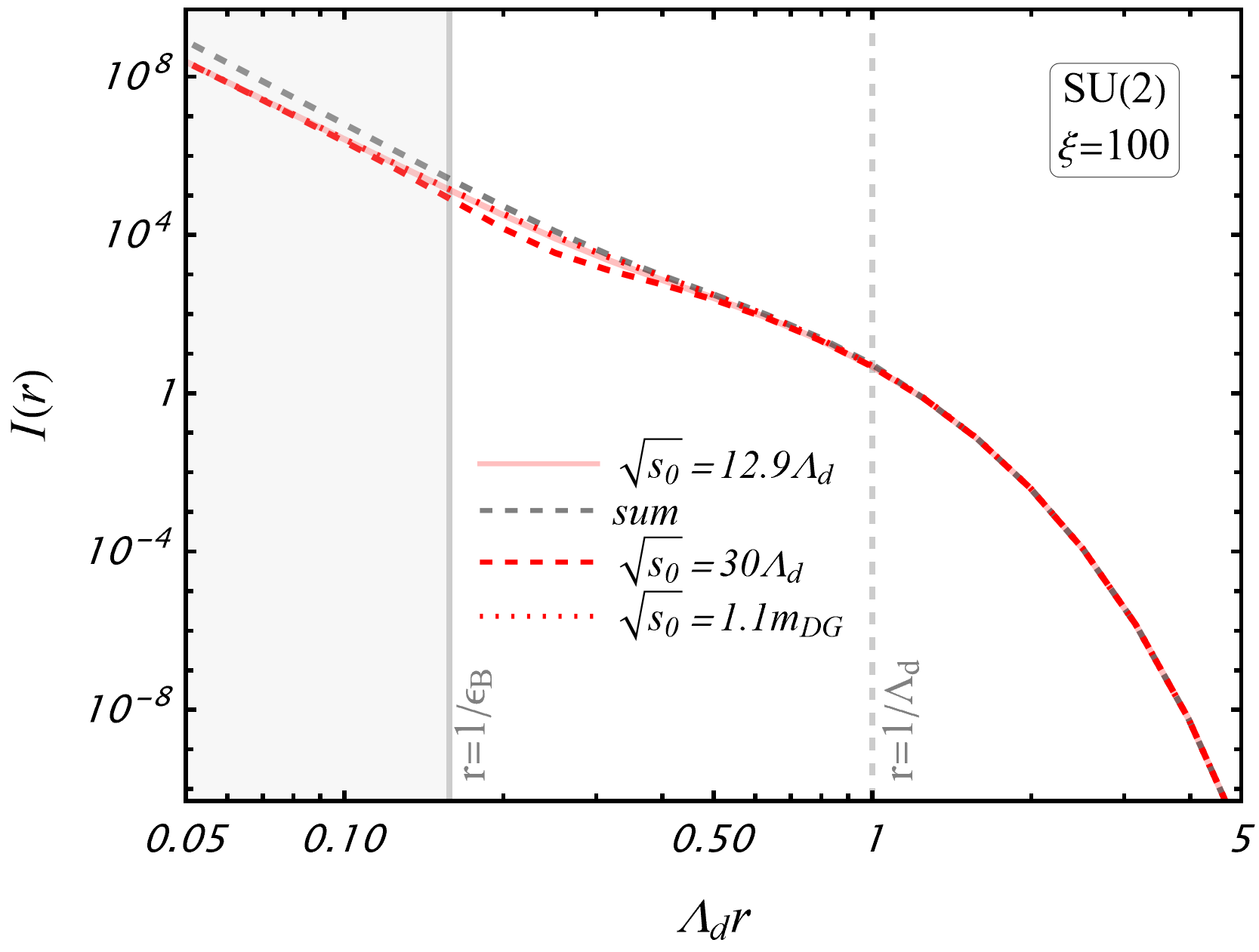}\includegraphics[width=0.32\linewidth]{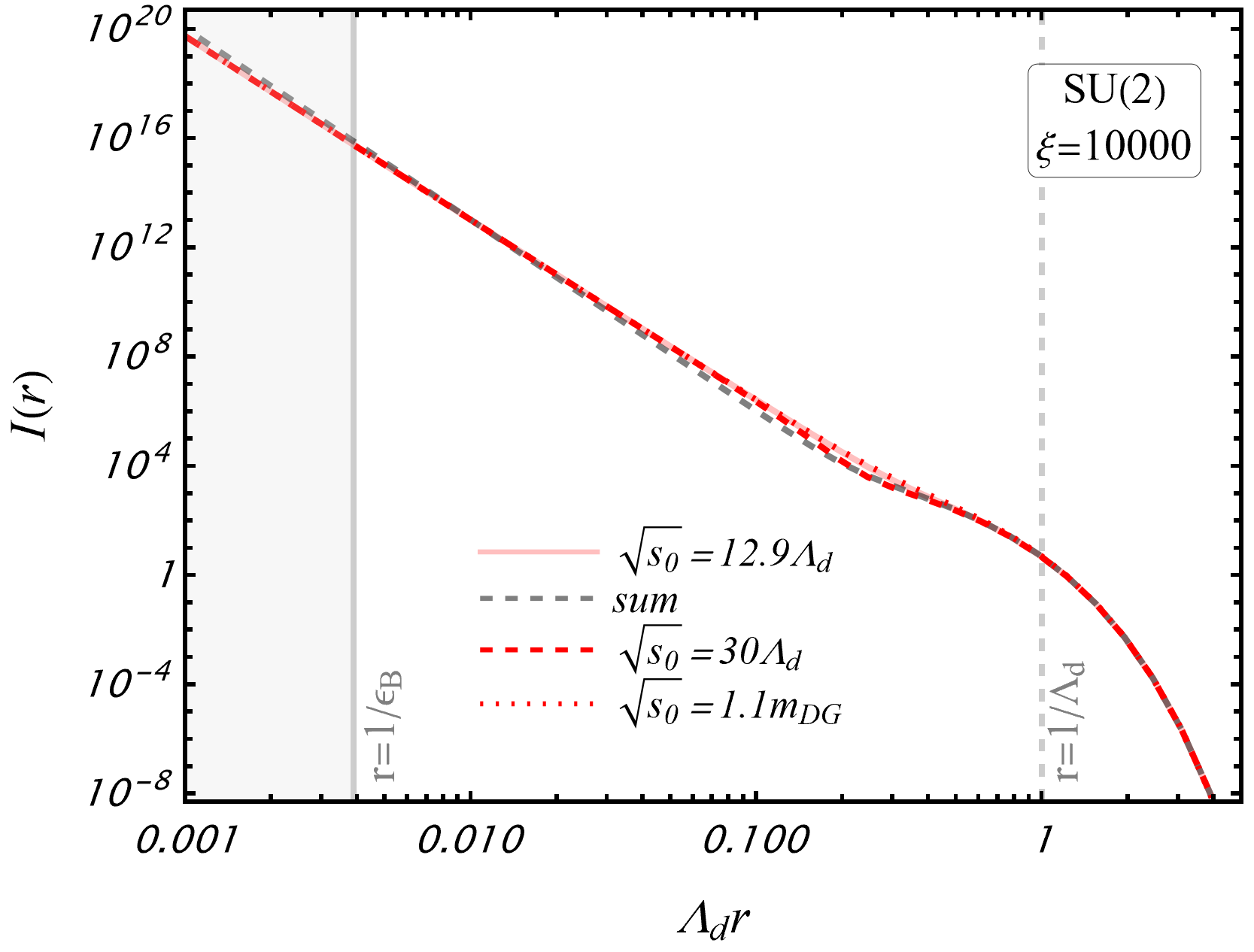}
             \includegraphics[width=0.32\linewidth]{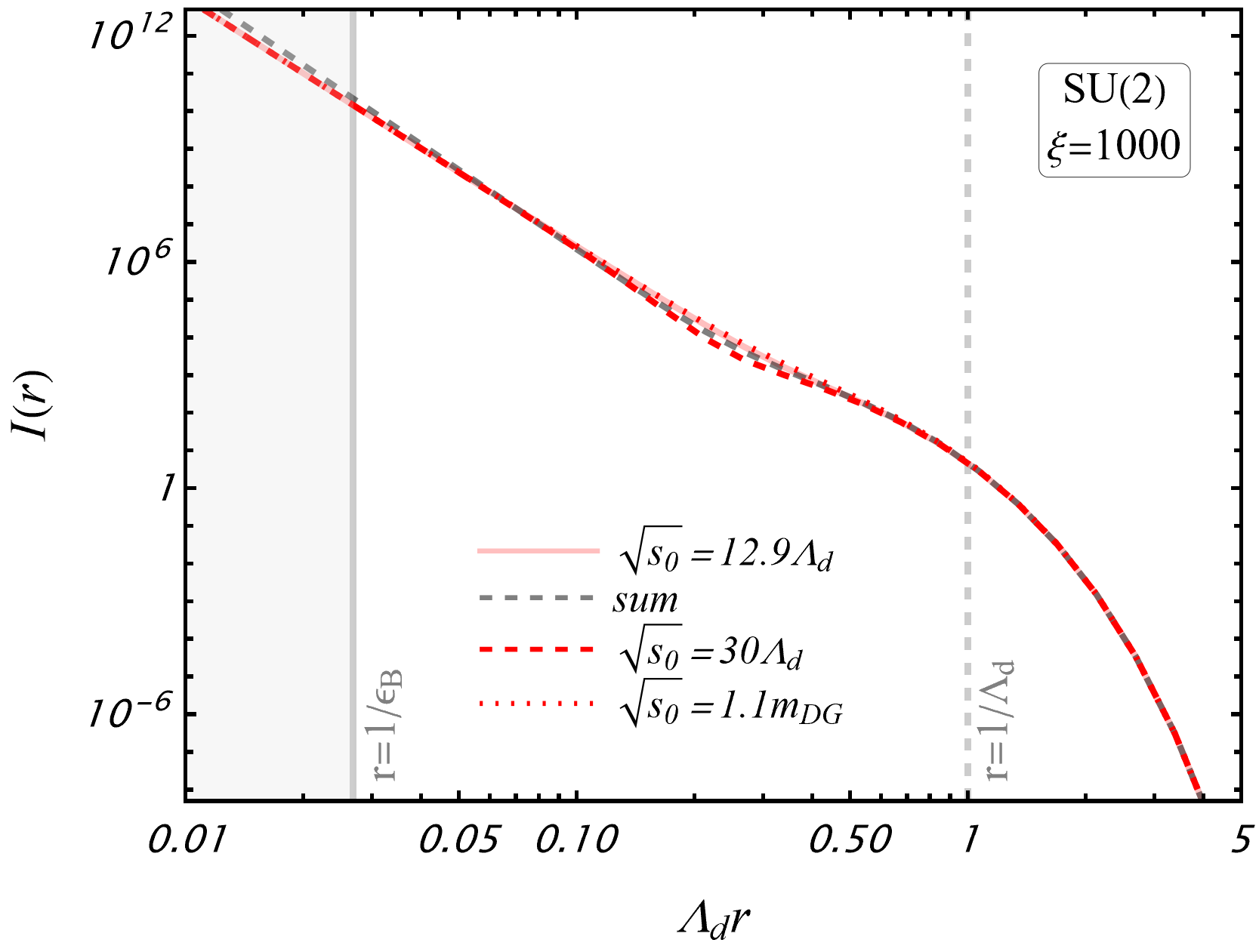}\includegraphics[width=0.32\linewidth]{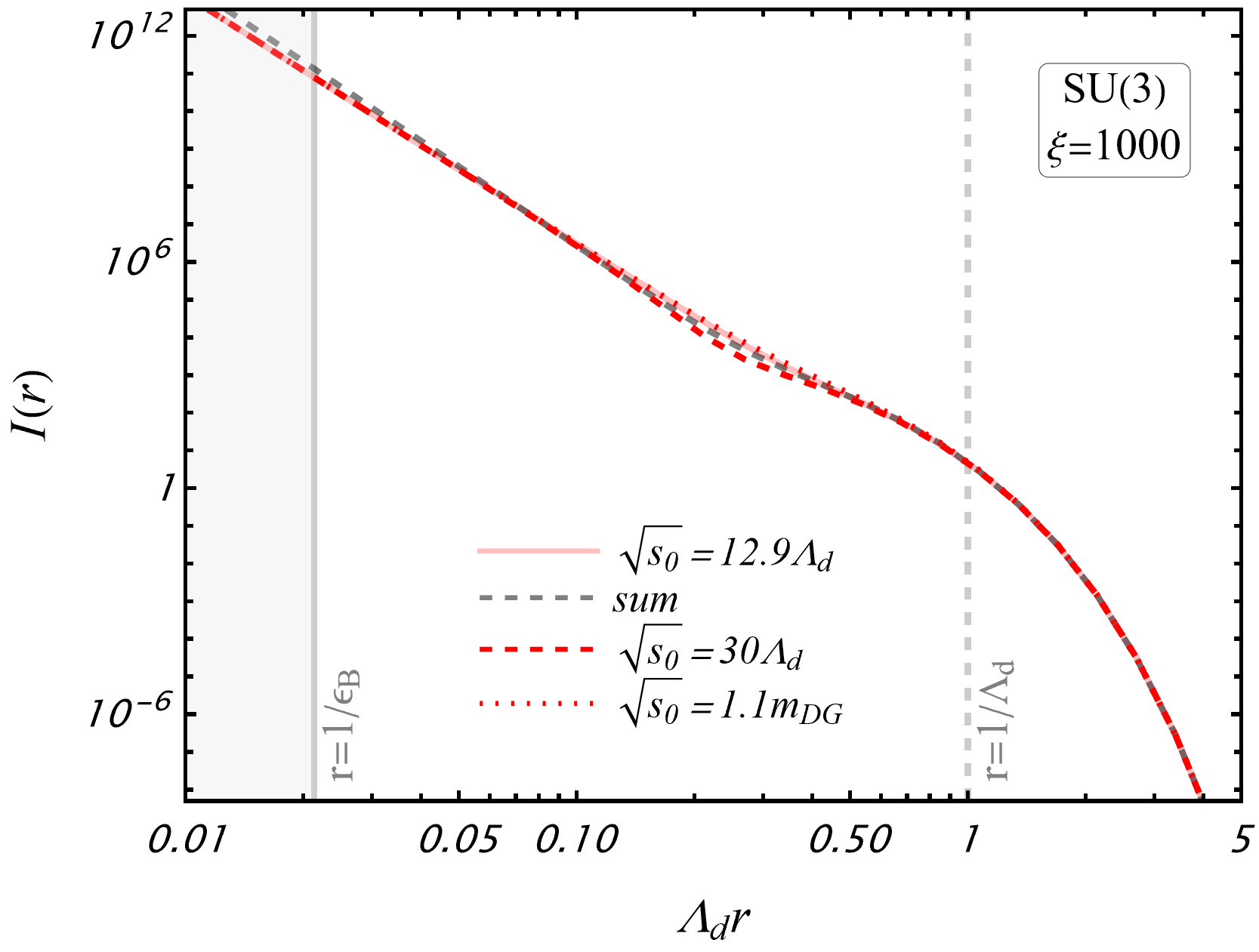}\includegraphics[width=0.32\linewidth]{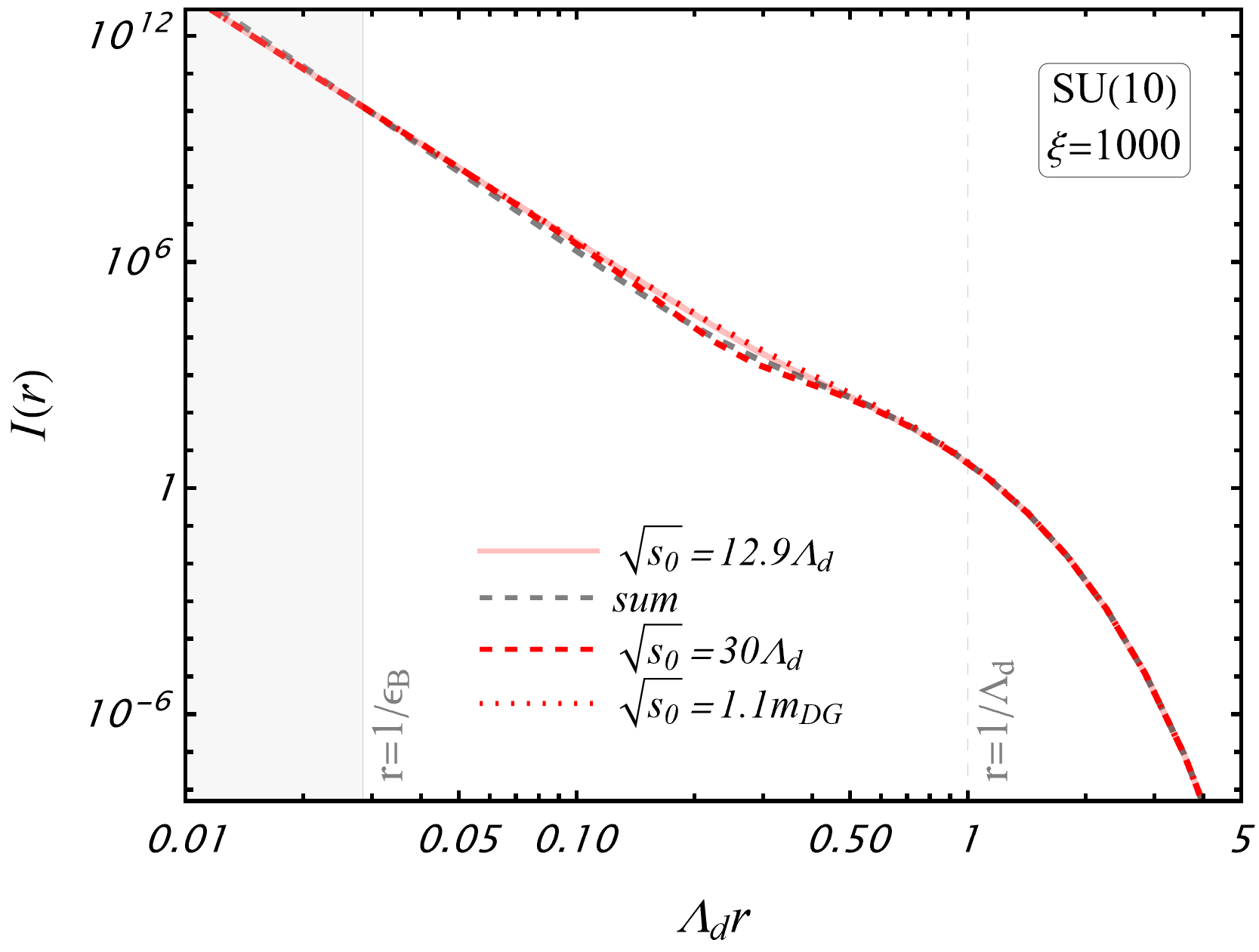} \caption{
 Top: Numerical evaluation of the spectral integral for $\SU(2)$ for  three different values of the hierarchy parameter (left to right) $
\xi = 10,\;100,\;10^4$. The dotted, solid, and dashed red curves correspond to $\sqrt{s_0}=1.1m_{\rm DG}$, $12\,\Lambda_d$, and $30\,\Lambda_d$, respectively. The dashed black curves show the approximation obtained from the sum of the Yukawa potential and the suppressed Casimir--Polder $r^{-7}$ interaction. The vertical dashed line marks $r=\Lambda_d^{-1}$, where the Casimir--Polder contribution becomes negligible. The shaded region corresponds to $r<\epsilon_B^{-1}$, where the spectral representation based solely on the leading multipole coefficient $d_2$ is no longer reliable and higher-order coefficients $d_n$ must be included. The near-indistinguishability of all curves in the region of validity demonstrates both the weak dependence on the matching scale $s_0$ and the accuracy of the Yukawa plus Casimir--Polder approximation. Bottom: same as top but for the three different numbers of colors (left to right) $N=2,3,10$ with $\xi=10^3$. Note that the conclusions for $\SU(2)$ also hold for   $\SU(3)$ and $\SU(10)$.
}
    \label{fig:spectral_vs_sum}
\end{figure}

The numerical results are shown in \cref{fig:spectral_vs_sum}.  Several values of $s_0$ are displayed together with the sum of the Yukawa and the Casimir--Polder potentials, where the latter is truncated at $r= \Lambda_d^{-1}$. Throughout the range of validity of the spectral representation of   \cref{eq:spectral_function_potential_def}, $ \epsilon_B^{-1}\lesssim r$, all curves are nearly indistinguishable. Therefore, the full spectral integral is accurately reproduced by the truncated sum of the two asymptotic contributions.

The figures also illustrate the relative importance of the two contributions. For moderate hierarchies,
\begin{equation}
    \xi\equiv \frac{m_Q}{\Lambda_d}
    \ll10^3\,,
\end{equation}
the Yukawa contribution dominates throughout the phenomenologically relevant region. For larger values of $\xi$, the Casimir--Polder interaction becomes increasingly important at intermediate distances, while the Yukawa contribution continues to control the long-distance behavior.

Overall, the numerical reconstruction confirms that the potential derived from the full spectral representation is accurately reproduced by the sum of a Casimir--Polder interaction and a Yukawa interaction throughout the region $\epsilon_B^{-1}\lesssim r\lesssim \Lambda_d^{-1}$,
while for $ r\gtrsim \Lambda_d^{-1}$,
the potential is well approximated by a pure Yukawa form. Furthermore, the dependence on the matching scale $s_0$ is negligible. This validates the approximation employed in the main text and shows that the full spectral reconstruction provides no appreciable correction to the simple potential obtained by summing the Yukawa and Casimir--Polder contributions.

\end{document}